\documentclass[a4paper,11pt]{article}
\pdfoutput=1 

\usepackage{jheppub} 

\usepackage{amsmath}
\usepackage{amssymb}
\usepackage{braket}
\usepackage{comment}
\usepackage{tensor}
\usepackage{color}
\usepackage{nccmath}
\usepackage{bigints}
\usepackage{pifont}
\usepackage{tikz}
\usepackage{tikz-cd}
\usepackage{simpler-wick}
\usepackage{soul}

\tikzset{
    partial ellipse/.style args={#1:#2:#3}{
        insert path={+ (#1:#3) arc (#1:#2:#3)}
    }
}
\newcommand*{\Scale}[2][4]{\scalebox{#1}{\ensuremath{#2}}}%

\newcommand{\cmark}{\ding{51}}%
\newcommand{\xmark}{\ding{55}}%

\newcommand{\rvline}{\hspace*{-\arraycolsep}\vline\hspace*{-\arraycolsep}}

\title{\boldmath Solving AdS$_3$ string theory at minimal tension: tree-level correlators}

\author[a]{Andrea Dei,}
\author[b,c]{Bob Knighton,}
\author[c]{Kiarash Naderi}
\affiliation[a]{Enrico Fermi Institute \& Kadanoff Center for Theoretical Physics,\\ \hspace*{0.3cm} University of Chicago, Chicago, IL 60637, USA}
\affiliation[b]{Department of Applied Mathematics \& Theoretical Physics, University of Cambridge,\\
\hspace*{0.3cm}Wilberforce Road, Cambridge CB3 0WA, United Kingdom}
\affiliation[c]{Institut f\"ur Theoretische Physik, ETH Z\"urich,\\ 
\hspace*{0.3cm} Wolfgang-Pauli-Strasse 27, 8093 Z\"urich, Switzerland}
\emailAdd{adei@uchicago.edu}
\emailAdd{rik23@cam.ac.uk}
\emailAdd{knaderi@phys.ethz.ch}

\abstract{We revisit the minimal tension ($k=1$) string theory on $\text{AdS}_3\times\text{S}^3\times\mathbb{T}^4$. We propose a new free-field description of the worldsheet theory and show how localization of string amplitudes emerges from the path integral. We exemplify our proposal by reproducing the worldsheet partition function of the $\mathfrak{psu}(1,1|2)_1$ WZW model and providing explicit expressions for spectrally-flowed vertex operators and DDF operators. We compute string correlators in the path integral formalism and obtain a precise tree-level match with correlation functions of the boundary symmetric orbifold. 
}

\begin{document}

\maketitle

\section{Introduction} 
\label{sec:introduction}

The AdS/CFT correspondence is often described as a strong-weak duality: theories of strongly coupled strings are dual to weakly coupled conformal field theories and vice versa the weakly coupled regime of string theory corresponds to the strongly coupled one on the boundary. The locution `strongly coupled', when referred to string theories, should not be intended as a synonym of `intractable' or `difficult to solve' and in the following we will frequently prefer the adjective `tensionless'.\footnote{Recall that the string coupling $\alpha'$ is inversely proportional to the string tension.} It is certainly true that in the extreme case of free boundary CFTs, formalisms based on a perturbative expansion in the bulk coupling $\alpha'$ may not be reliable and one may thus expect little control over the strongly coupled gravity dynamics. On the other hand, free CFTs feature a large collection of conserved currents and one should expect to recover some evidence of the associated symmetries also in the bulk. Indeed, the tensionless regime of string theory has been described as an `un-higgsed phase' of string theory, where additional symmetries become manifest~\cite{Gross:1988ue, Witten:1988zd, Moore:1993qe, Sagnotti:2011jdy}. More concretely, in the tensionless regime the leading Regge trajectory of AdS string theory gives rise to a tower of massless higher spin states, capturing the large collection of currents populating the boundary free CFT \cite{Sundborg:2000wp, Wittentalk, Mikhailov:2002bp, Sezgin:2002rt, Gaberdiel:2014cha, Gaberdiel:2015uca, Ferreira:2017pgt, Gaberdiel:2017oqg}. 

\smallskip

While supergravity techniques cannot be employed in the tensionless regime, the bulk dynamics may still be accessible via a worldsheet formulation of string theory. This is for instance the case of pure NS-NS AdS$_3$ string theory \cite{Teschner:1997ft, Giveon:1998ns, deBoer:1998gyt, Kutasov:1999xu, Teschner:1999ug, Maldacena:2000hw, Maldacena:2000kv, Maldacena:2001km}. In particular, the precise string background dual to the symmetric product orbifold has been identified and an exact incarnation of the AdS$_3$/CFT$_2$ duality has been formulated \cite{Eberhardt:2018ouy}, 
\begin{equation}
\begin{tikzpicture}[baseline = -0.6ex]
\node[inner sep=0pt] at (1.5,0.5)
   {\large{Pure NS-NS strings on}};
\node[inner sep=0pt] at (1.5,0)
{$\text{AdS}_3 \times \text{S}^3 \times \mathbb T^4 $}; 
   \node[inner sep=0pt] at (1.5,-0.5)
{with $k=1$}; 
   
\node[inner sep=0pt] at (5.2,-0.1)
   {$\Scale[2]{\iff} $};    

\node[inner sep=0pt] at (7.8,0)
   {{$\text{Sym}^K (\mathbb T^4)$ }};  
\end{tikzpicture}    \quad \ . 
\label{ads3/cft2}
\end{equation}
Although the symmetric orbifold of $\mathbb T^4$ is strictly speaking not a free field theory --- but an orbifold thereof --- it features a huge collection of conserved currents \cite{Gaberdiel:2014cha, Gaberdiel:2015mra}. The holographic pair \eqref{ads3/cft2} thus provides a concrete incarnation of tensionless strings being dual to free CFTs. Notice, that while it has been known for a long time that the CFT$_2$ dual to AdS$_3$ strings lives on the same moduli space of the symmetric product orbifold of $\mathbb T^4$, the duality \eqref{ads3/cft2} predicts the match of \emph{all} observables, whether they are protected or not.

\medskip

Before discussing the outcomes of the manuscript, let us briefly recap some important results in the study of tensionless AdS$_3$/CFT$_2$. The plethora of results recently derived effectively amounts to a proof of the duality: the all-loop perturbative spectrum was derived from the worldsheet and matched with the symmetric orbifold \cite{Gaberdiel:2018rqv, Eberhardt:2018ouy, Eberhardt:2019ywk}, string partition functions on BTZ black hole, thermal AdS$_3$, conical defects and wormhole geometries were computed and related to the corresponding quantities on the boundary \cite{Eberhardt:2020bgq, Eberhardt:2021jvj}. Boundary states have also been investigated \cite{Gaberdiel:2021kkp} and conformal perturbation theory off the symmetric product orbifold point has been recovered from the bulk \cite{Fiset:2022erp}. It was shown in \cite{Eberhardt:2019ywk, Eberhardt:2020akk, Dei:2020zui, Knighton:2020kuh} that correlators are delta-function localized to configurations where the worldsheet covers the boundary sphere, thus explaining the basic mechanism underlying the duality. Correlators of descendant fields were computed and matched with the boundary CFT expectation in~\cite{Bertle:2020sgd, Gaberdiel:2021njm}. 

\smallskip

Many of these results relied on a free-field realization of the worldsheet chiral algebra in terms of symplectic bosons and free fermions.  Specifically, the symplectic bosons and free fermions together generate the algebra $\mathfrak{u}(1,1|2)_1$, which is reduced to $\mathfrak{psu}(1,1|2)_1$ upon gauging a null $\mathfrak{u}(1)$ symmetry of the theory.\footnote{One should also pick an appropriate reality condition on the fields. Different reality conditions correspond to different signatures in the target space.} While the symplectic boson representation of $\mathfrak{psu}(1,1|2)_1$ is extremely useful, the gauge-fixing from $\mathfrak{u}(1,1|2)_1$ to $\mathfrak{psu}(1,1|2)_1$ comes with a few complications. First, the gauge-fixing itself is rather complicated, and the BRST structure is significantly more involved than the usual prescription in gauged WZW models \cite{Gaberdiel:2022als}. Furthermore, the prescription for calculating correlation functions involved the introduction of extra fields (so-called $W$-fields) which are vacua with respect to $\mathfrak{psu}(1,1|2)_1$ but not with respect to $\mathfrak{u}(1,1|2)_1$~\cite{Dei:2020zui}. Introducing these fields is necessary to obtain a sensible and non-vanishing result, but their physical significance is unclear. In summary, while the present formulation of the worldsheet theory allows for many explicit computations and accurate checks of the duality, it also contains various technical hurdles which hinder a direct derivation of correlation functions and seem to have no precise meaning from the boundary CFT$_2$ perspective. 

\medskip 

In this paper we explain how some of these difficulties can be bypassed. In Section~\ref{sec:worldsheet theory}, we consider the alternative free-field realization of \cite{Beem:2023dub}, which does not involve any gauging of null currents. This alternative free-field realization consists of one $\beta \gamma$ system and two $b c$ systems. After reviewing that these free fields generate the full chiral algebra $\mathfrak{psu}(1,1|2)_1$, we recover the representations previously derived in \cite{Eberhardt:2018ouy} in terms of symplectic bosons and free fermions. While we present this alternative free-field realization as being independent of the symplectic boson theory, in fact the two are intimately related, see \cite{Beem:2023dub}. We also explain its relation with the Wakimoto realization of $\mathfrak{sl}(2, \mathbb R)$, which has been exploited multiple times in the study of the AdS$_3$/CFT$_2$ duality (see for example \cite{Giveon:1998ns,Giveon:1999zm}). In Section~\ref{sec:worldsheet theory} we also explain the geometric interpretation of the alternative worldsheet fields we discuss: while the worldsheet $\beta \gamma$ system can be interpreted as position coordinate and associated derivatives on the boundary sphere, the two $b c$ systems realize on the worldsheet the spacetime $\mathcal N=2$ superspace coordinates. In \cite{Gaberdiel:2022als} the $\mathcal N=4$ algebra of Berkovits-Vafa-Witten \cite{Berkovits:1999im} was obtained in terms of the symplectic boson realization of $\mathfrak{psu}(1,1|2)_1$ we discussed above. In order to remove terms containing the null field mentioned above, the authors introduced additional ghosts and gauged them out. We conclude Section~\ref{sec:worldsheet theory} by explaining how, as suggested in \cite{Beem:2023dub}, in terms of the alternative free-field realization these additional ghost fields are no longer necessary and all $\mathcal N=4$ algebra generators can directly be written in terms of the newly introduced free fields. 

\smallskip

The utility of these alternative free fields is exemplified in Section~\ref{sec:The worldsheet spectrum}, where we rederive the spectrally-flowed characters of the $\mathfrak{psu}(1,1|2)_1$ WZW model, as well as the worldsheet partition function of \cite{Eberhardt:2018ouy}, recap the physical state conditions and rewrite explicit expressions for spacetime symmetry generators and DDF operators in our variables \cite{Eberhardt:2019qcl, Giveon:1998ns, Naderi:2022bus}. We explain in Section~\ref{sec:point at infinity} how localization of string amplitudes emerges from the path integral. It was noticed in \cite{Eberhardt:2019ywk} that in order for correlation functions on the worldsheet to develop poles only where vertex operators are inserted --- this is one of the basic axioms of 2D CFTs \cite{Goddard:1989dp, Gaberdiel:1998fs} --- it is necessary to introduce additional vertex operators dubbed `secret representations'. These fields behave as the vacuum for the $\mathfrak{psu}(1,1|2)_1$ generators but give rise to non-trivial representations when acted on by the worldsheet free fields. Section~\ref{sec:point at infinity} illustrates the geometric origin of the secret representations: when computing correlators on Euclidean AdS$_3$ they arise from the failure of the worldsheet coordinate $\gamma$ to cover the boundary sphere with a unique patch. 

\smallskip

The drastic simplification at the orbifold point of the boundary CFT$_2$ hints at a corresponding streamlining for the worldsheet bulk theory. Optimistically, since closed formulae for arbitrary CFT$_2$ correlators at the orbifold point are well-known \cite{Arutyunov:1997gt, Arutyunov:1997gi, Jevicki:1998bm, Lunin:2000yv, Lunin:2001pw, Pakman:2009zz, Roumpedakis:2018tdb, Dei:2019iym, Hikida:2020kil, Burrington:2022dii, Jia:2023ais} and the symmetric orbifold theory is essentially solved, one may suspect that also a perturbative solution of the bulk string theory may be within reach. Indeed, building on \cite{Eberhardt:2018ouy, Eberhardt:2019ywk, Dei:2020zui}, in Section~\ref{sec:correlation-functions} we compute from the worldsheet $n$-point functions of twisted ground states at tree level and exactly reproduce the dual symmetric orbifold expectation. The holographic match of higher-genus correlators will appear in \cite{higher-genus-paper}. This is the first time (non-protected) correlators of the symmetric orbifold are reproduced from the bulk in terms of the hybrid formalism of~\cite{Berkovits:1999im}.\footnote{Tree level four-point functions of $w$-twisted ground states were obtained in \cite{Dei:2022pkr} as a $k=1$ limit of bosonic AdS$_3$ string correlators.} We conclude in Section~\ref{sec:conclusions} with a discussion of some points deserving further investigation and a list of future directions. A number of appendices contain technical derivations or background material which will help the reader at various points in the main text. 

\section{The worldsheet theory}
\label{sec:worldsheet theory}

String theory on $\text{AdS}_3\times\text{S}^3 \times \mathbb T^4$ with pure NS-NS flux can be described in terms of the hybrid formalism \cite{Berkovits:1999im}, based on the Wess-Zumino-Witten model on the supergroup $\text{PSU}(1,1|2)$. This supergroup arises naturally in the description of the $\mathcal{N}=1$ superspace of $\text{AdS}_3\times\text{S}^3$, and captures the left- and right-super-isometries of the $\text{AdS}_3\times\text{S}^3$ superspace.\footnote{Strictly speaking, since $\pi_1(\text{SL}(2,\mathbb{R}))\cong\mathbb{Z}$, the above space is not simply-connected. We will therefore always implicitly assume that we are working with the universal covers $\widetilde{\text{SL}(2,\mathbb{R})}$ and $\widetilde{\text{PSU}(1,1|2)}$ so as to describe global $\text{AdS}_3$.} Since $\text{PSU}(1,1|2)$ is a matrix supergroup of dimension $6|8$, we can represent its elements faithfully as matrices. A convenient parametrization is the block decomposition
\begin{equation}
\text{PSU}(1,1|2)\cong
\begin{pmatrix}
\text{SL}(2,\mathbb{R}) & \rvline & \text{supercharges}\\
\hline
\text{supercharges} & \rvline & \text{SU}(2)
\end{pmatrix} \,.
\end{equation}
The WZW model on $\text{PSU}(1,1|2)$ at level $k$ is described in terms of the current algebra $\mathfrak{psu}(1,1|2)_k$, generated by six bosonic currents $J^a,K^a$ and eight fermionic currents $S^{\alpha\beta\gamma}$ on the worldsheet, satisfying the following OPEs\footnote{Here $a$ is a vector index taking values in $a \in \{ -, 3 , +\}$ while $\alpha, \beta, \gamma \in \{ -, + \}$.}

\begingroup
\allowdisplaybreaks
\begin{align}
J^a(z)J^b(w)&\sim\frac{k\widetilde{\kappa}^{ab}}{(z-w)^2}+\frac{\widetilde{f}\indices{^a^b_c}J^b(w)}{z-w}\,, \nonumber \\
K^a(z)K^b(w)&\sim\frac{k\kappa^{ab}}{(z-w)^2}+\frac{f\indices{^a^b_c}K^c(w)}{z-w}\,,\nonumber \\
J^a(z)S^{\alpha\beta\gamma}(w)&\sim\frac{(\widetilde{\sigma}^a)\indices{^\alpha_\delta}S^{\delta\beta\gamma}(w)}{z-w}\,, \label{eq:psu-current-algebra} \\
K^a(z)S^{\alpha\beta\gamma}(w)&\sim\frac{(\sigma^a)\indices{^\beta_\delta}S^{\alpha\delta\gamma}(w)}{z-w}\,, \nonumber \\
S^{\alpha\beta+}(z)S^{\gamma\delta-}(w)&\sim\frac{k\,\varepsilon^{\alpha\gamma}\varepsilon^{\beta\delta}}{(z-w)^2}+\frac{\varepsilon^{\alpha\gamma}(\sigma_a)\indices{^\beta^\delta}K^a(w)-\varepsilon^{\beta\delta}(\widetilde{\sigma}_a)^{\alpha\gamma}J^a(w)}{z-w}\,, \nonumber 
\end{align}
\endgroup
where $\widetilde{\kappa}^{ab}$, $\widetilde{f}\indices{^a^b_c}$ and $\kappa^{ab}$, $f\indices{^a^b_c}$ are the killing form and structure constants of $\mathfrak{sl}(2,\mathbb{R})$ and $\mathfrak{su}(2)$ respectively, and $\varepsilon^{+-}=-\varepsilon^{-+}=1$. See Appendix~\ref{app:psu-u-algebras} for our conventions. It was pointed out in \cite{Eberhardt:2018ouy} that at the minimal level $k=1$ a shortening condition occurs and the only allowed $\mathfrak{psu}(1,1|2)_1$ representations take the form  
\begin{equation} 
\begin{tabular}{ccc}  & $(\mathcal C_\lambda^{\frac{1}{2}}, \boldsymbol 2)$ & \\
$(\mathcal C_{\lambda+\frac{1}{2}}^{1}, \boldsymbol 1)$ & & $(\mathcal C_{\lambda+\frac{1}{2}}^0, \boldsymbol 1)$
\end{tabular}
\label{short-rep}
\end{equation}
where $\mathcal C^j_\lambda$ denotes an $\mathfrak{sl}(2, \mathbb R)$ continuous representation with spin $j$ and fractional part of $\lambda \mod 1$, while $\boldsymbol 1$ and $\boldsymbol 2$ stand for $\mathfrak{su}(2)$ representations of dimension $1$ and 2 respectively.

\subsection{An alternative free-field realization}
\label{sec:new-free-field-realization}

The shortening condition of the $\text{PSU}(1,1|2)$ model at $k=1$ hints at a simplification of the theory at minimal level. In \cite{Eberhardt:2018ouy,Dei:2020zui} it was pointed out that this simplification takes the form of a free-field realization of the current algebra $\mathfrak{psu}(1,1|2)_1$ in terms of two pairs of so-called symplectic bosons and two pairs of free fermions \cite{Gaiotto:2017euk}. As discussed in the Introduction, this comes with various technical difficulties. We discuss an alternative free-field realization of $\mathfrak{psu}(1,1|2)_1$ based on \cite{Beem:2023dub}, which does not require any gauging and avoids most of the complications listed above. 

Consider a commuting $\beta \gamma$ system with scaling dimensions $\Delta(\beta)=1$, $\Delta(\gamma)=0$ and OPEs
\begin{equation}
\beta(z)\gamma(w)\sim-\frac{1}{z-w} \,, 
\label{beta-gamma-OPE}
\end{equation}
and two anti-commuting $b c$ systems $(p_a,\theta^a)$, $a = 1, 2$, with conformal dimensions $\Delta(p_a)=1$, $\Delta(\theta^a)=0$ and OPEs
\begin{equation}
p_a(z)\theta^b(w)\sim\frac{\delta\indices{_a^b}}{z-w}\,.
\label{p-theta-OPE}
\end{equation}
Analogously, we should introduce anti-chiral versions $\overline{\beta},\overline{\gamma},\overline{p}_a,\overline{\theta}^a$ obeying analogous right-moving OPEs. All of these fields are free and can be described by the action
\begin{equation}\label{eq:free-field-action}
S=\frac{1}{2\pi}\int\left(\beta\overline{\partial}\gamma+p_a\overline{\partial}\theta^a+\overline{\beta}\partial\overline{\gamma}+\overline{p}_a\partial\overline{\theta}^a\right)\,,
\end{equation}
where summation over $a=1,2$ is understood. The stress tensor of the theory can be found by demanding that $\beta$ and $p_a$ are conformal primaries of weight $\Delta=1$, and  $\gamma$ and $\theta^a$ are primaries of weight $\Delta=0$. The (chiral) stress tensor is
\begin{equation} \label{eq:system-stress-tensor}
T_{\text{free}}=-\beta\partial\gamma-p_a\partial\theta^a
\end{equation}
and implies that the total central charge is $c=-2$. Note that normal ordering is understood. Let us now consider the currents\footnote{The right-moving currents are constructed by analogous combinations of the anti-chiral free fields.}
\begin{subequations}
\begin{align} \label{eq:sl2-currents}
J^+& =\beta\,, &  J^3 & =\beta\gamma+\tfrac{1}{2}(p_a\theta^a)\,, &  J^-& =(\beta\gamma)\gamma+(p_a\theta^a)\gamma\,,\\
K^+ & =p_2\theta^1\,, & K^3& =-\tfrac{1}{2}(p_1\theta^1)+\tfrac{1}{2}(p_2\theta^2)\,, & K^-& =p_1\theta^2\,, \label{eq:su2-currents}
\end{align}
and
\begin{align}
S^{+++} &=p_2\,, & S^{+-+}& =p_1 \,, & S^{-++} &=-\gamma p_2 \,, & S^{---}&=(\beta\gamma+p_a\theta^a)\theta^2 
 \,, \\
S^{++-}& =\beta\theta^1 \,, & S^{+--} &=-\beta\theta^2\,, &  S^{--+}&=-\gamma p_1 \,, & S^{-+-}& =-(\beta\gamma+p_a\theta^a)\theta^1  \,. \label{eq:psu-super-currents-d}
\end{align}
\label{new-free-field-realization}%
\end{subequations}
It can be checked using the free-field OPEs~\eqref{beta-gamma-OPE} and \eqref{p-theta-OPE} that these currents have conformal weight $\Delta=1$ and that indeed generate the current algebra $\mathfrak{psu}(1,1|2)_1$.\footnote{This and various computations in the manuscript are carried out with the help of the Mathematica package \cite{Thielemans:1991uw}.} In fact, bosonizing the free fields \eqref{beta-gamma-OPE} and \eqref{p-theta-OPE} and making appropriate field identifications, one can verify that eq.~\eqref{new-free-field-realization} is (up to a real form) the $\mathfrak{psl}(2|2)_1$ free field realization of \cite{Beem:2023dub}.  Notice that substituting the free-field realization~\eqref{new-free-field-realization} in the expression for the $\mathfrak{psu}(1,1|2)_1$ stress-tensor, 
\begin{align} \label{eq:psu-stress-tensor}
    T_{\mathfrak{psu}(1,1|2)_1} = &-(J^3 J^3) + \frac{1}{2} [(J^+ J^-) + (J^- J^+)] + (K^3 K^3) + \frac{1}{2} [(K^+ K^-) + (K^- K^+)] \nonumber \\ &+ \frac{1}{2} \epsilon^{\alpha\mu} \epsilon^{\beta \nu} \epsilon^{\gamma \rho} (S^{\alpha\beta\gamma} S^{\mu \nu \rho}) \,, 
\end{align}
one recovers the stress tensor \eqref{eq:system-stress-tensor},
\begin{equation} \label{eq:t-free-t-psu}
    T_{\mathfrak{psu}(1,1|2)_1} = T_{\text{free}} \,.
\end{equation}

In the following sections, it will prove useful to bosonize the free fields \eqref{beta-gamma-OPE} and \eqref{p-theta-OPE} as
\begin{subequations}
\begin{alignat}{2}
    \beta &= e^{\phi+i\kappa} \partial(i\kappa) \,, \qquad \quad &  \gamma & =e^{-\phi-i\kappa} \,, \\
	\theta^1 &= e^{i f_1} \,, \qquad \quad & p_1 & = e^{-i f_1} \,, \\
	\theta^2 &= e^{-i f_2} \,, \qquad \quad & p_2 & = e^{i f_2} \,,
\end{alignat}   
\label{new-free-fields-bosonization}%
\end{subequations}
where $\phi$ and $\kappa$ are free bosons with OPE
\begin{equation}
    \phi(z) \phi(w) \sim -\ln{(z-w)} \,, \qquad \kappa(z) \kappa(w) \sim -\ln{(z-w)} 
\end{equation}
and stress tensor
\begin{equation}
    T_{\phi,\kappa} = -\frac{1}{2} (\partial \phi)^2 - \frac{1}{2} (\partial \kappa)^2 + \frac{1}{2} \partial^2 \phi + \frac{1}{2} \partial^2(i\kappa) \,.
\end{equation}
The bosons $\phi$ and $\kappa$ account for a central charge $c(\phi,\kappa)=2$. In eq.~\eqref{new-free-fields-bosonization} we also introduced the free bosons
\begin{equation}
    f_i(z) f_j(w) \sim - \delta_{i,j} \ln{(z-w)} \,, \qquad i,j\in\{1,2\} \,, 
\end{equation}
whose stress-tensor
\begin{equation}
    T_{f_1,f_2} = -\frac{1}{2} \sum_{j=1}^2 (\partial f_j)^2 + \frac{1}{2} \partial^2(if_1) - \frac{1}{2} \partial^2(if_2) 
\end{equation}
accounts for a central charge $c(f_1, f_2)=-4$.

\subsection{Representations and spectral flow} \label{sec:representation}

We now study representations of the free fields \eqref{beta-gamma-OPE} and \eqref{p-theta-OPE} and see how they give rise to representations of $\mathfrak{psu}(1,1|2)_1$. 

\paragraph{Highest weight representations} The zero-mode algebra of the free-field theory is
\begin{equation}\label{eq:free-field-zero-mode-algebra}
[\beta_0,\gamma_0]=-1\,,\quad\{(p_a)_0,\theta^b_0\}=\delta\indices{_a^b}\,.
\end{equation}
We can find representations of this algebra by defining states $\ket{m}$ such that\footnote{Note that requiring $\mathfrak{psu}(1,1|2)_1$ currents to be periodic, forces to only consider the R-sector of $(\beta,\gamma)$ and $(p_a,\theta^a)$.}
\begin{equation}
\beta_0\ket{m}=m\ket{m+1}\,,\qquad\gamma_0\ket{m}=\ket{m-1}\,,\qquad (p_a)_0\ket{m}=0\,.
\end{equation}
Such a representation is labeled by the fractional part $\lambda + \frac{1}{2}$ of $m$ and the states form the Clifford module
\begin{equation}\label{eq:zero-mode-states}
\text{Span}\left\{\ket{m},\,\theta^a_0\ket{m},\,\theta^1_0\theta^2_0\ket{m}|\,m\in\mathbb{Z}+\lambda + \tfrac{1}{2}\right\}\,.
\end{equation}
The states of the form $\ket{m}$ are annihilated by $K^a_0$ and thus form a singlet with respect to $\mathfrak{su}(2)$. The $\mathfrak{sl}(2,\mathbb{R})$ generators on the other hand act as
\begin{equation}
J^3_0\ket{m}=m\ket{m}\,,\quad J^{+}_0\ket{m}=m\ket{m+ 1}\,,\quad J^-_0\ket{m}=m\ket{m-1}\,.
\end{equation}
Thus, the states $\ket{m}$ transform in the continuous representation $\mathcal{C}_{\lambda+\tfrac{1}{2}}^{j=0}$ under $\mathfrak{sl}(2,\mathbb{R})$. Checking the action of $K^a_0$ and $J^a_0$ on the other states we find that
\begin{equation}
\ket{m}\in\big(\mathcal{C}^{0}_{\lambda+\frac{1}{2}},\boldsymbol{1}\big)\,,\qquad\theta^a_0\ket{m}\in\big(\mathcal{C}^{\frac{1}{2}}_{\lambda},\boldsymbol{2}\big)\,,\qquad\theta^1_0\theta^2_0\ket{m}\in\big(\mathcal{C}^{1}_{\lambda + \frac{1}{2}},\boldsymbol{1}\big)\,,
\label{module}
\end{equation}
where the first and second entries determine the representation of $\mathfrak{sl}(2,\mathbb{R})$ and $\mathfrak{su}(2)$, respectively. We can group these three representations into a single representation of the zero mode algebra \eqref{eq:free-field-zero-mode-algebra}, which takes the form of the Clifford module \eqref{short-rep}. Thus, we see that the zero-mode representations of the free-field realization reproduce the short representations of $\mathfrak{psu}(1,1|2)$. Moreover, one can also check that all the short representation coefficients, which are spelled out in Appendix~\ref{app:algebras-and-reps}, are correctly accounted for. Promoting the zero mode representation to a highest-weight representation of the full mode algebra
\begin{equation}\label{eq:free-field-mode-algebra}
[\beta_{m},\gamma_n]=-\delta_{m+n,0}\,,\quad\{(p_a)_m,\theta^b_n\}=\delta\indices{_a^b}\delta_{m+n,0}\,,
\end{equation}
in turn reproduces the short representations of $\mathfrak{psu}(1,1|2)_1$. Following the notation of~\cite{Eberhardt:2018ouy}, we will refer to the affine representation built on \eqref{eq:zero-mode-states} as $\mathcal{F}_{\lambda}$. Note that there is a subtlety for $\lambda = \frac{1}{2} \mod 1$, since representations become reducible but indecomposable~\cite{Eberhardt:2018ouy}. Let us recap how this comes about. If $\lambda=\frac{1}{2} \mod 1$, then $m\in \mathbb{Z}$. The state $\ket{0}$ in \eqref{module} is annihilated by $J^3_0$, and therefore we can consistently take the quotient by states $\ket{m}$ with non-positive $m\in \mathbb{Z}$ in that representation. This shows that there are highest- and lowest-weight sub-representations of the affine representation $\mathcal{F}_{\lambda}$, which in \cite{Eberhardt:2018ouy} were denoted by $\mathcal{G}_{\pm}$. Moreover, following the same reasoning of Appendix~B of \cite{Eberhardt:2018ouy}, we can see that there are two trivial representations, $\ket{0}$ and $\theta^1_0 \theta^2_0 \ket{0}$.

\paragraph{Spectrally-flowed representations} The free-field realization admits a spectral-flow automorphism $\sigma^w$ which acts on the mode algebra as
\begin{equation}
\begin{aligned}
\sigma^w(\beta_n)& =\beta_{n-w}\,, & \qquad  \sigma^w(\gamma_n) &=\gamma_{n+w}\,, \\
\sigma^w((p_1)_n)& =(p_1)_{n-w}\,, & \qquad \sigma^w(\theta^1_n) &=\theta^1_{n+w}\,,
\end{aligned}
\label{eq:spectral-flow-free-fields}
\end{equation}
and which acts trivially on $p_2$ and $\theta^2$.\footnote{The choice of $p_1$ and $\theta^1$ as the $bc$ system which is affected by spectral flow may appear artificial. In fact, there is a continuum of possible choices, related by global $\text{SU}(2)$ transformations. However, only the choice made in 
eq.~\eqref{eq:spectral-flow-free-fields} is consistent with eq.~\eqref{stress-tensor-flow}. For it, eq.~\eqref{stress-tensor-flow} implies that the modes of any field $F$ on the worldsheet flow according to $\sigma^w(F_n)=F_{n+w(j-h)}$, where $j F = [K^3_0, F]$ and $h F = [J^3_0, F]$. Since $[K^3_0-J^3_0, p_2] = [K^3_0-J^3_0, \theta^2]=0$, we conclude that $p_2$ and $\theta^2$ do not flow.} The effect of $\sigma^w$ on the $\mathfrak{psu}(1,1|2)_1$ currents is
\begingroup
\allowdisplaybreaks
\begin{align}
\sigma^w(J^3_n)&=J^3_n+\frac{w}{2}\delta_{n,0}\,, & \sigma^w(J^{\pm}_n)&=J^{\pm}_{n\mp w}\,, \nonumber \\
\sigma^w(K^3_n)&=K^3_n+\frac{w}{2}\delta_{n,0}\,, & \sigma_w(K^{\pm}_n)&=K^{\pm}_{n\pm w}\,, \label{eq:psu-spectral-flow}
 \\
\sigma^w(S_n^{\alpha\beta+})& =S^{\alpha\beta+}_{n+\frac{1}{2}w(\beta-\alpha)}\,,  & \sigma^w(S_n^{\alpha\beta-})& =S^{\alpha\beta-}_{n+\frac{1}{2}w(\beta-\alpha)}\,, \nonumber
\end{align}
\endgroup
which agrees with the usual notion of spectral flow for $\mathfrak{psu}(1,1|2)_k$ at $k=1$ and implies
\begin{equation}
    \sigma^w(L_0) = L_0 + w(K^3_0 - J^3_0) \,,
    \label{stress-tensor-flow}
\end{equation}
where $L_0$ is the zero mode of the stress-tensor of $\mathfrak{psu}(1,1|2)_1$.

Composing the highest-weight representation $\mathcal{F}_{\lambda}$ with $\sigma^w$ yields a non-highest-weight representation $\sigma^w(\mathcal{F}_{\lambda})$. For example, the spectrally-flowed image $\ket{\psi}^{(w)}$ of a highest-weight state $\ket{\psi}$ will no longer be annihilated by all positive modes of the free fields, but will rather satisfy
\begin{align}
\beta_n\ket{\psi}^{(w)}&=0\,, \quad n>w\,, & \gamma_n\ket{\psi}^{(w)}& =0\,,  \quad n>-w\,,\\
(p_1)_n\ket{\psi}^{(w)}&=0\,, \quad n>w\,, & \theta^1_n\ket{\psi}^{(w)}& =0\,, \quad n>-w\,, \\
(p_2)_n\ket{\psi}^{(w)}&=0\,, \quad n>0 \,, & \theta^2_n\ket{\psi}^{(w)}& =0\,, \quad n>0\,.
\end{align}
Given the agreement with the representation theory of \cite{Eberhardt:2018ouy}, in our worldsheet theory we include the highest-weight representations $\mathcal{F}_{\lambda}$ as well as all spectrally-flowed images $\sigma^w(\mathcal{F}_{\lambda})$. This gives a worldsheet spectrum
\begin{equation}
\mathcal{H}=\bigoplus_{w\in\mathbb{Z}}\int_{0}^{1}\mathrm{d}\lambda\,\sigma^w(\mathcal{F}_{\lambda})\otimes\overline{\sigma^w(\mathcal{F}_{\lambda})}\,. \vspace{3pt}
\end{equation}

\paragraph{The $\boldsymbol x$ basis} As in any 2D CFT, a vertex operator $V(\phi,z)$ can be associated to each state $\phi$ on the worldsheet, 
\begin{equation}
    V(\phi, 0) \ket{0} = \phi \ . 
    \label{m-basis-V}
\end{equation}
While a priori vertex operators depend solely on the worldsheet coordinate $z$, we are eventually interested in studying dual CFT correlation functions, which explicitly depend on the spacetime insertion points. It thus proves useful to introduce a worldsheet chemical potential and define the so-called `$x$-basis' vertex operators
\begin{equation} \label{eq:x-conjugation}
    V(\phi;z,x) = e^{x J^+_0} V(\phi;z,0) e^{-x J^+_0} \ .
\end{equation}
This definition is natural from the spacetime point of view. Admittedly, $J^+_0 = \mathcal{L}_{-1}$ is the translation operator in the dual CFT and $x$ can then be interpreted as the position coordinate on the boundary sphere. 

While only in terms of vertex operators in the $x$-basis it is possible to reproduce correlation functions of the boundary CFT, this is one of the technical difficulties mentioned in the Introduction. In fact, correlation functions of spectrally flowed $x$-basis vertex operators are significantly more involved \cite{Maldacena:2001km, Minces:2005nb, Cagnacci:2013ufa, Eberhardt:2019ywk, Dei:2020zui, Dei:2021xgh, Dei:2021yom, Dei:2022pkr, Iguri:2022pbp, Bufalini:2022toj, Iguri:2023khc} than those computed in terms of the vertex operators \eqref{m-basis-V} \cite{Giribet:2000fy, Stoyanovsky:2000pg, Giribet:2001ft, Ribault:2005wp, Giribet:2005ix, Ribault:2005ms, Giribet:2005mc, Iguri:2007af}.

One of the benefits of the fields \eqref{beta-gamma-OPE} and \eqref{p-theta-OPE} is that they behave quite nicely when conjugated with $\exp(x J^+_0)$. In fact, only $\gamma$ transforms and does so in a simple way, 
\begin{equation}
e^{xJ^+_0}\gamma(z) e^{-xJ^+_0}=\gamma(z)-x\,.
\label{gamma-to-gamma-x}
\end{equation}

\subsection{Relation to the Wakimoto representation} \label{sec:wakimoto}

Readers familiar with the AdS$_3$ string theory literature may notice similarities of the free field realization \eqref{new-free-field-realization} with the Wakimoto realization of $\mathfrak{sl}(2, \mathbb R)_1$ \cite{Wakimoto:1986gf, DiFrancesco:1997nk}. Let us make this comparison precise.

The Wakimoto realization of $\mathfrak{sl}(2,\mathbb{R})_k$ is generated by the action \cite{Giveon:1998ns}
\begin{equation}\label{eq:wakimoto}
S=\frac{1}{2\pi}\int\mathrm{d}^2z\,\left(\frac{1}{2}\partial\Phi\,\overline{\partial}\Phi+\beta\overline{\partial}\gamma+\bar{\beta}\partial\overline{\gamma}-\frac{1}{4k}\beta\bar{\beta}e^{-Q_{\Phi}\Phi}-\frac{Q_{\Phi}}{4}R\Phi\right)\,,
\end{equation}
where $\beta$ is a $(1,0)$-form, $\gamma$ is a chiral scalar, and $\Phi$ is a non-chiral scalar satisfying the OPEs
\begin{equation} \label{eq:wakimoto-opes}
\beta(z)\gamma(w)\sim-\frac{1}{z-w}\,,\qquad \partial \Phi(z) \partial \Phi(w) \sim -\frac{1}{(z-w)^2} \,.
\end{equation}
Similar OPEs hold for the anti-holomorphic fields $\bar \beta, \bar \gamma$ and $\bar \partial \Phi$. The scalar $\Phi$ has a background charge $Q_{\Phi}=\sqrt{2/(k-2)}$, so that the central charge of the above action is
\begin{equation}
c(\beta,\gamma)+c(\Phi)=2+1+3Q_{\Phi}^2=\frac{3k}{k-2}=c(\mathfrak{sl}(2,\mathbb{R})_k)\,.
\end{equation}
The $\mathfrak{sl}(2,\mathbb{R})_k$ algebra is generated by the currents
\begin{equation} \label{eq:wakimoto-sl2r}
J^+=\beta\,,\qquad J^3=-\frac{\partial\Phi}{Q_{\Phi}}+\beta\gamma\,,\qquad J^-=-2\frac{(\partial\Phi\,\gamma)}{Q_{\Phi}}+(\beta\gamma)\gamma-(k-1)\partial\gamma\,,
\end{equation}
and the stress tensor is
\begin{equation}
T_{W}=-(\partial\Phi)^2-\frac{Q_{\Phi}}{2}\partial^2\Phi-\beta\partial\gamma\,.
\end{equation}

The theory we are interested in, of course, is not bosonic strings on $\text{AdS}_3$, but rather superstrings on $\text{AdS}_3\times\text{S}^3\times\mathbb{T}^4$, built in terms of the $\mathfrak{psu}(1,1|2)_k$ WZW model. Note, however, that at $k=1$ there is a conformal embedding
\begin{equation}
\mathfrak{sl}(2,\mathbb{R})_1\oplus\mathfrak{su}(2)_1\subset\mathfrak{psu}(1,1|2)_1 \,,
\end{equation}
meaning that the stress tensor of $\mathfrak{psu}(1,1|2)_1$ is recovered by the stress tensor of its bosonic subalgebra $\mathfrak{sl}(2,\mathbb{R})_1\oplus\mathfrak{su}(2)_1$.\footnote{A necessary and sufficient condition for a subalgebra $\widehat{\mathfrak{h}}\subset\widehat{\mathfrak{g}}$ to define a conformal embedding is that $c(\widehat{\mathfrak{h}})=c(\widehat{\mathfrak{g}})$ \cite{Goddard:1984vk}. Since
\begin{equation*}
c(\mathfrak{psu}(1,1|2)_k)-c(\mathfrak{sl}(2,\mathbb{R})_k)-c(\mathfrak{su}(2)_k)=-8\left(\tfrac{k^2-1}{k^2-4}\right) \,, 
\end{equation*}
we see that $\mathfrak{sl}(2,\mathbb{R})_k\oplus\mathfrak{su}(2)_k\subset\mathfrak{psu}(1,1|2)_k$ is a conformal embedding if and only if $k=1$.}
The existence of this conformal embedding suggests that it should be possible to study $\mathfrak{psu}(1,1|2)_1$ purely in terms of its bosonic subalgebra. Thus, we are motivated to study the theory $\mathfrak{sl}(2,\mathbb{R})_1\oplus\mathfrak{su}(2)_1$.

Note that, for $k>2$, the background charge $Q_{\Phi}$ is real, whereas for $k=2$ it diverges. However, for $k<2$, there is a qualitative change in the behavior of $\Phi$, since its background charge becomes imaginary. Specifically, for $k=1$ we have
\begin{equation}
Q_{\Phi}=\sqrt{-2}\,,\qquad c(\Phi)=1+3Q_{\Phi}^2=-5\,.
\end{equation}

A free-field realization of $\mathfrak{su}(2)_1\oplus\mathfrak{u}(1)$ can be constructed in terms of a pair of fermionic first-order systems $(b_a,c^a)$ ($a=1,2$) satisfying the OPEs
\begin{equation}
b_a(z)c^b(w)\sim\frac{\delta\indices{_a^b}}{z-w}
\label{su2-u1-bc}
\end{equation}
and with conformal weights $\Delta(b_a)=\Lambda$, $\Delta(c^a)=1-\Lambda$. These free fields generate four currents
\begin{equation}
K^3=-\frac{1}{2}(b_1c^1-b_2c^2)\,,\qquad K^+=b_2c^1\,,\qquad K^-=b_1c^2 \,,
\label{su(2)bc-currents}
\end{equation}
\begin{equation}
J=\frac{1}{\sqrt{2}}(b_1c^1+b_2c^2)\,.
\end{equation}
The currents $K^a$ generate $\mathfrak{su}(2)_1$, while $J$ commutes with $K^a$ and thus generates a disjoint $\mathfrak{u}(1)$. If we define $J=i\partial\varphi$, then $\varphi$ obeys the OPE
\begin{equation}
\varphi(z)\varphi(w)=-\log(z-w)
\end{equation}
and has background charge 
\begin{equation}
    Q_{\varphi}=-\sqrt{-2}(1-2\Lambda) \,.
    \label{Q-varphi}
\end{equation}
For $\Lambda=\tfrac{1}{2}$, the background charge \eqref{Q-varphi} vanishes and we recover the `usual' free-field construction of $\mathfrak{su}(2)_1\oplus\mathfrak{u}(1)$, which was also used in \cite{Eberhardt:2019ywk}. On the other hand, for $\Lambda=1$, the background charge reads
\begin{equation}
    Q_{\varphi} = \sqrt{-2} = Q_\Phi 
\end{equation}
and exactly equals the background charge of $\Phi$. We thus understand that for $\Lambda =1$ we can identify the extra current $\partial \varphi$ generated by the $bc$ system \eqref{su2-u1-bc} with the $\partial \Phi$ current entering the Wakimoto representation of $\mathfrak{sl}(2,\mathbb R)_1$, 
\begin{equation}
J= i \, \partial \varphi = i \, \partial \Phi = \frac{1}{\sqrt 2} (b_1 c^1 + b_2 c^2)    \,.
\label{Wakimoto-identification}
\end{equation}
As a consequence, it is no longer necessary to gauge away any current and one can realize $\mathfrak{sl}(2,\mathbb{R})_1\oplus\mathfrak{su}(2)_1$ as we did in eqs.~\eqref{eq:sl2-currents} and \eqref{eq:su2-currents}. In fact, using eq.~\eqref{Wakimoto-identification}, and making the identifications $b_a = p_a$ and $c^a = \theta^a$, eqs.~\eqref{eq:wakimoto-sl2r} and \eqref{su(2)bc-currents} respectively reproduce eqs.~\eqref{eq:sl2-currents} and \eqref{eq:su2-currents}. 

Thus, we can group together the Wakimoto scalar $\Phi$ and the $\mathfrak{su}(2)_1$ WZW model into two pairs $(p_a,\theta^a)$ of fermions with $\Delta(p_a)=1$ and $\Delta(\theta^a)=0$ (along with their right-moving counterparts). We conclude that we can write the `free-field' part of the Wakimoto representation of $\mathfrak{sl}(2,\mathbb{R})_1\oplus\mathfrak{su}(2)_1$ as
\begin{equation}
S_{\text{free}}=\frac{1}{2\pi}\int\left(\beta\overline{\partial}\gamma+\bar{\beta}\partial\overline{\gamma}+p_a\overline{\partial}\theta^a+\overline{p}_a\partial\overline{\theta}^a\right)\,.
\end{equation}
This is precisely the free-field realization of $\mathfrak{psu}(1,1|2)_1$ discussed in this section.

One may ask about the interaction term which appears in \eqref{eq:wakimoto}, yet which does not appear in the free field realization \eqref{new-free-field-realization}. Written in terms of the free fields, the interaction Lagrangian takes the form
\begin{equation}
\mathcal{L}_{\text{int}}=-\frac{1}{4k}\beta\bar{\beta}e^{-Q_{\Phi}\Phi}\stackrel{k=1}{=}-\frac{1}{4}\beta\bar{\beta}(\theta^1\theta^2\overline{\theta}^1\overline{\theta}^2)\,.
\end{equation}
In principle, it seems that one should include this term in the $\mathfrak{psu}(1,1|2)_1$ Lagrangian in order to obtain a complete worldsheet description of the model. It is not entirely clear why the $k=1$ theory seems to be completely consistent without the introduction of this interaction term. See also \cite{McStay:2023thk} for a discussion on this point. One explanation is that $\mathcal{L}_{\text{int}}$ is not a physical state in the gauge-fixed worldsheet theory. As we will explain below, and as outlined in Appendix \ref{app:hybrid}, the worldsheet physical spectrum is specified by a double cohomology of operators $G^+_0,\widetilde{G}^+_0$. However, the interaction Lagrangian $\mathcal{L}_{\text{int}}$ is not annihilated by $G^+_0$. Thus, from the point of view of the gauge-fixed theory, $\mathcal{L}_{\text{int}}$ does not represent a well-defined deformation of the free action $S_{\text{free}}$. Nonetheless, we admit that this reasoning is rather post-hoc and we will instead take the action $S_{\text{free}}$ as the definition of our worldsheet theory, regardless of its connection with the sigma model description of $\text{AdS}_3\times\text{S}^3$.

We should note that the relationship between the $\mathfrak{psu}(1,1|2)_1$ WZW model and the Wakimoto representation of $\mathfrak{sl}(2,\mathbb{R})_1$ \textit{without} the interaction term has been discussed in the literature before, see \cite{Naderi:2022bus,McStay:2023thk}. The above discussion sharpens these observations to include the full $\mathfrak{psu}(1,1|2)_1$ theory, and not just that of the $\mathfrak{sl}(2,\mathbb{R})_1$ subalgebra.

\subsection{Spacetime interpretation of the free fields} \label{sec:dual-coordinates}

The form of the metric of Euclidean global $\text{AdS}_3$ in Poincar\'{e} coordinates, 
\begin{equation}
\mathrm{d}s^2=\frac{\mathrm{d}r^2+\mathrm{d}\gamma\mathrm{d}\overline{\gamma}}{r^2} \,,
\end{equation}
suggests that $\gamma$ and $\bar \gamma$ should be interpreted as the position coordinates $x$ and $\bar x$ in the boundary CFT$_2$ \cite{Giveon:1998ns}. This intuition was strengthened in \cite{Eberhardt:2019ywk}, where  it was observed that at $k=1$ the field $\gamma$, when inserted inside correlation functions, behaves as the covering map from the worldsheet to the boundary sphere. Notice that in order for a field on the worldsheet to be holographically dual to the boundary position coordinate $x$, one would expect that: 
\begin{itemize}
\item it has spacetime conformal dimension $h = -1$;
\item it is a singlet under the $\mathcal R$-symmetry and hence commutes with $K^3_0$ on the worldsheet. 
\end{itemize}
As one can easily check, $\gamma$ satisfies both of these requirements.

Now, let us interpret the worldsheet fields $\theta^1$ and $\theta^2$. The worldsheet admits global fermionic symmetries $\delta^1,\delta^2$ such that
\begin{equation}
\delta^a\gamma=\theta^a\,,\qquad\delta^a\theta^b=0\,,\qquad\delta^ap_b=-\delta\indices{^a_b}\beta\,,\qquad\delta^a\beta=0\,,
\label{delta-N=2}
\end{equation}
see eqs.~\eqref{new-free-field-realization}. In particular, we have $\delta^1=-S^{++-}_0$ and $\delta^2=S^{+--}_0$. Indeed, these symmetries form a subset of the global $\mathfrak{psu}(1,1|2)$ algebra. 
Given that the zero modes of $\mathfrak{psu}(1,1|2)_1$ are the wedge modes of the small $\mathcal{N}=4$ superconformal algebra in two dimensions, and the fact that the partner of $\gamma$ under these transformations is $\theta^a$, it is natural to interpret $\theta^a$ as representing the boundary $\mathcal{N}=2$ superspace coordinates $\vartheta,\bar{\vartheta}$. Indeed, notice that by the same logic of the discussion above, the proposed fields $\theta^a$
\begin{itemize}
    \item have spacetime conformal dimension $-\frac{1}{2}$;
    \item transform as a doublet with respect to the spacetime $\mathfrak{su}(2)$ which is identified with the $\mathfrak{su}(2)$ of $\mathfrak{psu}(1,1|2)_1$, see \cite{Gaberdiel:2021njm}. 
\end{itemize}
We note, however, that these statements should only be considered true at the level of the $\mathfrak{psu}(1,1|2)_1$ WZW model and not in the full string theory. In string theory, after imposing the physical state conditions, we are not aware of a way to construct supersymmetry generators which commute with physical state conditions, and for which $\gamma,\theta^a$ have the simple interpretation of spacetime superspace coordinates.

In the case of $\text{AdS}_3\times\text{S}^3\times\mathbb{T}^4$, the boundary CFT$_2$ has the structure of a (small) $\mathcal N =4$ superconformal field theory. In particular, it contains an $\mathcal N=2$ subalgebra which acts on the (holomorphic) superspace coordinates $(x, \vartheta, \bar \vartheta)$ as \cite{Blumenhagen:2013fgp}
\begin{subequations}
\begin{align}
    \delta x &= \xi(x) + \frac{1}{2} \vartheta \, \bar \epsilon(x) + \frac{1}{2} \bar \vartheta \, \epsilon(x) \,, \\
    \delta \vartheta &= \epsilon(x) + \frac{1}{2} \vartheta \, \partial_x \xi(x) - \vartheta \, \alpha(x) + \frac{1}{2} \vartheta \, \bar \vartheta \,  \partial_x \epsilon(x) \,, \\
    \delta \bar \vartheta &= \bar \epsilon(x) + \frac{1}{2} \, \bar \vartheta \, \partial_x \xi(x) + \bar \vartheta \, \alpha(x) - \frac{1}{2} \vartheta \, \bar \vartheta \, \partial_x \bar \epsilon(x) \,.  
\end{align}
\label{delta-x-etc}%
\end{subequations}
The Grassmann even functions $\xi(x)$ and $\alpha(x)$ respectively parametrize infinitesimal translations and infinitesimal U(1) $\mathcal R$-charge rotations. Similarly, the Grassmann odd functions $\epsilon(x)$ and $\bar \epsilon(x)$ parametrize infinitesimal superconformal transformations generated by the two $\mathcal N=2$ supercharges.

To strengthen the observation that $\gamma,\theta^1,\theta^2$ represent spacetime superspace coordinates, we study Noether's charges associated to the local $\mathcal{N}=2$ transformations. To illustrate this, let us set 
\begin{equation}
    \epsilon = 0 \,, \qquad \bar{\epsilon}=0 \,, \qquad \xi = - \eta \gamma^{n+1} \,, \qquad \alpha = 0 \,,
\end{equation}
with $\eta$ a constant bosonic variable in eqs.~\eqref{delta-x-etc}. We find 
\begin{align}
    \delta \gamma &= - \eta \gamma^{n+1} \ , \nonumber \\
    \delta \theta^1 &= - \eta \frac{n+1}{2} \gamma^n \theta^1 \ , \\
    \delta \theta^2 &= - \eta \frac{n+1}{2} \gamma^n \theta^2 \nonumber \ .
\end{align}
The charge implementing this variation is
\begin{subequations}
\begin{equation} \label{eq:charge-stress-tensor}
    \mathcal{L}_n = \oint dz [(\beta \gamma) \gamma^n + \frac{n+1}{2} \gamma^n (p_a \theta^a)] \ .
\end{equation}
This is exactly the DDF operator associated to the spacetime stress-tensor written in our variables, as we will explicitly see in Section~\ref{sec:ddf}. For instance, setting $n=-1,0,1$ gives the zero modes of $\mathfrak{sl}(2,\mathbb{R})_1$ currents, see eq.~\eqref{eq:sl2-currents}. While using this approach we cannot fix the correct normal ordering, we have anticipated the correct prescription in eq.~(\ref{eq:charge-stress-tensor}). A similar procedure also works for the Cartan generator $\mathcal{J}$ of the $\mathcal{R}$-symmetry currents, and the supercurrents of an $\mathcal{N}=2$ subalgebra. In fact, in addition to the stress-tensor, one obtains the following Noether's charges associated to the local spacetime transformations
\begin{equation} \label{eq:charge-r-symmetry}
	\mathcal{J}_n = \oint dz K^3 \gamma^n \,,
\end{equation}
\begin{equation}
	\underline{\mathcal{G}}^+_r = -2 \oint dz \Big[ - p_2 \gamma^{r+\frac{1}{2}} \Big] + \oint dz \Big[\theta^1 (\beta \gamma^{r+\frac{1}{2}}) + 2 (r+\tfrac{1}{2}) (\theta^1 K^3) \gamma^{r-\frac{1}{2}} \Big] \,,
 \label{underline-G-+}
\end{equation}
\begin{equation}
	\underline{\mathcal{G}}^-_r = \oint dz \Big[ p_1 \gamma^{r+\frac{1}{2}} \Big] -\frac{1}{2} \oint dz \Big[ -\theta^2 (\beta \gamma^{r+\frac{1}{2}}) + 2 (r+\tfrac{1}{2}) (\theta^2 K^3) \gamma^{r-\frac{1}{2}} \Big] \,.
  \label{underline-G--}
\end{equation}
\label{N=2-psu112}%
\end{subequations}
Indeed, one can check that the operators \eqref{N=2-psu112} form an $\mathcal N=2$ algebra with $c=6$. The underline in $\underline{\mathcal{G}}^+_r$ and $\underline{\mathcal{G}}^-_r$ emphasizes that the Noether charges in eqs.~\eqref{underline-G-+} and \eqref{underline-G--} do \textit{not} agree with the DDF operators we will obtain in Section~\ref{sec:ddf}. In fact, the Noether charges in eqs.~\eqref{underline-G-+} and \eqref{underline-G--}  do not commute with the physical state conditions of the hybrid formalism. This is expected, since as already mentioned, the symmetry \eqref{delta-N=2} is a symmetry of the $\text{PSU}(1,1|2)_1$ non-linear sigma model and \emph{not} a symmetry of the full AdS$_3 \times \text{S}^3 \times \mathbb T^4$ string theory. 

\subsection[The worldsheet \texorpdfstring{$\mathcal{N}=4$}{N=4} algebra]{\boldmath The worldsheet \texorpdfstring{$\mathcal{N}=4$}{N=4} algebra}
\label{sec:worldsheet-theory}

In the RNS formulation of string theory on a group manifold $G$ \cite{Gepner:1986wi}, one introduces worldsheet fermions that transform in the adjoint representation of $G$, see e.g.~\cite{Ferreira:2017pgt} for a review. Although this makes the worldsheet supersymmetry manifest, the spacetime supersymmetry is hidden and is realized after GSO projection. On the other hand, Green-Schwarz like formulations of string theory make spacetime supersymmetry manifest \cite{Green:1987sp}. There exists a third option: a hybrid formulation that makes the supersymmetry of $6$ flat directions manifest, while the other directions are described in the RNS formulation \cite{Berkovits:1999im}. We will sometimes refer to this worldsheet model as the Berkovits, Vafa and Witten formalism. In the same paper, the analysis is extended to $\text{AdS}_3 \times \text{S}^3 \times \mathbb T^4$. In the hybrid formalism, it is natural to describe the physical states in a cohomological language. More specifically, the worldsheet has a small $\mathcal{N}=4$ topologically twisted algebra, and one of the supercurrents is the BRST current. For a detailed review of this formalism see \cite{Gerigk:2012lqa,Gaberdiel:2022als}. In the following we briefly sketch the hybrid formalism on $\text{AdS}_3 \times \text{S}^3 \times \mathbb T^4$ while a more rigorous discussion can be found in Appendix~\ref{app:hybrid}.

Let us now elaborate on the background that we are interested in, i.e.\ $\text{AdS}_3 \times \text{S}^3 \times \mathbb{T}^4$ with pure NS-NS flux. The supergroup that describes strings moving on $\text{AdS}_3 \times \text{S}^3$ is $\text{PSU}(1,1|2)$ with the affine algebra $\mathfrak{psu}(1,1|2)_k$. In other words, strings moving on this background are described by a WZW model on $\mathfrak{psu}(1,1|2)_k$ and $\sqrt{k}$ is proportional to the radius of $\text{AdS}_3$ and $\text{S}^3$ in units of the string length. As we briefly mentioned above, in order to describe superstrings moving on $\text{AdS}_3 \times \text{S}^3 \times \mathbb{T}^4$, we need additional ingredients. The worldsheet theory has holomorphic (left-moving) and anti-holomorphic (right-moving) sectors. For brevity, here we focus only on the holomorphic sector and we combine them in Section~\ref{sec:correlation-functions}, where we compute correlation functions. The holomorphic sector consists of
\begin{itemize}
\item the $\mathfrak{psu}(1,1|2)_k$ WZW model, 
\item the free bosons $\rho$ and $\sigma$, 
\item a topologically twisted $\mathcal N=4$ superconformal theory on $\mathbb{T}^4$.
\end{itemize}
Similarly for the anti-holomorphic sector. From now on we focus on $k=1$, which is the relevant value of the level in our worldsheet theory. We reviewed the $\mathfrak{psu}(1,1|2)_k$ model at level $k=1$ in Section~\ref{sec:new-free-field-realization}. The free bosons $\rho$ and $\sigma$ have OPEs
\begin{subequations} \label{eq:rho-sigma}
\begin{equation}
    \rho(z) \rho(w) \sim - \ln{(z-w)} \,, \qquad \sigma(z) \sigma(w) \sim - \ln{(z-w)} \,,
\end{equation}
and stress-tensor
\begin{equation}
    T_{\rho\sigma}=-\frac{1}{2} [(\partial\rho)^2 + (\partial \sigma)^2] + \frac{3}{2} \partial^2(\rho+i\sigma) \,.
\end{equation}
\end{subequations}
The topologically twisted $\mathbb T ^4$ sector, which is reviewed in Appendix~\ref{app:topologically twisted algebra}, consists of $4$ free bosons and $4$ free fermions on $\mathbb{T}^4$, with a  topologically twisted stress-tensor. We denote the generators of this algebra on the worldsheet by a subscript $C$. In particular, note that the Cartan generator of the $\mathcal{R}$-symmetry currents is $J_C$, which is bosonized as
\begin{equation}
    J_C = \tfrac{1}{2} \partial (iH) \ , \qquad H(z) H(w) \sim -2 \ln{(z-w)} \,.
\end{equation}

The full worldsheet theory has the structure of a topologically twisted (small) $\mathcal{N}=4$ superconformal algebra with $c=6$ on the worldsheet \cite{Berkovits:1999im}. We denote the stress-tensor by $T$, $\mathfrak{su}(2)_1$ $\mathcal{R}$-symmetry currents by $J$, $J^{++}$, $J^{--}$ and the supercurrents by $G^{\pm}$ and $\widetilde{G}^{\pm}$. The explicit form of these generators can be found in eq.~(\ref{eq:n=4-worldsheet}) while in Table~\ref{tab:twisted-N=4}  we recap their conformal dimensions and whether or not they are conformal primaries. We will discuss the physical state conditions in Section~\ref{sec:ddf}, but let us anticipate the basic idea: the zero mode of the generator $G^+$ is the BRST charge which defines the physical states. For the exact physical state conditions see eq.~\eqref{eq:physical}.

\begin{table}
    \centering
    \begin{tabular}{c|cccccccc}
        & $T$ & $J$ & $J^{++}$ & $J^{--}$ & $G^+$ & $G^-$ & $\widetilde G^+$ & $\widetilde G^-$ \\
        \hline
        Weight & $2$ & $1$ & $0$ & $2$ & $1$ & $2$ & $1$ & $2$ \\
        Primary & \cmark & \xmark & \cmark & \cmark & \cmark & \cmark & \cmark & \cmark \\
    \end{tabular}
    \caption{We list the generators of the topologically twisted $\mathcal N=4$ algebra with their conformal weight. In the third line of the table, a check mark or a cross respectively denotes whether or not the corresponding generator is a conformal primary.}
    \label{tab:twisted-N=4}
\end{table}
It is important to note that, as suggested in \cite{Beem:2023dub}, in the formulation \eqref{new-free-field-realization} of $\mathfrak{psu}(1,1|2)_1$, since \emph{no} additional gauging is needed, one obtains an honest worldsheet topologically twisted $\mathcal{N}=4$ algebra with $c=6$. Let us elaborate on this: once we have a $\mathfrak{psu}(1,1|2)_k$ algebra, the construction of \cite{Berkovits:1999im} ensures that there exists a worldsheet $\mathcal{N}=4$ topologically twisted algebra with $c=6$. However, for the case of $k=1$, as we mentioned in the Introduction, the realization of $\mathfrak{psu}(1,1|2)_1$ in terms of symplectic bosons and free fermions involved an additional gauging with respect to a null-current, usually denoted by $Z$. Hence, the anti-commutation relations of $\mathfrak{u}(1,1|2)_1$ differed from the ones of $\mathfrak{psu}(1,1|2)_1$ by additional terms proportional to $Z$, which after gauging indeed gave a realization of $\mathfrak{psu}(1,1|2)_1$. This `contamination' of $Z$-terms implied that the worldsheet did not manifestly have an $\mathcal{N}=4$ topologically twisted algebra with $c=6$, see \cite{Gaberdiel:2022als}. In that paper, this issue was resolved by introducing additional ghosts and gauging them out. In our case, as suggested in \cite{Beem:2023dub}, these complications are absent since we begin by considering the alternative $\mathfrak{psu}(1,1|2)_1$ realization \cite{Beem:2023dub}, and therefore we get an exact $\mathcal{N}=4$ topologically twisted algebra on the worldsheet.


\section{The worldsheet spectrum}
\label{sec:The worldsheet spectrum}

In this section, we compute the full spectrum of the physical states on $\text{AdS}_3\times\text{S}^3\times\mathbb{T}^4$. While this already appeared in \cite{Eberhardt:2018ouy, Eberhardt:2020bgq}, we hope to convince the reader of the utility of the free-field realization \eqref{new-free-field-realization} in simplifying various steps of that computation. We also collect some clarifications about the contribution of ghosts to the partition function, which appeared in~\cite{Gaberdiel:2021kkp}. Furthermore, we discuss the physical state conditions on the worldsheet, and rewrite expressions for spectrally-flowed vertex operators following \cite{Naderi:2022bus}. Finally, we write a set of DDF operators based on \cite{Giveon:1998ns,Eberhardt:2019qcl,Naderi:2022bus}, which allows an exact realization of the single-particle spectrum of the dual CFT on the worldsheet.

\subsection{Spectrally-flowed characters}

The primary ingredient is the computation of spectrally-flowed $\mathfrak{psu}(1,1|2)_1$ characters, which, as we will see, are easily calculated in our language. The spectrum of the $\mathfrak{psu}(1,1|2)_1$ WZW model has the form
\begin{equation}
\mathcal{H}=\bigoplus_{w\in\mathbb{Z}}\int_{0}^{1}\mathrm{d}\lambda\,\sigma^w(\mathcal{F}_{\lambda})\otimes\overline{\sigma^w(\mathcal{F}_{\lambda})}\,.
\end{equation}
We would like to compute the partition function
\begin{equation}
\begin{split}
Z_{\mathfrak{psu}(1,1|2)_1}(t,z;\tau)&=\text{Tr}_{\mathcal{H}}\left[q^{L_0}\overline{q}^{\overline{L}_0}x^{J^3_0}\overline{x}^{\overline{J}^3_0}y^{K^3_0}\overline{y}^{\overline{K}^3_0}\right]\\
&=\sum_{w\in\mathbb{Z}}\int_{0}^{1}\mathrm{d}\lambda\left|\text{ch}(\sigma^w(\mathcal{F}_{\lambda}))(t,z;\tau)\right|^2\,,
\end{split}
\end{equation}
where $q=e^{2\pi i\tau}$, $x=e^{2\pi it}$, and $y=e^{2\pi iz}$, and we have defined the characters of the representations $\sigma^w(\mathcal{F}_{\lambda})$ as
\begin{equation}
\text{ch}(\sigma^w(\mathcal{F}_{\lambda}))(t,z;\tau)=\text{Tr}_{\sigma^w(\mathcal{F}_{\lambda})}\left[q^{L_0}x^{J^3_0}y^{K^3_0}\right]\,.
\end{equation}

As discussed in the previous section, the highest-weight representations $\mathcal{F}_{\lambda}$ of the free-field algebra are generated by descendants of states of the form $\ket{m}$ such that
\begin{equation}
\beta_0\ket{m}=m\ket{m+1}\,,\qquad\gamma_0\ket{m}=\ket{m-1}\,,\qquad (p_a)_0\ket{m}=0\,.
\end{equation}
Clearly, since the bosons and fermions decouple, this character factorizes into a character from the $\beta\gamma$ system and a character from the $p_a \theta^a$ system. The $\beta\gamma$ character is easily calculated as follows: the $J^3_0$ eigenvalue of $\ket{m}$ is $m$, while $L_0\ket{m}=K^3_0\ket{m}=0$, and so the $\beta\gamma$ character is simply
\begin{equation}
\sum_{m\in\mathbb{Z}+\lambda}\frac{x^{m}}{\eta(\tau)^2}\,,
\end{equation}
where each factor of $1/\eta(\tau)$ accounts for the oscillator modes of one of the bosons. Meanwhile, the fermionic character can be computed by noting that the fermions have the following charges with respect to $J^3_0$ and $K^3_0$:
\begin{equation}
\begin{split}
[J^3_0,p_1]=\tfrac{1}{2}\,,&\qquad[K^3_0,p_1]=-\tfrac{1}{2}\,,\\
[J^3_0,p_2]=\tfrac{1}{2}\,,&\qquad[K^3_0,p_2]=\tfrac{1}{2}\,,
\end{split}
\end{equation}
while $\theta^1$ and $\theta^2$ have charges opposite to $p_1$ and $p_2$, respectively. Thus, the character of the $p_a \theta^a$ system is simply the character of two pairs of free fermions with flavors $\frac{t\pm z}{2}$ in the R-sector,\footnote{The fermions are in the R-sector since they have integer conformal weight.} namely
\begin{equation}
\frac{\vartheta_2(\frac{t+z}{2};\tau)\vartheta_2(\frac{t-z}{2};\tau)}{\eta(\tau)^2}\,.
\end{equation}
Putting the bosonic and fermionic contributions together gives the character
\begin{equation}
\text{ch}(\mathcal{F}_{\lambda})(t,z;\tau)=\sum_{m\in\mathbb{Z}+\lambda}x^m\frac{\vartheta_2(\frac{t+z}{2};\tau)\vartheta_2(\frac{t-z}{2};\tau)}{\eta(\tau)^4}\,.
\end{equation}

For the spectrally-flowed characters, we note that spectral flow acts on $J^3_0,K^3_0,L_0$ as
\begin{equation}\label{eq:spectral-flow-quantum-numbers}
\sigma^w(J^3_0)=J^3_0+\frac{w}{2}\,,\qquad\sigma^w(K^3_0)=K^3_0+\frac{w}{2}\,,\qquad\sigma^w(L_0)=L_0+w(K^3_0-J^3_0)\,.
\end{equation}
This means that, since $\mathcal{F}_{\lambda}$ and $\sigma^w(\mathcal{F}_{\lambda})$ are isomorphic as vector spaces, we can replace a trace over $\sigma^w(\mathcal{F}_{\lambda})$ with a trace over $\mathcal{F}_{\lambda}$, so long as we appropriately shift the quantum numbers as in \eqref{eq:spectral-flow-quantum-numbers}. Thus, we have
\begin{equation}
\begin{split}
\text{Tr}_{\sigma^w(\mathcal{F}_{\lambda})}\left[q^{L_0}x^{J^3_0}y^{K^3_0}\right]&=\text{Tr}_{\mathcal{F}_{\lambda}}\left[q^{L_0+w(K^3_0-J^3_0)}x^{J^3_0+\frac{w}{2}}y^{K^3_0+\frac{w}{2}}\right]\\
&=x^{\frac{w}{2}}y^{\frac{w}{2}}\text{ch}(\mathcal{F}_{\lambda})(t-w\tau,z+w\tau;\tau)\,.
\end{split}
\end{equation}
Therefore, we can immediately write down the spectrally-flowed characters in terms of the unflowed characters. We find
\begin{equation}
\begin{split}
\text{ch}(\sigma^w(\mathcal{F}_{\lambda}))(t,z;\tau)&=x^{\frac{w}{2}}y^{\frac{w}{2}}\sum_{m\in\mathbb{Z}+\lambda}x^{m}q^{-wm}\frac{\vartheta_2(\frac{t+z}{2};\tau)\vartheta_2(\frac{t-z}{2}-w\tau;\tau)}{\eta(\tau)^4}\\
&=x^{w}q^{-\frac{w^2}{2}}\sum_{m\in\mathbb{Z}+\lambda}x^{m}q^{-wm}\frac{\vartheta_2(\frac{t+z}{2};\tau)\vartheta_2(\frac{t-z}{2};\tau)}{\eta(\tau)^4}\,,
\end{split}
\end{equation}
where we have used the theta function identity
\begin{equation}
\vartheta_2(z-w\tau;\tau)=q^{-\frac{w^2}{2}}e^{2\pi iwz}\vartheta_2(x;\tau)\,,\qquad w\in\mathbb{Z}\,.
\end{equation}
Shifting $m\to m-w$ in the sum then gives
\begin{equation}\label{eq:spectrally-flowed-character}
\text{ch}(\sigma^w(\mathcal{F}_{\lambda}))(t,z;\tau)=q^{\frac{w^2}{2}}\sum_{m\in\mathbb{Z}+\lambda}x^{m}q^{-wm}\frac{\vartheta_2(\frac{t+z}{2};\tau)\vartheta_2(\frac{t-z}{2};\tau)}{\eta(\tau)^4}\,.
\end{equation}
This agrees precisely with the spectrally-flowed characters of $\mathfrak{psu}(1,1|2)_1$ found in \cite{Eberhardt:2018ouy}. Thus, we conclude that the spectrum of spectrally-flowed highest-weight states in our free worldsheet theory agrees with that of the $\mathfrak{psu}(1,1|2)_1$ WZW model, suggesting that they are indeed equivalent.

\subsection{The worldsheet partition function}

For completeness, we include the derivation of the full worldsheet partition function from the above characters. This subsection is essentially a summary of Section 5 of \cite{Eberhardt:2018ouy} and contains no new results. We break the computation of the worldsheet partition function into the $\mathfrak{psu}(1,1|2)_1$ spectrum, the $\mathbb{T}^4$ spectrum, and the contribution from the $(\rho,\sigma)$ ghosts of the hybrid formalism.

\paragraph{\boldmath The $\mathfrak{psu}(1,1|2)_1$ spectrum} As mentioned above, the spectrum of the $\mathfrak{psu}(1,1|2)_1$ which is part of the worldsheet theory takes the form
\begin{equation}
\bigoplus_{w\in\mathbb{Z}} \int_{0}^{1}\mathrm{d}\lambda \, \sigma^w(\mathcal{F}_{\lambda})\otimes\overline{\sigma^w(\mathcal{F}_{\lambda})}\,.
\end{equation}
Thus, we can write the partition function of the $\mathfrak{psu}(1,1|2)_1$ model in terms of the characters calculated above. We find
\begin{equation}
Z_{\mathfrak{psu}(1,1|2)_1}=\int_{0}^{1}\mathrm{d}\lambda\sum_{w\in\mathbb{Z}}\left|\text{ch}(\sigma^w(\mathcal{F}_{\lambda}))(t,z;\tau)\right|^2\,.
\end{equation}
In order to make the computation simpler, we can use Poisson resummation on \eqref{eq:spectrally-flowed-character} to write $\text{ch}(\sigma^w(\mathcal{F}_{\lambda}))$ as a formal sum of delta functions:
\begin{equation}
\text{ch}(\sigma^w(\mathcal{F}_{\lambda}))(t,z;\tau)=q^{\frac{w^2}{2}}\sum_{m\in\mathbb{Z}}e^{2\pi im\lambda}\delta(t-w\tau+m)\frac{\vartheta_2(\frac{t+z}{2};\tau)\vartheta_2(\frac{t-z}{2};\tau)}{\eta(\tau)^4}\,.
\end{equation}
Hence, the partition function reads
\begin{equation}
\begin{split}
Z_{\mathfrak{psu}(1,1|2)_1}&=\int_{0}^{1}\mathrm{d}\lambda\sum_{w,m,n\in\mathbb{Z}}|q|^{w^2}e^{2\pi i(m-n)\lambda}\left|\frac{\vartheta_2(\frac{t+z}{2};\tau)\vartheta_2(\frac{t-z}{2};\tau)}{\eta(\tau)^4}\right|^2\\
&\hspace{5cm}\times\delta(t-w\tau+m)\delta(\overline{t}-w\overline{\tau}+n)\\
&=\sum_{w,m\in\mathbb{Z}}|q|^{w^2}\left|\frac{\vartheta_2(\frac{t+z}{2};\tau)\vartheta_2(\frac{t-z}{2};\tau)}{\eta(\tau)^4}\right|^2\delta^{(2)}(t-w\tau+m)\,.
\end{split}
\end{equation}
Crucially, we see that the integral localizes onto the subset of the moduli space for which the worldsheet holomorphically wraps the boundary of $\text{AdS}_3$. This is a direct result of the fact that integrating out $\beta$ in the free-field theory results in a delta functional $\delta(\overline{\partial}\gamma)$ in the path integral, which demands that the path integral localizes onto holomorphic covering maps.

\paragraph{\boldmath The $\mathbb{T}^4$ spectrum} The $\mathbb{T}^4$ spectrum of the worldsheet theory is calculated in a straightforward manner. The $\mathbb{T}^4$ sigma model on the worldsheet consists of four free bosons and two pairs of topologically-twisted fermions. The topological twist means that we should treat the fermions in the Ramond-sector, and so we have
\begin{equation}
Z_{\mathbb{T}^4}=\left|\frac{\vartheta_2(0;\tau)^2}{\eta(\tau)^6}\right|^2\Theta_{\mathbb{T}^4}(\tau)\,.
\end{equation}
Here, $\Theta_{\mathbb{T}^4}(\tau)$ is the Narain theta function
\begin{equation}
\Theta_{\mathbb{T}^4}=\sum_{(p,\overline{p})\in\Gamma_{4,4}}q^{\frac{p^2}{2}}\overline{q}^{\frac{\overline{p}^2}{2}}\,,
\end{equation}
where $\Gamma_{4,4}$ is the Narain lattice of momentum and winding modes.

\paragraph{The ghost contribution} In the hybrid formalism of strings on $\text{AdS}_3\times\text{S}^3$, there are two types of ghosts, known as $\rho$ and $\sigma$ \cite{Berkovits:1999im}. The $\sigma$ ghost is nothing more than the usual free boson entering the bosonization of the $(b,c)$ conformal ghosts of superstring theory. The role of $\rho$ is less well understood but it essentially comes from the bosonization of the superconformal ghost of superstring theory combined with the compact directions.

The partition function of the $(b,c)$ system is found by noting that the $(b,c)$ system removes two bosonic oscillators from the set of physical states on the worldsheet. Thus,
\begin{equation}
Z_{\sigma}=|\eta(\tau)^2|^2\,.
\end{equation}
The $\rho$ ghost contribution can be found using the following logic. The field $\rho$ is a scalar with central charge $c(\rho) = 1+3Q_\rho^2=28$, see Appendix~\ref{app:hybrid}. If we were to introduce a set of topologically-twisted fermions $\Psi^{\dagger}$ and $\Psi$ of conformal weights $h(\Psi^{\dagger})=1$ and $h(\Psi)=0$, then the combinations
\begin{equation}
\widetilde{\beta}=e^{\rho}\partial\Psi\,,\qquad\widetilde{\gamma}=\Psi^{\dagger}e^{-\rho}
\end{equation}
define a bosonic first-order system with $h(\widetilde{\beta})=2$ and $h(\widetilde{\gamma})=-1$. These are commuting versions of the conformal $(b,c)$ ghosts of bosonic string theory, and as such remove two topologically twisted fermions from the set of physical states. Since we had to introduce a pair $(\Psi^{\dagger},\Psi)$ of topologically-twisted fermions to obtain $(\widetilde{\beta},\widetilde{\gamma})$ from $\rho$, we conclude that the $\rho$ ghost removes two `pairs' of topologically twisted fermions, i.e.
\begin{equation}
Z_{\rho}=\left|\frac{\eta(\tau)^2}{\vartheta_2(0;\tau)^2}\right|^2\,.
\label{Z-rho}
\end{equation}

\paragraph{Assembling all contributions} The full string partition function takes the form
\begin{equation}
\begin{split}
Z_{\text{string}}&=Z_{\mathfrak{psu}(1,1|2)_1} \, Z_{\mathbb{T}^4}Z_{\sigma} \, Z_{\rho}\\
&=\sum_{w,m\in\mathbb{Z}}|q|^{w^2}\left|\frac{\vartheta_2(\tfrac{t+z}{2};\tau)\vartheta_2(\tfrac{t-z}{2};\tau)}{\eta(\tau)^6}\right|^2\Theta_{\mathbb{T}^4}(\tau)\delta^{(2)}(t-w\tau+m)\\
&=\sum_{w,m\in\mathbb{Z}}|x|^{w}\left|\frac{\vartheta_2(\tfrac{z}{2}+\frac{w\tau}{2}-\frac{m}{2};\tau)\vartheta_{2}(-\tfrac{z}{2}+\frac{w\tau}{2}-\frac{m}{2};\tau)}{\eta(\tau)^6}\right|^2\Theta_{\mathbb{T}^4}(\tau)\delta^{(2)}(t-w\tau+m)\,,
\end{split}
\end{equation}
where in the last line we used the delta function to substitute $t=w\tau-m$. In order to simplify the theta functions, let us introduce the notation
\begin{equation}
\vartheta_{(0,0)}=\vartheta_3\,,\qquad\vartheta_{(0,1/2)}=\vartheta_2\,,\qquad\vartheta_{(1/2,0)}=\vartheta_4\,,\qquad\vartheta_{(1/2,1/2)}=\vartheta_1\,.
\end{equation}
The Jacobi theta functions satisfy the quasi-periodicity property
\begin{equation}
\vartheta_{(\alpha,\beta)}\left(\pm\frac{z}{2}+\frac{w\tau}{2}-\frac{m}{2};\tau\right)=q^{-\frac{w^2}{8}}y^{\pm\frac{w}{2}}e^{-i\pi w\alpha}\vartheta_{(\alpha+b/2,\beta+w/2)}\left(\pm\frac{z}{2};\tau\right)\,.
\end{equation}
Thus, we can write the string partition function as
\begin{equation}
Z_{\text{string}}=\sum_{w,m\in\mathbb{Z}}|x|^{\frac{w}{2}}\left|\frac{\vartheta_{(m/2,1/2+w/2)}(-\frac{z}{2};\tau)\vartheta_{(m/2,1/2+w/2)}(\frac{z}{2};\tau)}{\eta(\tau)^6}\right|^2\Theta_{\mathbb{T}^4}(\tau)\,\delta^{(2)}(t-w\tau+m)\,.
\end{equation}
As was argued in \cite{Eberhardt:2018ouy}, the string partition function then reduces, after taking the physical state condition, to the single-particle partition function of $\text{Sym}^K(\mathbb{T}^4)$ in the large $K$ limit.

\subsection{The physical states on the worldsheet} 
\label{sec:physical-states}

The previous section demonstrated that the free-field realization of $\mathfrak{psu}(1,1|2)_1$ allows us to compute the worldsheet partition function and reproduce the partition function of the symmetric orbifold. We will now argue that this conclusion can be refined by directly matching the physical states in the worldsheet theory with the states in the dual symmetric orbifold. This has been done using the bosonization of symplectic bosons in \cite{Naderi:2022bus}. Here we discuss how it can be reformulated in terms of the free-field realization \eqref{new-free-field-realization}. 

In the hybrid formalism, the physical states are defined as states $\phi$ that satisfy
\begin{equation} \label{eq:physical}
    G^+_0 \phi = \widetilde{G}^+_0 \phi = (J_0 - \tfrac{1}{2}) \phi = T_0 \phi = 0 \ , \qquad \phi \sim \phi+G^+_0 \widetilde{G}^+_0 \psi \,.
\end{equation}
In other words, the physical states are defined as the double cohomology of the worldsheet $\mathcal{N}=4$ superconformal algebra which we discussed in Section~\ref{sec:worldsheet-theory}. Note that the generators in eq.~(\ref{eq:physical}) have explicit expressions in terms of $\mathfrak{psu}(1,1|2)$ generators and ghosts as in eqs.~(\ref{eq:n=4-worldsheet}). The task is therefore finding the physical states associated to certain spacetime states. In fact, the spectrum of the symmetric product orbifold $\text{Sym}^K(\mathbb T^4)$ is generated by acting with (fractional) modes of four free bosons and four free fermions on $w$-twisted ground states subject to the orbifold invariance conditions.\footnote{This means that the left and the right conformal dimensions $h$ and $\bar{h}$ should satisfy $h-\bar{h}\in\mathbb{Z}$ (if both bosonic or both fermionic) or $h-\bar{h}\in \mathbb{Z}+\frac{1}{2}$ (if one fermionic and the other bosonic), see e.g.~\cite{David:2002wn}.} The same mechanism can be mimicked from the bulk in two steps. First, one identifies the worldsheet operators dual to $w$-twisted ground states. Second, to each mode of the spacetime bosons and fermions, one associates a DDF operator \cite{DelGiudice:1971yjh} on the worldsheet \cite{Giveon:1998ns, Eberhardt:2019qcl, Naderi:2022bus}. The whole bulk spectrum is then generated by applications of these DDF operators on the worldsheet states dual to $w$-twisted ground states of the symmetric orbifold, subject to the orbifold invariance conditions which arise from diagonal modular invariance on the worldsheet \cite{Eberhardt:2018ouy}. We will analyze DDF operators in Section~\ref{sec:ddf}, while we discuss here how the symmetric orbifold $w$-twisted ground states are built from the worldsheet.

Let us begin by reviewing some properties of symmetric orbifold $w$-twisted ground states of $\mathbb{T}^4$. For $w$ odd, the ground state $\sigma_w$ is a singlet with respect to the $\mathfrak{su}(2)$  $\mathcal R$-symmetry of $\mathbb{T}^4$ and has conformal dimension
\begin{equation} \label{eq:st-odd-w}
    h_w = \frac{w^2-1}{4w} \,.
\end{equation}
Due to fermion zero modes, for $w$ even there are two $\mathfrak{su}(2)$ $\mathcal R$-symmetry singlets and one $\mathcal R$-symmetry doublet. The two ground states in the doublet, which we denote by $\sigma_w^{\pm}$, have conformal dimension\footnote{Note that the two singlets also have the same conformal dimension as in eq.~(\ref{eq:st-even-w}), see e.g.\ \cite{Gaberdiel:2015uca, Gaberdiel:2018rqv}.}
\begin{equation} \label{eq:st-even-w}
    h_w^{\pm} = \frac{w}{4} \,.
\end{equation}
The worldsheet physical states dual to symmetric orbifold $w$-twisted ground states were identified in \cite{Dei:2020zui}. For $w$ odd, they take the following form
\begin{equation}
    \Omega_{w} = \Phi_w e^{2\rho+i\sigma+iH} \,,
        \label{odd-w-ground-state}
\end{equation}
while for $w$ even they read
\begin{equation}
    \Omega_{w}^{\pm} = \Phi_w^{\pm} e^{2\rho+i\sigma+iH} \,. 
    \label{even-w-ground-state}
\end{equation}
In \eqref{even-w-ground-state} the $\pm$ emphasizes that $\Omega_{w}^{\pm}$ is a doublet with respect to the $\mathfrak{su}(2) \subset \mathfrak{psu}(1,1|2)_1$, which is dual to the $\mathfrak{su}(2)$ $\mathcal R$-symmetry in spacetime \cite{Gaberdiel:2021njm}. In \cite{Dei:2020zui}, the ground states \eqref{odd-w-ground-state} and \eqref{even-w-ground-state} where expressed in terms of symplectic bosons. As explained in Appendix~\ref{app:ours-to-symp}, following the bosonization of \cite{Naderi:2022bus}, one can rewrite these fields in terms of the fields $f_1$, $f_2$, $\phi$ and $\kappa$, which bosonize the fields \eqref{beta-gamma-OPE} and \eqref{p-theta-OPE} introduced in Section~\ref{sec:new-free-field-realization}. For $w$ odd, we get
\begin{equation} \label{eq:w-odd}
    \Phi_w(z,x=0) = \exp{\Big[\frac{w+1}{2}(if_1-if_2)+(m_w+w)\phi+i m_w \kappa \Big]} \,,
\end{equation}
with
\begin{equation} \label{eq:m-w-odd}
    m_w = -\frac{(w-1)^2}{4w} \,.
\end{equation}
It is straightforward to show that this state satisfies eq.~(\ref{eq:physical}) and is therefore physical. In particular, it has vanishing worldsheet conformal dimension and the spacetime conformal dimension agrees with \eqref{eq:st-odd-w}. For $w$ even, we have
\begin{equation} \label{eq:w-even}
    \Phi^{\pm}_w(z,x=0) = \exp{\Big[\frac{\pm (if_1+if_2)}{2} \Big]} \exp{\Big[\frac{w+1}{2}(if_1-if_2)+(m^{\pm}_w+w)\phi+i m^{\pm}_w \kappa\Big]} \,,
\end{equation}
where
\begin{equation} \label{eq:m-w-even}
    m^{\pm}_w = - \frac{w-2}{4} \,.
\end{equation}
Again, these states are physical, have worldsheet weight $0$, and the spacetime weight agrees with eq.~(\ref{eq:st-even-w}).

\subsection{Delta-function operators} 
\label{sec:delta-function}

In Section~\ref{sec:correlation-functions} we will compute correlation functions of $w$-twisted ground states from the worldsheet. A key observation in that calculation is that the vertex operators that we wrote down in the previous section can be written in another useful form which involves delta-functions. The main advantage of this approach is twofold. First, the vertex operators associated to $w$-twisted ground states now have a tractable form in the $x$ basis, and second they allow us to compute correlation functions in the path integral formalism. In this section, we take a short detour, review this construction following \cite{Witten:2012bh} and eventually write down the vertex operator associated to the $w$-twisted ground states in this language. The reader already familiar with this language or mainly interested in directly knowing the final result may skip this section and jump to eqs.~\eqref{eq:w-odd-twisted-delta} and \eqref{eq:w-even-twisted-delta}.

An interesting feature of $\beta\gamma$ systems comes from the existence of an infinite tower of vacua obeying the property \cite{Friedan:1985ge}
\begin{equation}
\begin{split}
\beta_n\ket{0}^{(w)}=0\,,&\quad n>-w-\Lambda\,,\\
\gamma_n\ket{0}^{(w)}=0\,,&\quad n\geq w+\Lambda\,.
\end{split}
\end{equation}
Here $w\in\mathbb{Z}$ labels the distinct vacua. In the bosonized language (see Section~\ref{sec:new-free-field-realization}), these states can be written as the vertex operators $e^{w\phi}$, and can be thought of as the image of the trivial representation under the automorphism
\begin{equation}
\beta_{n}\to\beta_{n-w}\,,\quad\gamma_n\to\gamma_{n+w}\,.
\end{equation}
This automorphism plays the role of spectral flow in the $\mathfrak{psu}(1,1|2)_1$ model, and the states $\ket{0}^{(w)}$ play the role of spectrally-flowed ground states, as we discussed in Section~\ref{sec:representation}.

While the states $\ket{0}^{(w)}$ are readily written down in the bosonized theory in terms of the local operator $e^{w\phi}$, we are interested in expressing these states purely in terms of the $\beta\gamma$ system. The appropriate expression is in terms of local delta-function operators \cite{Witten:2012bh}:
\begin{equation}\label{eq:delta-function-cases}
e^{w\phi}=
\begin{cases}
\delta_w(\gamma)\,,\quad w>0\,,\\
\delta_{-w}(\beta)\,,\quad w<0\,,
\end{cases}
\end{equation}
where we have introduced the notation
\begin{equation} \label{eq:delta-product-formula}
\delta_w(\gamma):=\prod_{i=0}^{w-1}\delta(\partial^i\gamma)\,,
\end{equation}
and similarly for $\beta$. What this notation means is that if one inserts the operator $\delta_{w}(\gamma)$ at a point $z$ on the worldsheet, the field $\gamma$ will be restricted to have a zero of degree $w$ at $z$. This is reflected in the OPE
\begin{equation}
\gamma(y)e^{w\phi}(z)\sim\mathcal{O}((y-z)^{w})\,,
\label{gamma-delta-ope}
\end{equation}
so that $e^{w\phi}(z)$ indeed imposes a zero of degree $w$ on $\gamma$.

The conformal dimensions of delta function operators can be read off using classical reasoning. If $\gamma$ has conformal weight $1-\Lambda$, then $\delta_{w}(\gamma)$ should have conformal weight
\begin{equation}
\begin{split}
\Delta(\delta_{w}(\gamma))&=\sum_{i=0}^{w-1}\Delta(\delta(\partial^i\gamma))=-\sum_{i=0}^{w-1}(1-\Lambda+i)\\
&=-\frac{w(w+1-2\Lambda)}{2}=-\frac{w(w+Q_\phi)}{2}\,,
\end{split}
\end{equation}
which is indeed the conformal weight of $e^{w\phi}$ with $Q_\phi=1-2\Lambda$, see Appendix~\ref{app:beta-gamma}. Furthermore, since $\beta$ has weight $\Lambda$, $\delta_{w}(\beta)$ has weight
\begin{equation}
\Delta(\delta_{w}(\beta))=-\sum_{i=0}^{w-1}(\Lambda+i)=-\frac{w(w-Q_\phi)}{2}=\Delta(e^{-w\phi})\,,
\end{equation}
so that negative exponents of $\phi$ naturally map to delta functions of $\beta$, not $\gamma$. In particular, in our case $\Lambda=1$ and so the conformal dimension of $\delta_w(\gamma)$ is
\begin{equation}
    \Delta(\delta_w(\gamma)) = -\frac{w(w-1)}{2} \,.
\end{equation}

As we will see in Section~\ref{sec:correlation-functions}, an important advantage of using the delta-function notation is that it allows us to perform the computation of correlation functions using path integrals. Let us then briefly discuss the role of delta function operators from a path integral perspective. For a field $\varphi$, we define the delta function operator $\delta(\varphi)$ as the formal integral
\begin{equation}\label{eq:delta-function-integral-form}
\delta(\varphi)(z)=\int\frac{\mathrm{d}\zeta}{2\pi}\,\exp\left(i\zeta\varphi(z)\right)\,,
\end{equation}
where $\zeta$ is some Lagrange multiplier.\footnote{If $\varphi$ is a primary of weight $\Delta$, we should formally treat $\zeta$ as having weight $-\Delta$, as to make sense of the exponential.} Thus, we can write
\begin{equation} \label{eq:fourier-deltas}
\begin{split}
\delta_{w}(\gamma)(0)&=\prod_{i=0}^{w-1}\int\frac{\mathrm{d}\zeta_i}{2\pi}\,\exp\left(i\zeta_i\partial^i\gamma(0)\right)\,,\\
\delta_{w}(\beta)(0)&=\prod_{i=0}^{w-1}\int\frac{\mathrm{d}\zeta_i}{2\pi}\,\exp\left(i\zeta_i\partial^i\beta(0)\right)\,.
\end{split}
\end{equation}
These expressions can be argued to be identical to the definitions \eqref{eq:delta-function-cases} by computing their OPEs with other fields in the theory \cite{Witten:2012bh}. For example, eq.~\eqref{gamma-delta-ope} can be checked by making use of eqs.~\eqref{eq:fourier-deltas}. Further checks and examples can be found in Appendix~\ref{app:beta-gamma}. 

The role of the operator $\delta(\beta)$ in the path integral can also be read off using the above representation. If we consider the correlator $\Braket{\delta(\beta)(0)\mathcal{O}}$ for some operator $\mathcal{O}$, we can write it in the path integral as
\begin{equation}
\begin{split}
\Braket{\delta(\beta)(0)\mathcal{O}}&=\int\mathcal{D}(\beta,\gamma)e^{-S[\beta,\gamma]}\int\frac{\mathrm{d}\zeta}{2\pi}\,e^{i\zeta\beta(0)}\,\mathcal{O}\\
&=\int\frac{\mathrm{d}\zeta}{2\pi}\,\int\mathcal{D}(\beta,\gamma)e^{-S_{\text{eff}}[\beta,\gamma,\zeta]}\mathcal{O}\,,
\end{split}
\end{equation}
with
\begin{equation}
S_{\text{eff}}=\frac{1}{2\pi}\int_{\Sigma}\left(\beta\overline{\partial}\gamma-2\pi i\zeta\beta\,\delta^{(2)}(z)\right)\,.
\end{equation}
The equations of motion found upon varying $\beta$ are
\begin{equation}
\overline{\partial}\gamma=2\pi i\zeta\delta^{(2)}(z)\,,
\end{equation}
which is equivalent to saying that $\gamma$ has a pole at $z=0$ of residue $\zeta$.

With this preparation at hand, now we are ready to write the vertex operators in this language. In particular, we are interested in states of the form $e^{(m+w)\phi+im\kappa}$, see e.g.\ eq.~\eqref{eq:w-odd}. We propose that this state can be represented in the form
\begin{equation}\label{eq:delta-function-m}
e^{(m+w)\phi+im\kappa}= \left(\frac{\partial^w\gamma}{w!}\right)^{-m} e^{w \phi} = \left(\frac{\partial^w\gamma}{w!}\right)^{-m}\delta_{w}(\gamma)\,,
\end{equation}
where we used eq.~\eqref{eq:delta-function-cases} and the standard radial normal ordering is implied. In Appendix~\ref{app:beta-gamma} we provide a derivation of this equation. Having this, the state $\Phi_w$ for $w$ odd in eq.~\eqref{eq:w-odd} becomes
\begin{equation} \label{eq:w-odd-twisted-delta}
    \Phi_w(z,x) = \exp{\Big[\frac{w+1}{2}(if_1-if_2)\Big]} \Big( \frac{\partial^w \gamma}{w!} \Big)^{-m_w} \delta_w(\gamma(z)-x) \,,
\end{equation}
where $m_w$ is defined in eq.~\eqref{eq:m-w-odd}. Note that using eq.~\eqref{gamma-to-gamma-x} we have explicitly calculated the conjugation with $J^+_0=\beta_0$ in eq.~\eqref{eq:x-conjugation}. For $w$ even, the states $\Phi_w^{\pm}$ in eq.~\eqref{eq:w-even} become
\begin{equation} \label{eq:w-even-twisted-delta}
    \Phi_w^{\pm}(z,x) = \exp{\Big[\frac{\pm (if_1+if_2)}{2} \Big]} \exp{\Big[\frac{w+1}{2}(if_1-if_2)\Big]} \Big( \frac{\partial^w \gamma}{w!} \Big)^{-m_w^{\pm}} \delta_w(\gamma(z)-x) \,,
\end{equation}
where $m_w^{\pm}$ is defined in eq.~\eqref{eq:m-w-even}.

\subsection{Spacetime symmetry generators and DDF operators} \label{sec:ddf}

In the symmetric product orbifold $\text{Sym}^K(\mathbb T^4)$ four free bosons together with four free fermions realize the small $\mathcal N=4$ superconformal algebra \eqref{eq:n=4-commutation-relations}, see \cite{David:2002wn} for a review. A number of references \cite{Giveon:1998ns, Ito:1998vd, Andreev:1999nt, Kutasov:1999xu, Ashok:2009jw,  Eberhardt:2019qcl, Naderi:2022bus} explained how this algebra emerges in the bulk and can be constructed in terms of $\mathfrak{psu}(1,1|2)_1$ generators, Wakimoto and ghost fields on the worldsheet. Since the symmetry algebra of the boundary CFT is built out of free bosons and fermions, one might speculate that also on the worldsheet it may be constructed purely in terms of free fields. In fact, in \cite{Naderi:2022bus} a bosonization of the symplectic boson free-field realization \eqref{old-free-field-realization} was exploited to express symmetry generators and DDF operators \cite{DelGiudice:1971yjh} purely in terms of free fields on the worldsheet and directly in the hybrid formalism. In this section, we will rewrite the DDF operators of \cite{Naderi:2022bus} in the free-field variables of Section~\ref{sec:new-free-field-realization} and we will observe an improvement at the level of decoupling of an additional null current: since the symplectic bosons and free fermions \eqref{sympl-bosons-algebra} actually realize $\mathfrak{u}(1,1|2)_1$ --- instead of just $\mathfrak{psu}(1,1|2)_1$, see Appendix \ref{app:ours-to-symp} --- the spacetime $\mathcal N=4$ algebra in \cite{Naderi:2022bus} was realized on the worldsheet after gauging the null current $Z$. As a consequence, when expressed in terms of free fields on the worldsheet, the (anti-)commutation relations of $\mathcal N = 4$ generators contained additional terms of the rough form 
\begin{equation}
    \mathcal Z_n = \oint \text dt \, Z(t) \, \gamma(t)^{-n} \,. 
    \label{mathcal-Z}%
\end{equation}
For instance, instead of \eqref{G+-G--spacetime-N=4}, the super Virasoro generators of the spacetime CFT obeyed
\begin{equation}
    \{ \mathcal G^+_r, \mathcal G^-_s \} = (r^2 -\tfrac{1}{4}) \, \delta_{r+s,0} \, \mathcal I + (r-s) \, \mathcal J_{r+s} + \mathcal L_{r+s} + (r+s+1) \,\mathcal Z_{r+s}  \,. 
\end{equation}
Since physical states on the worldsheet are annihilated by the action of the null field $Z$, the presence of terms like \eqref{mathcal-Z} is strictly speaking not a problem. On the other hand, they have no interpretation in the spacetime theory and it would be desirable to realize spacetime generators and DDF operators in terms of free fields on the worldsheet \emph{without} any need of modding out null currents.  

For the DDF operators of the bosonic generators of the spacetime $\mathcal{N}=4$ algebra we get \cite{Naderi:2022bus,Eberhardt:2019qcl,Giveon:1998ns},
\begin{subequations}
\begin{equation} 
 \mathcal{L}_n = \oint \text dz \left[ (\beta \gamma) \gamma^n + \mfrac{n+1}{2} \gamma^n (p_a \theta^a) \right] \,, \qquad \mathcal{I} = \oint \text dz \, \gamma^{-1} \partial \gamma \,, 
\label{eq:ddf-stress-tensor}
\end{equation}
 \begin{equation} \label{eq:ddf-r-symmetry}
 \mathcal{J}^{--}_n = \oint \text dz \, p_1 \theta^2 \gamma^n \,, \quad \mathcal{J}_n = \oint \mfrac{\text dz}{2}\left[ (p_2\theta^2)-(p_1\theta^1) \right] \gamma^n \,, \quad \mathcal{J}^{++}_n = \oint \text dz \,  p_2 \theta^1 \gamma^n \,, \ 
 \end{equation}
 while for the spacetime supercharges we obtain 
 \begingroup
\allowdisplaybreaks
 \begin{equation} \label{eq:ddf-supercurrents-1}
\widetilde{\mathcal{G}}^-_r = \oint \text dz \, p_1 \gamma^{r+\frac{1}{2}} \,,  \qquad \qquad   \mathcal{G}^+_r = - \oint \text dz \, p_2 \gamma^{r+\frac{1}{2}} \,,
 \end{equation}
 and
 \begin{align}
 \mathcal{G}^-_r &= \oint \text dz \Bigl[ -\theta^2 (\beta \gamma^{r+\frac{1}{2}}) - (r+\tfrac{1}{2}) \theta^2(p_1\theta^1) \gamma^{r-\frac{1}{2}} -p_1 (e^{-\rho} \, G^-_C \, - e^{-\rho-i\sigma}) \gamma^{r+\frac{1}{2}} \Bigr] \,, \\
\widetilde{\mathcal{G}}^+_r &= \oint \text dz \, \Bigl[ \theta^1 (\beta \gamma^{r+\frac{1}{2}}) + (r+\tfrac{1}{2}) \theta^1 (p_2\theta^2) \gamma^{r-\frac{1}{2}} -p_2 (e^{-\rho} \, G^-_C - e^{-\rho-i\sigma}) \gamma^{r+\frac{1}{2}} \Bigr] \,. 
 \end{align}
 \endgroup
 \label{DDF}%
\end{subequations}
They exactly satisfy the spacetime small $\mathcal{N}=4$ algebra with $c=6$, see eqs.~\eqref{eq:n=4-commutation-relations}, without any need to gauge an additional null current $Z$. Finally, for the spacetime free bosons and fermions we have, see \cite{Naderi:2022bus}
\begin{subequations}
\begin{align}
\partial \bar{\mathcal{X}}^j_n & = \oint \text dz \, \gamma^n \partial \bar{X}^j \,, &  \partial \mathcal{X}^j_n &= \oint dz \Big[\gamma^n \partial X^j + n \gamma^{n-1} e^{\rho+i H^j} \theta^2 \theta^1 \Big] \,, \\
\Psi^{+,j}_r &= \oint \text dz \, \theta^1 \, \gamma^{(r-\frac{1}{2})} \, e^{\rho+i H^j} \,, & \Psi^{-,j}_r &= - \oint \text dz \, \theta^2 \, \gamma^{(r-\frac{1}{2})} \, e^{\rho+i H^j} \,.
\end{align}
\label{DDF-spacetime-bosons-and-fermions}%
\end{subequations}
We have adopted the convention that the spacetime theory is generated by $4$ free bosons $\partial \mathcal{X}^j$ and $\partial \bar{\mathcal{X}}^j$ where $j\in\{1,2\}$, and $4$ free fermions $\Psi^{\alpha,j}$ where $j\in\{1,2\}$ and $\alpha\in\{+,-\}$. They satisfy the (anti-)commutation relations spelled out in eqs.~\eqref{eq:bos-ferm-t4}. Note that we are using the same notation for the DDF operators and the modes of the spacetime fields associated to the bosons and fermions. As one can directly check, (anti-)commutation relations of the operators in eq.~\eqref{DDF} exactly reproduce the spacetime symmetry algebra \eqref{eq:n=4-commutation-relations}, while the DDF operators \eqref{DDF-spacetime-bosons-and-fermions} obey 
\begin{align}
[\partial \mathcal X^i_n , \partial \bar{\mathcal X}^j_m] & = n \, \delta_{ij} \, \mathcal I \, \delta_{n+m,0} \,,  \\
\{ \Psi^{\alpha, j}_r, \Psi^{\beta, l}_s \} & = \epsilon^{\alpha \beta} \, \epsilon^{l j} \, \mathcal I \, \delta_{r+s,0} \,,
\end{align}
as expected.


\section{\boldmath The path integral on Euclidean \texorpdfstring{$\text{AdS}_3$}{AdS3}}
\label{sec:point at infinity}

\subsection{Compactifying the conformal boundary}

In the next section, we will turn our attention to reproducing symmetric orbifold correlation functions in the free-field worldsheet theory. Unlike the calculation of the partition function, however, computations of dual CFT correlators introduce an additional subtlety which we will explain in this section.

As discussed previously, one should think of the fields $(\gamma,\theta^a)$ as maps from the worldsheet into the left-moving $\mathcal{N}=2$ superspace of the boundary of $\text{AdS}_3$. In Lorenzian signature, the boundary of $\text{AdS}_3$ is an infinite cylinder, and spectrally-flowed states correspond to worldsheets which wrap the boundary cylinder $w$ times.

The benefit of the infinite cylinder (or equivalently the punctured complex plane) is that it admits a global coordinate chart. This means that one can treat $\gamma,\overline{\gamma}$ as scalar fields on the worldsheet, and interpret their values as coordinates on the $\text{AdS}_3$ boundary. In the computation of sphere correlation functions in the boundary CFT, however, we find it much more convenient to move to Euclidean signature, for which the boundary is $\mathbb{CP}^1$. In this case, thinking of $\gamma,\overline{\gamma}$ as coordinates breaks down since $\mathbb{CP}^1$ does not possess a single global chart. Thus, we must modify the free-field theory if we want to study correlation functions in the spacetime CFT.

\begin{figure}
\centering
\begin{tikzpicture}
\begin{scope}
\draw[smooth,thick] (0,1) to[out=30,in=150] (2,1) to[out=-30,in=210] (3,1) to[out=30,in=150] (5,1) to[out=-30,in=30] (5,-1) to[out=210,in=-30] (3,-1) to[out=150,in=30] (2,-1) to[out=210,in=-30] (0,-1) to[out=150,in=-150] (0,1);
\draw[smooth,thick] (0.4,0.1) .. controls (0.8,-0.25) and (1.2,-0.25) .. (1.6,0.1);
\draw[smooth,thick] (0.5,0) .. controls (0.8,0.2) and (1.2,0.2) .. (1.5,0);
\draw[smooth,thick] (3.4,0.1) .. controls (3.8,-0.25) and (4.2,-0.25) .. (4.6,0.1);
\draw[smooth,thick] (3.5,0) .. controls (3.8,0.2) and (4.2,0.2) .. (4.5,0);
\fill (0,0.25) circle (0.05);
\draw[thick, fill = blue, fill opacity = 0.05] (0,0.25) [partial ellipse = 0:360:0.3 and 0.15];
\fill (2,0.5) circle (0.05);
\draw[thick, dashed, fill = blue, fill opacity = 0.05] (2,0.5) [partial ellipse = 0:360:0.3 and 0.2];
\fill (3,-0.5) circle (0.05);
\draw[thick, fill = blue, fill opacity = 0.05] (3,-0.5) [partial ellipse = 0:360:0.3 and 0.2];
\fill (5,-0.25) circle (0.05);
\draw[thick, dashed, fill = blue, fill opacity = 0.05] (5,-0.25) [partial ellipse = 0:360:0.3 and 0.15];
\node[above] at (4,2) {$\gamma^{-1}(U)$};
\draw[thick, -latex] (4,2) -- (4,0.75);
\node[above] at (0,2.25) {$\gamma^{-1}(V)$};
\draw[thick, -latex] (0,2.25) -- (0,0.35);
\draw[thick, -latex] (6,0) -- (7.5,0);
\node[above] at (6.75,0) {$\gamma$};
\end{scope}
\begin{scope}[xshift = 10cm]
\fill[blue, opacity = 0.05] (0,0) [partial ellipse = 30:150:2 and 2];
\draw[thick] (0,0) circle (2);
\draw[thick, fill = blue, fill opacity = 0.05] (0,1) [partial ellipse = 0:-180:1.73 and 0.4];
\draw[thick, dashed] (0,1) [partial ellipse = 0:180:1.72 and 0.4];
\node at (0,-1) {$U$};
\node at (0,1) {$V$};
\fill (0,2) circle (0.05);
\node[above] at (0,2) {$\infty$};
\end{scope}
\end{tikzpicture}
\caption{The field $\gamma$ is a map from the worldsheet to the Riemann sphere $\mathbb{CP}^1$. Crucially, $\gamma$ is only a free field locally, i.e.~in the preimage of a local patch of $\mathbb{CP}^1$. We can treat the global theory by gluing together two coordinate patches $U,V$ on $\mathbb{CP}^1$ and treating the worldsheet theory differently on the preimages of $U,V$ under $\gamma$. Shrinking $V$ to contain only the point at infinity effectively results in introducing points on the worldsheet where $\gamma$ is allowed to have a pole, despite no vertex operator being inserted there.}
\label{fig:coordinate-patches}
\end{figure}
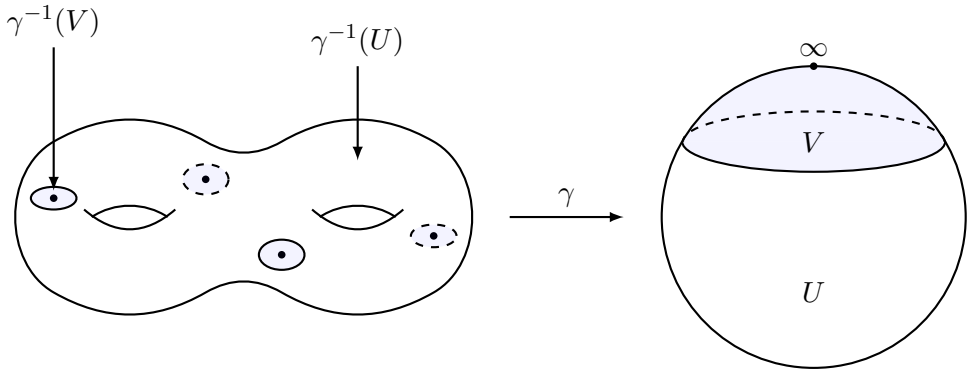

In order to consider the `global' theory, one would need to consider two coordinate patches $U,V$ on $\mathbb{CP}^1$. The field $\gamma$ restricted to the sets $\gamma^{-1}(U)$ and $\gamma^{-1}(V)$ can be treated as free fields, since they can be taken to be coordinate functions in some open set of the complex plane (see Figure \ref{fig:coordinate-patches}). The worldsheet theory then needs to be carefully treated so that $\gamma|_{\gamma^{-1}(U)}$ and $\gamma|_{\gamma^{-1}(V)}$ agree on the overlaps of $\gamma^{-1}(U)$ and $\gamma^{-1}(V)$. This is technically possible (see \cite{Nekrasov:2005wg} for a detailed exposition on this topic), but in practice quite cumbersome.

An alternative approach was suggested by \cite{Frenkel:2005ku}. We take $U$ to be $\mathbb{CP}^1\setminus\{\infty\}$ so that $\gamma$ can be taken to live in the complex plane. In order to include the point at infinity, we introduce a `fictitious' vertex operator $D$ (which is a singlet with respect to $\mathfrak{psu}(1,1|2)_1$, see below) which has the OPE
\begin{equation}
\gamma(z)D(\lambda)\sim\mathcal{O}\left(\frac{1}{z-\lambda}\right)\,.
\end{equation}
If the operator $D$ is inserted at the point $\lambda$ on the worldsheet, $\gamma$ will have a pole there, and thus we can effectively include the point at infinity into the range of the field $\gamma$.

Since the field $D$ should not correspond to a state in the dual CFT, but rather just be an artifact of chosing a single coordinate system on $\mathbb{CP}^1$, $D$ should be uncharged with respect to the boundary conformal algebra. Specifically, it should be a singlet with respect to the algebra $\mathfrak{psu}(1,1|2)_1$. Furthermore, in order to induce a simple pole with $\gamma$, it should lie in the $w=-1$ spectrally-flowed sector of the worldsheet theory. A natural candidate for such an operator is\footnote{If we restrict to states of the form $e^{ia f_1+ibf_2+c\phi+id\kappa}$, then $D$ is the unique operator having a simple pole with $\gamma$, worldsheet dimension $\Delta=1$, spacetime dimension $h=0$, and which transforms as a singlet under $\mathfrak{su}(2)_1$.}
\begin{equation} \label{eq:d-secret}
D=p_1p_2\delta(\beta)=e^{-if_1+if_2}\delta(\beta)\,.
\end{equation}
Indeed, (the integral of) $D$ is a singlet with respect to $\mathfrak{psu}(1,1|2)_1$, as can be checked by computing its OPEs with the generators defined in Section \ref{sec:new-free-field-realization}.\footnote{Note that although the actions of $J^-$, $S^{---}$ and $S^{-+-}$ on $D$ are non-zero, they are total derivatives, for example
\begin{equation*}
J^-(z)D(w)\sim\partial_w\left(\frac{p_1p_2e^{-2\phi-i\kappa}(w)}{z-w}\right)+\cdots\,.
\end{equation*}
Since, as we will discuss in a moment, we integrate $D$ over the whole worldsheet, these total derivatives do not contribute.}

The worldsheet conformal dimension of $D$ is $(1,1)$, and so it classically defines a marginal deformation on the worldsheet. In fact, $D$ is exactly marginal at the quantum level, as can be seen as follows. The physical state on the worldsheet in the $w=-1$ spectrally-flowed sector, according to Section \ref{sec:physical-states}, is given by
\begin{equation}
\Omega_{w=-1}=e^{2\rho+i\sigma+iH}e^{i\kappa}\,.
\end{equation}
This means that it satisfies eq.~\eqref{eq:physical}. In order to understand the deformation better, we consider the $\mathcal{N}=2$ primary fields following \cite{Berkovits:1994vy}, i.e.~we consider state $\widetilde{\Omega}_{w=-1}$ such that
\begin{equation}
    \Omega_{w=-1}=\widetilde{G}^+_0 \widetilde{\Omega}_{w=-1} \,, \qquad \widetilde{\Omega}_{w=-1} = e^{i\kappa} e^{\rho+i\sigma} \,.
\end{equation}
Now note that $\widetilde{\Omega}_{w=-1}$ has both zero conformal dimension and charge with respect to the worldsheet $\mathcal{N}=2$ algebra, and so is a BPS state. Therefore, if we deform the theory by a super descendant of it, the $\mathcal{N}=4$ structure remains unchanged \cite{Berkovits:1994vy,Blumenhagen:2013fgp}.\footnote{Note that $\widetilde{\Omega}_{w=-1}$ is a singlet with respect to $\mathcal{R}$-symmetry generators $J$ and $J^{\pm\pm}$ and therefore it indeed preserves the $\mathcal{N}=4$ structure \cite{Berkovits:1994vy}.} In fact, the prescription for deforming the theory is considering the following insertion \cite{Berkovits:1994vy}
\begin{equation}
    D = G^-_{-1} G^+_0 \widetilde{\Omega}_{w=-1} = \widetilde{G}^-_{-1} \Omega_{w=-1} \ ,
\end{equation}
where both equalities can be directly checked. This shows that $D$ is exactly marginal.

Since $D$ is exactly marginal, we propose that the correct worldsheet theory for studying dual CFT correlation functions in $\text{AdS}_3$ is governed by the action
\begin{equation}
S=\frac{1}{2\pi}\int(\beta\overline{\partial}\gamma+\overline{\beta}\partial\overline{\gamma}+p_a\overline{\partial}\theta^a+\overline{p}_a\partial\overline{\theta}^a+D\overline{D})\,.
\end{equation}
Note that $D$ in this description plays a role analogous to a screening operator in the Coulomb gas description of the $\text{SL}(2,\mathbb{R})$ WZW model \cite{Gerasimov:1990fi,Giribet:1999ft,Giribet:2001ft, DiFrancesco:1997nk}. Specifically, the screening operator $\mathcal{S}_-$ of \cite{Giribet:2001ft} at $k=1$ takes the form (translating into our fields)
\begin{equation}
\mathcal{S}_-=p_1p_2\beta^{-1}\,.
\end{equation}
The operators $D$ and $\mathcal{S}_-$ share many features: they have the same worldsheet and spacetime scaling dimensions, are both marginal, and both have simple poles in the OPE with $\gamma$. As we will see in Section \ref{sec:localization}, $D$ has a much more natural interpretation in the path integral formalism, and it is thus our proposal that $D$ is the `correct' screening operator to include.

\subsection{As a `secret' representation}

In \cite{Eberhardt:2019ywk}, a similar idea was presented. In their analysis, they considered correlation functions of the $\text{SL}(2,\mathbb{R})$ WZW model, and found that certain solutions to the local $\mathfrak{sl}(2,\mathbb{R})_k$ Ward identities satisfied the property
\begin{equation}
\Braket{\gamma(z)\prod_{i=1}^{n}V^{w_i}_i(x_i,z_i)}=\Gamma(z)\Braket{\prod_{i=1}^{n}V^{w_i}_{i}(x_i,z_i)}\,,
\end{equation}
where $\gamma(z)$ is the Wakimoto field in $\mathfrak{sl}(2,\mathbb{R})_k$ and $\Gamma:\Sigma\to\mathbb{CP}^1$ is a branched covering map from the worldsheet to the boundary.\footnote{Here, we do not notationally differentiate between the $\gamma$ of the free field realization of $\mathfrak{psu}(1,1|2)_1$ and that of the Wakimoto representation since, as we saw in Section \ref{sec:worldsheet theory}, these are essentially the same field.} However, such a map $\Gamma$ must have poles at points $z=\lambda_a$ on the worldsheet which are distinct from the points $z_i$. This would seem to imply that $\gamma$ has a nontrivial OPE with a field inserted at $z=\lambda_a$, despite no such field being inserted into the correlator.

The authors of \cite{Giribet:2019new,Eberhardt:2019ywk} proposed the existence of a field $\varphi$ which is a singlet with respect to $\mathfrak{sl}(2,\mathbb{R})_k$ but which was not a singlet with respect to the Wakimoto variables (in their parlence, such a field would live in a `secret' representation of the Wakimoto algebra). Their state was the lift of the state $\ket{m=\tfrac{k}{2},j=\tfrac{k}{2}}^{(w=-1)}$ in the discrete representation $\sigma^{-1}(\mathcal{D}^{j=\tfrac{k}{2}})$ to the Wakimoto variables. For $k=1$, their definition translated into the free field realization \eqref{new-free-field-realization} would give
\begin{equation}
\varphi=e^{i\kappa}\,.
\end{equation}
While this does not reproduce the state $D$ described above, we note that
\begin{equation}
D= Q_{-1}\varphi = \widetilde G^-_{-1} \Omega_{w=-1}\,,
\label{D=Qphi}
\end{equation}
where $Q=p_1p_2\partial\gamma$ is the field defined in Appendix~\ref{app:hybrid}. Thus, while not exactly the same state as the secret representation of \cite{Eberhardt:2019ywk}, our field $D$ is directly related to it. Notice that the presence of $Q_{-1}$ in \eqref{D=Qphi} is quite natural. In fact, as discussed in \cite{Dei:2020zui} and reviewed in section \ref{sec:correlation-functions}, in the hybrid formalism, the physical operators are dressed by the action of $Q_{-1}$ when inserted in correlation functions.

In the conventions of \cite{Eberhardt:2019ywk}, the OPE of the free boson entering the Wakimoto representation with the secret representation field has a simple pole with residue $1$. Taking into account the different normalization of $\partial \Phi$ and translating to our conventions, this reads
\begin{equation}
    \frac{i}{\sqrt 2} \partial \Phi(z) D(\lambda) \sim \frac{D(\lambda)}{z-\lambda} \,.
    \label{dPhi-D-OPE}
\end{equation}
Comparing eq.~\eqref{eq:sl2-currents} with eq.~\eqref{eq:wakimoto-sl2r} we get
\begin{equation}
    \frac{i}{\sqrt{2}} \partial \Phi = \frac{1}{2} (p_a \theta^a) = \frac{1}{2} \partial(-if_1+if_2) \,,
\end{equation}
and it is easy to check that eq.~\eqref{dPhi-D-OPE} is satisfied, in agreement with the result of \cite{Eberhardt:2019ywk}.
As we will see shortly, the crucial difference between $\varphi$ and $D$ is that we will consider the effect of $D$ integrated over the worldsheet, whereas $\varphi$ was taken in \cite{Eberhardt:2019ywk} to lie at specific insertion points. However, although the locations of the operator $D$ are integrated over the worldsheet, these integrals will wind up localizing to a finite set of points, corresponding to the poles of a holomorphic covering map. Thus, while our prescription is a priori different from that of \cite{Eberhardt:2019ywk}, the end result is effectively the same, and we wind up with extra field insertions at a discrete set of points on the worldsheet. Hence, we will still sometimes refer to $D$ as the `secret representation', although it is not precisely the state considered in \cite{Eberhardt:2019ywk}.

\subsection{Correlation functions and holomorphic maps} 
\label{sec:localization}

Since $D$ is a marginal operator on the worldsheet, we can use it to implement a deformation of the free field theory. That is, we can modify the action by adding the term
\begin{equation}
\xi\int_{\Sigma}D\overline{D}=\xi\int_{\Sigma} Q_{-1}\overline{Q}_{-1}\varphi\overline{\varphi}\,.
\end{equation}
In conformal perturbation theory, correlation functions of the deformed theory can be computed as 
\begin{equation}
\Braket{\cdots}_{\xi}=\sum_{N=0}^{\infty}\frac{\xi^N}{N!}\Braket{\left(\prod_{a=1}^{N}\int_{\Sigma}\mathrm{d}^2\lambda_a\,D\overline{D}\right)\cdots} \,, 
\label{conf-pert}
\end{equation}
where the dots denote an arbitrary collection of vertex operators. On the right-hand-side of \eqref{conf-pert}, our correlation functions will then be computed by integrating over configurations of $\gamma$ for which $\gamma$ has a pole at $\lambda_a$. The $1/N!$ factor will cancel the symmetry corresponding to permutations of the $\lambda_a$'s, and the resulting path integral will include all possible maps $\gamma:\Sigma\to\mathbb{CP}^1$. The order of $\xi$ appearing in any correlator will read off the number of poles of $\gamma$, i.e.~the number of pre-images of infinity on the worldsheet. Thus, the power of $\xi$ will read off the number of times that the worldsheet $\Sigma$ wraps the boundary sphere of Euclidean $\text{AdS}_3$. In a sense, the parameter $\mu=-\log{\xi}$ then acts as a chemical potential for wrapping a string around the asymptotic boundary of $\text{AdS}$.

Let us investigate these correlation functions further. Consider a (not necessarily local) operator $\mathcal{O}$ which is a function purely of $\gamma$ and the fermions $p_a,\theta^a$, but not of $\beta$. Furthermore, let us assume that $\mathcal{O}$ has a definite charge $q(\mathcal{O})$ with respect to the current $p_a\theta^a$. Then the correlation function
\begin{equation}\label{eq:perturbative-correlator}
\Braket{\mathcal{O}}_{\xi}=\sum_{N=0}^{\infty}\frac{\xi^N}{N!}\Braket{\left(\prod_{a=1}^{N}\int_{\Sigma}\mathrm{d}^2\lambda_a\,(D\overline{D})(\lambda_a)\right)\mathcal{O}}
\end{equation}
has at most one nonvanishing contribution, namely that for which
\begin{equation}\label{eq:N-conservation-general}
2N+q(\mathcal{O})=2g-2\,,
\end{equation}
which is required by the anomalous charge conservation of $p_a\theta^a$. Assuming that $\mathcal{O}$ factorizes into a $\beta \gamma$ contribution  $\mathcal{O}_{\gamma}$ and a $p_a \theta^a$ contribution $\mathcal{O}_{p,\theta}$, we can compute the correlation function of the bosonic and fermionic sectors separately. The bosonic correlator will then be defined by the path integral
\begin{equation}
\frac{\xi^N}{N!}\int_{\Sigma^{N}}\mathrm{d}^{2N}\lambda\int\mathcal{D}\beta\,\mathcal{D}\gamma\,\prod_{a=1}^{N}\delta(\beta(\lambda_a))\,\mathcal{O}_{\gamma}\,e^{-S[\beta,\gamma]}\,,
\label{path-int-bosonic}
\end{equation}
where we made us of eq.~\eqref{eq:d-secret}. The presence of delta function operators should not surprise the reader, since these are usual in string theory path integral computations, see for example \cite{Verlinde:1987sd}. Now, as noted in Section~\ref{sec:delta-function}, we can write the delta functions formally as
\begin{equation}
\delta(\beta(\lambda_a))=\int\frac{\mathrm{d}\zeta_a}{2\pi}e^{i\zeta_a\beta(\lambda_a)}\,.
\end{equation}
Inserting these delta functions thus has the effect of modifying the $\beta,\gamma$ system action, 
\begin{equation}
S[\beta,\gamma]\to\frac{1}{2\pi}\int_{\Sigma}\beta\left(\overline{\partial}\gamma-2\pi i\sum_{i=1}^{n}\zeta_a\delta^{(2)}(z,\lambda_a)\right)\,, 
\end{equation}
and therefore, upon integrating out $\beta$, the bosonic contribution \eqref{path-int-bosonic} to the path integral can be written as
\begin{equation}
\frac{\xi^N}{(2\pi)^NN!}\int\mathrm{d}^{2N}\lambda\,\mathrm{d}^N\zeta\,\int\mathcal{D}\gamma\,\mathcal{O}_{\gamma}\,\delta\left(\overline{\partial}\gamma-2\pi i\sum_{a=1}^{N}\zeta_a\,\delta^{(2)}(z,\lambda_a)\right)\,.
\end{equation}
The delta functional restricts the functional integral over complex functions $\gamma$ on $\Sigma$ to those which are meromorphic with poles at $z=\lambda_a$ and residues $\zeta_a$.\footnote{If $\gamma\sim \zeta_a/(z-\lambda_a)$, then $\overline{\partial}\gamma\sim 2\pi i\zeta_a\delta^{(2)}(z,\lambda_a)$.} Integrating over the locations and residues of the poles then extends this integral to the space
\begin{equation}\label{eq:fn-dimension}
\mathcal{F}_N=\left\{\gamma\text{ meromorphic on }\Sigma\text{ with }N\text{ poles}\right\}\,.
\end{equation}
The space $\mathcal{F}_N$ can equivalently be thought of as the space of holomorphic maps $\gamma:\Sigma\to\mathbb{CP}^1$ of degree $N$. This space is in fact finite dimensional with complex dimension
\begin{equation}
\text{dim}_{\mathbb{C}}(\mathcal{F}_N)=2N+1-g\,.
\end{equation}
Equivalently, $\text{dim}_{\mathbb{C}}(\mathcal{F}_N)=g-1-q(\mathcal{O})$ by \eqref{eq:N-conservation-general}. For $g=0$, the dimension of $\mathcal{F}_N$ is found by noting that a meromorphic function $\gamma$ on the sphere with $N$ poles is necessarily a rational function $\gamma=Q_N/P_N$ with $Q_N,P_N$ polynomials of degree $N$. Such a rational function has $2N+1$ independent coefficients.\footnote{For $g\neq 0$, the dimension of $\mathcal{F}_N$ can be computed via the Riemann-Roch theorem: the tangent space $\text{T}_{\gamma}\mathcal{F}_N$ is given by $\text{H}^0\left(\gamma^*\text{T}\mathbb{CP}^1\right)$. By Riemann-Roch, we have
\begin{equation*}
\dim\text{H}^0(\gamma^*\text{T}\mathbb{CP}^1)-\dim\text{H}^1(\gamma^*\text{T}\mathbb{CP}^1)=\text{deg}(\gamma^*\text{T}\mathbb{CP}^1)+1-g=2N+1-g\,.
\end{equation*}
Assuming $\text{H}^1(\gamma^*\text{T}\mathbb{CP}^1)\cong 0$, this gives the desired dimension of $\mathcal{F}_N$. If $\text{H}^1(\gamma^*\text{T}\mathbb{CP}^1)$ is nontrivial, then there will exist a nontrivial correction term to \eqref{eq:fn-dimension}.} The bosonic part of the correlator \eqref{eq:perturbative-correlator} is then given by the finite dimensional integral
\begin{equation}
\left(\frac{\xi}{2\pi}\right)^N\int_{\mathcal{F}_N}\mathrm{d}\gamma\,\mathcal{O}_{\gamma}\,.
\end{equation}

The localization of the infinite-dimensional integral over $\beta,\gamma$ to the finite dimensional integral over the space $\mathcal{F}_N$ of holomorphic maps $\gamma:\Sigma\to\mathbb{CP}^1$ is a hallmark of the $k=1$ theory, first proposed in \cite{Pakman:2009zz}, and demonstrated in \cite{Eberhardt:2019ywk,Eberhardt:2020akk,Dei:2020zui,Knighton:2020kuh} through careful analyses of worldsheet Ward identities. In the formalism presented in this paper, we explain the mechanism leading to localization from the path integral perspective.

As we will see in the next section, not only are these path integral manipulations useful to show from a path integral perspective why correlation functions localize, but we will use them to compute correlation functions of spectrally-flowed states. We will find that they precisely reproduce the correlators of twisted-sector states in the dual symmetric orbifold CFT$_2$.


\section{Spacetime correlation functions from the worldsheet}
\label{sec:correlation-functions}

In this section we study correlation functions of the physical states on the worldsheet. Specifically, we will observe that the free field realization discussed in Section~\ref{sec:worldsheet theory} and the delta function identities of Section~\ref{sec:delta-function} allow to carry out the path integral and calculate correlators of spectrally-flowed states. We will see that the spacetime correlation functions of symmetric orbifold twisted-sector ground states are reproduced exactly from the worldsheet.

\subsection{Symmetric orbifold correlation functions} 
\label{sec:symmetric-orbifold}

Let us start by reviewing closed-form formulae for $n$-point functions of the boundary CFT$_2$. The reader already familiar with this material, may want to skip this section and continue reading from Section~\ref{sec:outline-correlators-computation}. In the symmetric product orbifold of a seed CFT $X$ with central charge $c$, 
\begin{equation}
    \text{Sym}^K(X) = \frac{(X)^{\otimes K}}{S_K} \,,
\end{equation}
correlators of states $\mathcal O_{w_i}$ in the single cycle $w_i$-twisted sector can be computed following the covering space method of \cite{Hamidi:1986vh, Lunin:2000yv}. In a nutshell, correlators of the base space $\mathbb{CP}^1$, where fields develop non-trivial monodromies, are related to correlators on a (possibly higher-genus) covering space $\Sigma$, where all monodromies are lifted. The holomorphic map
\begin{equation}
\begin{aligned}
    \Gamma: \quad  & \Sigma && \longrightarrow && \mathbb{CP}^1 \\
    & z && \longmapsto && x
\end{aligned}
\end{equation}
from the covering space $\Sigma$ to the base space $\mathbb{CP}^1$ is called the covering map. Let us review some of its properties, which will be important in the following. Near the ramification points $z_i$, the covering map obeys the Taylor expansion
\begin{equation}
\Gamma(z) \sim x_i + a_i^\Gamma (z-z_i)^{w_i} + \mathcal O( (z-z_i)^w) \,, \qquad z \to z_i 
\label{covering-map-Taylor}
\end{equation}
and has 
\begin{equation}
    N= 1-g + \frac{1}{2} \sum_{i=1}^n (w_i-1)
    \label{Riemann-Hurwitz}
\end{equation} 
simple poles with residues $\xi_a^\Gamma$,
\begin{equation}
    \Gamma(z) \sim \frac{\xi_a^\Gamma}{z-\lambda_a^\Gamma} \,, \qquad z \to \lambda_a^\Gamma \,, \qquad a = 1, \dots \,, N \,. 
    \label{poles-Gamma}
\end{equation}
In eq.~\eqref{Riemann-Hurwitz}, $g$ denotes the genus of the covering space $\Sigma$. We are interested here in symmetric orbifold correlators in the large $K$ expansion. The leading contribution corresponds to $g=0$ and to the covering space being a sphere, $\Sigma = \mathbb{CP}^1$. In order to distinguish it from the base space sphere, we will frequently denote the covering space sphere by $\Sigma$ instead of $\mathbb{CP}^1$.  

The derivative of the covering map has double poles at $z=\lambda_a^\Gamma$ and zeros of order $w_i-1$ at $z=z_i$. At genus zero, it can thus be written as
\begin{equation}
\partial \Gamma(z) = C^\Gamma \, \frac{\prod_{i=1}^n (z-z_i)^{w_i-1}}{\prod_{a=1}^N(z-\lambda_a^\Gamma)^2} \,,
\label{dGamma}
\end{equation}
where $C^\Gamma$ is a non-trivial function of the points $z_i,x_i$ and can be computed by integrating eq.~\eqref{dGamma}. One finds 
\begin{equation}
    C^\Gamma = (x_i - x_j) \Biggl( \int_{z_i}^{z_j} \text dz \, \frac{\prod_{k=1}^n (z-z_k)^{w_k-1}}{\prod_{a=1}^N(z-\lambda_a^\Gamma)^2} \Biggr)^{-1} \,,
    \label{C-int}
\end{equation}
for any $i, j \in \{ 1 \,, \dots \,, n\}$ with $i \neq j$. It is important to notice that also $a_i^\Gamma$ entering eq.~\eqref{covering-map-Taylor} as well as $\xi_a^\Gamma$ and $\lambda_a^\Gamma$ in eq.~\eqref{poles-Gamma} are non-trivial functions of the ramification points $z_i$. In fact, comparing eqs.~\eqref{covering-map-Taylor} and \eqref{dGamma} we can express $a_i^\Gamma$ as 
\begin{equation} \label{eq:symm-aj}
    a^{\Gamma}_i = \frac{C^\Gamma}{w_i} \frac{\prod_{j\neq i} (z_i-z_j)^{w_j-1}}{\prod_a (z_i-\lambda_a^\Gamma)^2} \,. 
\end{equation}
Similarly, deriving eq.~\eqref{poles-Gamma} and comparing with \eqref{dGamma} we find 
\begin{equation} \label{eq:symm-xi}
    \xi^{\Gamma}_a = -C^\Gamma \frac{\prod_i (\lambda_a^\Gamma-z_i)^{w_i-1}}{\prod_{b \neq a} (\lambda_a^\Gamma-\lambda_b^\Gamma)^2} \,.
\end{equation}

We are now ready to discuss correlators of the symmetric product orbifold. At genus zero they take the form \cite{Hamidi:1986vh, Lunin:2000yv, Lunin:2001pw, Pakman:2009zz, Roumpedakis:2018tdb, Dei:2019iym, Hikida:2020kil, Knighton:2023xzg}\footnote{Here we restrict to connected correlators. This corresponds to the covering space being connected.}
\begin{multline}
    \left\langle \prod_{i=1}^n \mathcal O_{w_i}(x_i) \right\rangle_{\mathbb{CP}^1} =  K^{1-\frac{n}{2}} \, \prod_{i=1}^{n} w_i^{-\frac{c(w_i^2+1)}{12}} \, \sum_{\Gamma: \Sigma \to \mathbb{CP}^1} |C^\Gamma| \prod_{i=1}^n |a^{\Gamma}_i|^{-2h_i+\frac{c}{12}(w_i-1)} \\
    \times \, \prod_{a=1}^N |\xi^{\Gamma}_a|^{-\frac{c}{6}} \left\langle \prod_{i=1}^n \widetilde{\mathcal O}_{w_i}(z_i) \right\rangle_{\Sigma} \ .
    \label{eq:sym-correlators}
\end{multline}
Let us recap the various definitions entering this formula. The left-hand-side is a correlator of single cycle vertex operators in the $w_i$-twisted sector, with insertion points taking values on the base space sphere, $x_i \in \mathbb{CP}^1$. In the right-hand-side, the sum over $\Gamma$ runs over all the covering maps from $\Sigma$ to $\mathbb{CP}^1$ with ramification points $z_1, \dots, z_n$ and ramification indices $w_1, \dots,  w_n$. The various terms multiplying the covering space correlator in the right-hand-side may be interpreted as the conformal factors due to the conformal transformation relating the covering to the base space. We already introduced $a_j^\Gamma$, $\xi_a^\Gamma$ and $C^\Gamma$ respectively in eqs.~\eqref{covering-map-Taylor}, \eqref{poles-Gamma} and \eqref{dGamma}.\footnote{Let us briefly comment on the factor $|C^\Gamma|$ which enters eq.~\eqref{eq:sym-correlators} and is sometimes overlooked in the literature. Notice that under base space dilatations, $C^\Gamma$ scales as the $x_i$, see eq.~\eqref{C-int}. Using the Riemann-Hurwitz formula one sees that the presence of $C^\Gamma$ is necessary to ensure the correct scaling of the right-hand-side of \eqref{eq:sym-correlators}. More rigorously, one can recover this factor by tracing the various factors of $C^\Gamma$ entering the covering space method derivation outlined in Section~2 of \cite{Hikida:2020kil} and reviewed in \cite{Knighton:2023xzg}.} We denoted by $c$ the central charge of the seed theory $X$, which in the case of $X=\mathbb T^4$ reads $c=6$. The states $\widetilde{\mathcal O}_{w_i}$ are obtained by lifting $\mathcal O_{w_i}$ to the covering space. The base space conformal dimensions $h_i$ of $\mathcal O_{w_i}$ is related to the covering space conformal dimension $\Delta_i$ of $\widetilde{\mathcal O}_{w_i}$ as 
\begin{equation}
    h_i = \frac{c}{24} \frac{w_i^2-1}{w_i} + \frac{\Delta_i}{w_i} \,. 
    \label{h-Delta-relation}
\end{equation}
Notice that the correlator in the right-hand-side of \eqref{eq:sym-correlators} is computed on the covering surface $\Sigma$, i.e.~$z_1, \dots, z_n \in \Sigma$. 

In the rest of this section we will focus on correlators of $w_i$-twisted ground states $\sigma_{w_i}$ of the symmetric orbifold of $\mathbb{T}^4$. As was already mentioned in Section~\ref{sec:The worldsheet spectrum}, $\sigma_{w_i}$ has base space conformal dimension 
\begin{equation}
    h_i = \frac{w_i^2-1}{4 w_i} \,. 
\end{equation}
From \eqref{h-Delta-relation} we see that when lifted to the covering space they have conformal dimensions 
\begin{equation}
    \Delta_i = 0 \,,
\end{equation}
and the vertex operator $\widetilde{\mathcal O}_{w_i}$ in this case is simply the identity operator. Eq.~\eqref{eq:sym-correlators} then simplifies to  
\begin{equation}
    \left\langle \prod_{i=1}^n \sigma_{w_i}(x_i) \right\rangle_{\mathbb{CP}^1} = K^{1-\frac{n}{2}} \, \prod_{i=1}^{n} w_i^{-\frac{(w_i^2+1)}{2}} \sum_{\Gamma: \Sigma \to \mathbb{CP}^1} |C^\Gamma| \prod_{i=1}^n |a^{\Gamma}_i|^{-2h_i+\frac{w_i-1}{2}} \prod_{a=1}^N |\xi^{\Gamma}_a|^{-1} \,.
   \label{eq:spacetime-correlator}
\end{equation}

\subsection{Outline of the calculation}
\label{sec:outline-correlators-computation}

The genus expansion of symmetric orbifold correlators resembles the genus expansion of string theory and in the large $K$ limit the string coupling and $K$ are identified as
\begin{equation}
    g_s \sim \frac{1}{\sqrt K} \,. 
\end{equation}
Our task is now to reproduce eq.~\eqref{eq:spacetime-correlator} from tree-level string theory. Before going into the details of the calculation of the worldsheet correlators in the hybrid formalism, let us provide an outline. The correlation functions we want to compute are of the rough form
\begin{equation}
\Braket{\prod_{i=1}^{n}\Phi_{w_i}(x_i,z_i)}\,,
\end{equation}
where $\Phi_w$ enter the expression for the worldsheet operators dual to the twisted-sector ground states in the symmetric orbifold, see eqs.~\eqref{odd-w-ground-state}-\eqref{eq:m-w-even}. These states have the form
\begin{equation}
\Phi_{w}=\Phi_w^{\text{ferm}}\left(\frac{\partial^{w}\gamma}{w!}\right)^{-m}\delta_{w}(\gamma-x)\,,
\end{equation}
where $\Phi_{w}^{\text{ferm}}$ depends only on the fermions $\theta^1,\theta^2$ and $m$ is defined in terms of $w$ through equation \eqref{eq:m-w-odd} for $w$ odd and $\eqref{eq:m-w-even}$ for $w$ even: see eqs.~\eqref{eq:w-odd-twisted-delta} and \eqref{eq:w-even-twisted-delta}. Focusing only on the $\beta,\gamma$ system, we are thus tasked with calculating the correlator
\begin{equation}
\Braket{\prod_{i=1}^{n}\left(\frac{\partial^{w_i}\gamma(z_i)}{w_i!}\right)^{-m_i}\delta_{w_i}(\gamma(z_i)-x_i)}_{\beta,\gamma}\,.
\end{equation}
Let us for the moment keep the genus $g$ generic.

As described in Section~\ref{sec:localization}, since we are considering correlators on global Euclidean $\text{AdS}_3$, we have to deform the free-field action by the operator $D$, which represents the point at infinity in $\mathbb{CP}^1\cong\partial\text{AdS}_3$. In particular, the number of such insertions will be determined by the overall charge conservation of the $p_a,\theta^a$ systems, see eq.~\eqref{eq:N-conservation-general}. The path integral we will need to calculate then becomes (again, focusing only on the $\beta \gamma$ dependence)\footnote{As we will discuss below, this expression will be slightly modified when we consider the correlators in the hybrid formalism.}
\begin{equation}\label{eq:schematic-path-integral}
\frac{1}{N!}\int\mathrm{d}^{2N}\lambda\,\int\mathcal{D}(\beta,\gamma)\,\prod_{a=1}^{N}\delta(\beta(\lambda_a))\prod_{i=1}^{n}\left(\frac{\partial^{w_i}\gamma(z_i)}{w_i!}\right)^{-m_i}\delta_{w_i}(\gamma(z_i)-x_i)\,.
\end{equation}
Below, we will see that at tree level the fermionic charge conservation will require $N$ to be given by the Riemann-Hurwitz formula
\begin{equation}
N=1-g+\sum_{i=1}^{n}\frac{w_i-1}{2}\,, 
\end{equation}
with $g=0$. In fact, this also holds for higher genus \cite{higher-genus-paper}. For this value of $N$, the argument of Section~\ref{sec:localization} tells us that, upon integrating out $\beta$, the path integral over $\gamma$ will reduce to the space $\mathcal{F}_N$ of holomorphic maps $\gamma:\Sigma\to\mathbb{CP}^1$ with degree $N$. This space has dimension
\begin{equation}
\text{dim}\,\mathcal{F}_N=2N+1-g=\sum_{i=1}^{n}w_i-(n+3g-3)\,.
\end{equation}
Moreover, the delta functions in \eqref{eq:schematic-path-integral} will impose $\sum_{i=1}^{n}w_i$ conditions on $\gamma$ such that the dimension of the path integral will become $-(n+3g-3)=-\text{dim}(\mathcal{M}_{g,n})$. This means that, unless the moduli of the worldsheet are precisely tuned to a zero-dimensional sublocus of $\mathcal{M}_{g,n}$, the path integral will vanish.

When the path integral does not vanish, the field $\gamma$ will be constrained to be a holomorphic map $\Gamma:\Sigma\to\mathbb{CP}^1$ of degree $N$ which satisfies
\begin{equation}
\Gamma(z)\sim x_i+a_i^{\Gamma}(z-z_i)^{w_i}+\cdots
\end{equation}
near the insertion points $z_i$. Such a map is a \textit{ramified covering map} of the boundary sphere. As discussed above, these maps are intrinsically linked to the computation of correlation functions of twist fields in the symmetric orbifold CFT.

In order to compare the correlation functions on the worldsheet to those of the symmetric orbifold, however, we will need to know more than that the worldsheet path integral localizes; we will need to know the Jacobian that arises upon integration over $\mathcal{M}_{g,n}$. Since we are interested in planar contributions to the symmetric orbifold, let us now specialize to genus $g=0$ on the worldsheet. In Appendix \ref{app:path-integral}, we will argue that for any operator $\mathcal{O}(\gamma)$ depending only on $\gamma$ (as well as the moduli of the worldsheet), the $\beta,\gamma$ path integral gives
\begin{equation} \label{eq:measure}
\begin{split}
&\frac{1}{N!}\int\mathrm{d}^{2N}\lambda\,\int\mathcal{D}(\beta,\gamma)\,\prod_{a=1}^{N}\delta(\beta(\lambda_a))\delta_{w_i}(\gamma(z_i)-x_i)\,\mathcal{O}(\gamma)\\
&\hspace{1.5cm}=z_{12}^{-1}z_{23}^{-1}z_{13}^{-1}\sum_{\Gamma}(C^{\Gamma})^{2}\prod_{i=1}^{n}(a_i^{\Gamma})^{-\frac{w_i+1}{2}}\mathcal{O}(\Gamma)\,\delta^{(n-3)}(\Gamma)\,,
\end{split}
\end{equation}
where the sum is over all holomorphic covering maps $\Gamma:\Sigma\to\mathbb{CP}^1$ satisfying $\Gamma(z_i)=x_i$ for $i=1,2,3$ (such maps always exist) and $z_{ij}=z_i-z_j$. The delta function $\delta^{(n-3)}(\Gamma)$ is given by demanding $\Gamma(z_i)=x_i$ for $i\geq 4$, specifically
\begin{equation}
\delta^{(n-3)}(\Gamma):=\prod_{i=4}^{n}\delta(z_i-\Gamma^{-1}(x_i))\,.
\end{equation}
Using this identity, the calculation of correlators in the worldsheet theory becomes a simple one involving Wick contractions of the various fermions and ghosts. The result, after integrating over the moduli space $\mathcal{M}_{g,n}$, will reproduce the form \eqref{eq:spacetime-correlator} of the spacetime correlation functions.

\subsection{String correlators in the hybrid formalism}

In this section and the next, we discuss our calculation of the tree-level correlation function in string theory. Let us review how tree-level string correlation functions are computed in the hybrid formalism. Consider $n$ physical vertex operators $V_j$ with $j=1, \dots n$, which satisfy eq.~\eqref{eq:physical}. The associated string correlator on the sphere reads~\cite{Berkovits:1999im,Berkovits:1994vy}\footnote{We thank Cassiano Daniel for related discussions and pointing out eq.~\eqref{eq:hybrid-correlators} to us.} 
\begin{equation}
    \mathcal I = \int \mathrm{d}^2 z_4 \cdots \mathrm{d}^2 z_n \, I(z_1, z_2 , z_3, z_4, \dots, z_n)
    \label{eq:hybrid-correlators-integrated}
\end{equation}
where $z_1$, $z_2$ and $z_3$ are fixed and\footnote{We define the picture number of $V_j$ as $P_j V_j = (\partial \rho)_0 V_j$ and assume that the sum of picture numbers of $V_j$ reads $\sum_j P_j = -2n$. In fact, if this is not the case, the correlator of $V_j = \phi_j e^{2 \rho + i \sigma + i H}$ trivially vanishes \cite{Dei:2020zui, Gaberdiel:2021njm}.}
\begin{equation} \label{eq:hybrid-correlators}
    I(z_1, \dots, z_n) = \Big|\Big\langle \widetilde{V}_1(x_1,z_1) \, V_2(x_2,z_2) \, [G^+_0 \widetilde{V}_3](x_3,z_3) \prod_{j=4}^n [G^-_{-1} G^+_0 \widetilde{V}_j](x_j,z_j) \Big\rangle\Big|^2 \,.
\end{equation}
The vertex operator $\widetilde V_j$ is related to $V_j$ by
\begin{equation}
    \widetilde{V}_j = -(e^{-\rho-i H})_0 V_j \,, \qquad  V_j = \widetilde{G}^+_0 \widetilde V_j \,, \qquad \widetilde{G}^+=e^{\rho+i H} \,.
\end{equation}
Using\footnote{Eqs.~\eqref{eq:g+j++} and \eqref{eq:comparison-hybrid-u} can be checked using the Jacobi identity and imposing $\rho$ and $\sigma$ charge conservation for vertex operators of the form $V_j = \phi_j e^{n\rho+i\sigma+ik_1 H^1+ik_2 H^2}$, where $\phi_j$ is a state built out of $\mathfrak{psu}(1,1|2)_1$ generators and $\mathbb{T}^4$ worldsheet bosons. The physical state condition \eqref{eq:physical} requires $n=k_1+k_2$. Eq.~\eqref{eq:comparison-hybrid-u} only holds inside correlation functions.}
\begin{equation} \label{eq:g+j++}
   G^+_0 \widetilde V_j = -G^+_0 (e^{-\rho-iH})_0 V_j = -(e^{i\sigma})_{1} \widetilde{G}^-_{-1} V_j 
\end{equation}
and
\begin{equation} \label{eq:comparison-hybrid-u}
    G^-_{-1} G^+_0 \widetilde{V}_j = \widetilde{G}^-_{-1} V_j \,, 
\end{equation}
eq.~\eqref{eq:hybrid-correlators} can equivalently be written as
\begin{equation}
    I = \Big|\Big\langle [(e^{-\rho-iH})_0 V_1](x_1,z_1) \, V_2(x_2,z_2) \,  [(e^{i\sigma})_{1} \widetilde G^-_{-1} V_3](x_3, z_3) \prod_{j=4}^n [\widetilde G^-_{-1} V_j](x_j,z_j) \Big\rangle\Big|^2 \,.
    \label{eq:hybrid-correlators-2}
\end{equation}

Strictly speaking, eqs.~\eqref{eq:hybrid-correlators} and \eqref{eq:hybrid-correlators-2} are only valid for $n \geq 3$. In fact, for $n=2$, the right-hand-side of eq.~\eqref{eq:hybrid-correlators-2} reduces to 
\begin{equation}
   \Big|\Big\langle [(e^{-\rho-iH})_0 V_1](x_1,z_1) V_2(x_2,z_2) \Big\rangle\Big|^2 \,, 
\end{equation}
which vanishes since the background charge of $\sigma$ (which is $3$) is not saturated. In order to define $n$-point functions with $n \geq 2$, in \cite{Dei:2020zui,Gaberdiel:2021njm,Fiset:2022erp} a slightly different prescription was used and correlators were defined as 
\begin{equation} \label{eq:iprime}
    \mathcal I^{\prime} = \int \mathrm{d}^2 z_4 \cdots \mathrm{d}^2 z_n\, \Big|\Big\langle [(e^{i\sigma})_0 (e^{-\rho-iH})_0 V_1](x_1,z_1) V_2(x_2,z_2) \prod_{j=3}^n [\widetilde{G}^-_{-1} V_j](x_j,z_j) \Big\rangle\Big|^2 \,.
\end{equation}
We see that eq.~\eqref{eq:iprime} differs from eqs.~\eqref{eq:hybrid-correlators-integrated} and \eqref{eq:hybrid-correlators-2} only for the position where $e^{i\sigma}$ is inserted. Notice that all the results derived in \cite{Dei:2020zui,Gaberdiel:2021njm,Fiset:2022erp} for $n$-point correlators with $n \geq 3$ would not get modified should one repeat their computations using eq.~(\ref{eq:hybrid-correlators}) in place of \eqref{eq:iprime}. In fact, in \cite{Dei:2020zui,Gaberdiel:2021njm,Fiset:2022erp} the authors always effectively considered ratios of correlation functions in their calculations, for which the precise location of $e^{i \sigma}$ does not matter.

\subsection[Tree-level correlator of $w$-twisted ground states from the worldsheet]{Tree-level correlator of $\boldsymbol w$-twisted ground states from the worldsheet} \label{sec:sphere}

Now we proceed to calculate the correlation function in eq.~\eqref{eq:hybrid-correlators-integrated}. As we saw previously in Section~\ref{sec:localization}, the correlation functions of $\beta$ $\gamma$ localize. For this reason, it is enough to focus on the left-moving part of the correlation function, as the right-moving piece is similar. The final answer is then an integral over the product of the left and the right components. To be specific, we consider the $n$-point function of the $w$-twisted ground states for $w$ odd, i.e.\ we set
\begin{equation}
    V_i = \Omega_{w_i} \,, \quad i\in\{1,\cdots,n\} \,,
\end{equation}
see eqs.~\eqref{odd-w-ground-state} and \eqref{eq:w-odd-twisted-delta}. The calculation allowing a mixture of $w$ being odd or even is almost identical, as the expressions for the vertex operators are similar, see eq.~\eqref{eq:w-even}. We will comment on this at the end of this section. Recall that we should also include secret representations inside the correlation functions, as discussed in Section~\ref{sec:point at infinity}, and we will do this explicitly in the following. We begin by calculating the unintegrated correlation function in eq.~(\ref{eq:hybrid-correlators}). We split the left-moving part of the unintegrated correlation function into three different contributions:
\begin{itemize}
\item the one involving $\rho$, $\sigma$ and $H$, denoted by $I_{\text{rest}}$;
\item the one involving $f_1$ and $f_2$, denoted by $I_{\text{f}}$;
\item the one involving $\beta$ and $\gamma$, denoted by $I_{\text{b}}$.
\end{itemize}
Since the $\beta\gamma$ correlation function localizes to the covering maps, when we compute the other contributions, we already assume that they are localized on the corresponding locus. Explicitly putting the secret representations, the unintegrated correlation function then is
\begin{equation} \label{eq:i-total}
    I (z_1, \dots, z_n) = |I_{\text{rest}} I_{\text{f}} I_{\text{b}}|^2 \,.
\end{equation}

\paragraph{The contribution of $\boldsymbol{\rho, \sigma}$ and $\boldsymbol H$} Let us start with the first contribution $I_{\text{rest}}$. It equals
\begin{equation}
    I_{\text{rest}}=\left< e^{\rho+i\sigma}(z_1) e^{2\rho+i\sigma+iH}(z_2) e^{i\sigma}(z_3)\right> = (z_1-z_2)^{-1} (z_2-z_3) (z_3-z_1) \,.
    \label{Irest}
\end{equation}

\paragraph{The contribution of $\boldsymbol{f_1}$ and $\boldsymbol{f_2}$} The second part is more complicated and reads
\begin{equation}
    I_{\text{f}}=\left< \prod_{a=1}^N e^{-if_1+if_2}(\lambda_a)  e^{\frac{w_1+1}{2}(if_1-if_2)}(z_1) e^{\frac{w_2+1}{2}(if_1-if_2)}(z_2) \prod_{j=3}^n e^{\frac{w_j-1}{2}(if_1-if_2)}(z_j) \right> \,,
\end{equation}
where we have used the bosonization of our fields (see Section~\ref{sec:new-free-field-realization}) and eq.~\eqref{eq:d-secret}, 
\begin{equation}
    D = e^{-if_1 + if_2} \delta(\beta) \,.
\end{equation}
The background charge conservation for $f_1$ and $f_2$ imposes that
\begin{equation} \label{eq:N-degree-covering}
    N = 1 + \sum_{j=1}^n \frac{w_j-1}{2} \,,
\end{equation}
which fixes $N$ as the degree of the covering map. This correlator can be calculated using Wick contractions as all the vertex operators are exponentials of free bosons. It is equal to
\begin{align}
    I_{\text{f}}&=(z_1-z_2)^{w_1+w_2} \prod_{a=1}^N \prod_{j=1}^n (z_j-\lambda_a)^{-(w_j-1)} \prod_{a<b} (\lambda_a-\lambda_b)^2 \prod_{i<j} (z_i-z_j)^{\frac{(w_i-1)(w_j-1)}{2}} \nonumber \\ 
    & \hspace{1cm} \times \, \prod_{a=1}^N (z_1-\lambda_a)^{-2} (z_2-\lambda_a)^{-2} \prod_{j=3}^n (z_1-z_j)^{w_j-1} (z_2-z_j)^{w_j-1}  \,.
\end{align}
Note that using eqs.~(\ref{eq:symm-aj}) and (\ref{eq:symm-xi}) this can also be written as
\begin{equation} \label{eq:If-expression}
    I_{\text{f}} = (C^{\Gamma})^{-\frac{3}{2}} \Big(\prod_{a=1}^N [\xi^{\Gamma}_a]^{-\frac{1}{2}} \prod_{j} [a_j^{\Gamma}]^{\frac{w_j-1}{4}} \Big) \times a_1^{\Gamma} a_2^{\Gamma} (z_1-z_2)^2 \,,
\end{equation}
where we neglected $w_j$-dependent constant factors.

\paragraph{The contribution of $\boldsymbol \beta$ and $\boldsymbol \gamma$} Let us now focus on the $\beta \gamma$ correlator. The unintegrated correlator is
\begin{equation} \label{eq:ib-sphere}
    I_{\text{b}}=\left< \prod_{a=1}^N \delta(\beta) \prod_{i=1}^2\Big( \frac{\partial^{w_i} \gamma}{w_i!} \Big)^{-m_{w_i}} \delta_w(\gamma(z_i)-x_i) \prod_{j=3}^n \Big( \frac{\partial^{w_j} \gamma}{w_j!} \Big)^{-m_{w_j}+1} \delta_w(\gamma(z_j)-x_j) \right> \,.
\end{equation}
As already mentioned above, we will compute $I_{\text b}$ as a path integral. Particular attention should be devoted to various Jacobian factors that arise when integrating $\delta_w(\gamma(z)-x)$ and similar operators. While in Appendix~\ref{app:path-integral} we rigorously compute these factors by relating $I_{\text{b}}$ to correlators of a topological string theory, let us present a heuristic argument which should at least convince the reader that these Jacobians are actually relevant. 

Let us consider the set of meromorphic functions of $z$. Recall that $\delta_w(f(z)-x)$ denotes an object that inside the path integral forces $f(z)-x$ to have a zero of order $w$ and can be written as
\begin{equation} \label{eq:delta-meromorphic}
    \delta_w(f(z)-x) = \delta(f(z)-x) \prod_{j=1}^{w-1} \delta((\partial^j f)(z)) \,.
\end{equation}
Now let us consider the following path integral
\begin{equation}
    R = \int \mathcal{D} f \,\delta_w(f(z)-x) G(f(t)) \,,
    \label{R=}
\end{equation}
where $G$ is a functional of meromorphic functions. One would naively expect
\begin{equation}
    R \stackrel{?}{=} \sum_{f^{*}} G(f^{*}(t)) \,, 
    \label{R?}
\end{equation}
where $f^{*}-x$ is a meromorphic function with a zero of order $w$ around $z$.\footnote{Here we are assuming there are finitely many functions satisfying this. In other words, we are assuming there are additional constraints on $f^{*}$ due to special form of the functional $G$. We will see that this is the relevant case for us.} However, scaling the coordinates as $z \rightarrow \eta z$ and $t \rightarrow \eta t$, one realizes that since the scaling dimension of $\delta_w(f(z)-x)$ is $d=-\frac{w(w-1)}{2}$, the scaling dimensions of \eqref{R=} and \eqref{R?} do not match for $w>1$. This suggests that one is neglecting Jacobian factors of the form 
\begin{equation} \label{eq:jacobian}
    \Big(\partial^{w} f(z)\Big)^{-\frac{w-1}{2}} \,, 
\end{equation}
which would indeed make the scaling dimensions even. While this argument is not a rigorous derivation of such Jacobains, it suggests to make for the integrated correlator an ansatz of the form
\begin{multline} 
\label{eq:ansatz}
   \frac{1}{N!} \int \text d^{2N} \lambda_a \int \text d^2 z_4 \dots \text d^2 z_n \, |I_{\text{b}} I_{\text{f}} I_{\text{rest}}|^2  = |z_1-z_2|^{-2} |z_2-z_3|^{-2} |z_1-z_3|^{-2} \\
   \times \, \sum_{\Gamma} A_{\Gamma} |C^{\Gamma}|^{2 \alpha_C} |a^{\Gamma}_1|^{-2} |a^{\Gamma}_2|^{-2} \, |I_{\text{f}} I_{\text{rest}}|^2 \prod_{j=1}^n |a_j^{\Gamma}|^{2\alpha_{w_j}} \,,
\end{multline}
and fix the various free parameters in the ansatz by imposing the correct scaling behaviour. Let us explain the various terms entering \eqref{eq:ansatz}. We already observed that the various delta functions in \eqref{eq:ib-sphere} force $\gamma$ to be a covering map with ramification points $z_i$ and ramification indices $w_i$. The sum in \eqref{eq:ansatz} is thus running over such covering maps. Notice that $a_j^{\Gamma}$ can be written as 
\begin{equation}
    a_j^{\Gamma} = \frac{\partial^w \Gamma}{w!}
\end{equation}
and hence factors of the form \eqref{eq:jacobian} are accounted for by $\alpha_{w_j}$, which is a function of $w_j$. We already encountered $C^{\Gamma}$ in the computation of the fermionic correlator and we allow for its presence also in our ansatz: $\alpha_C$ is an arbitrary ($w_i$-independent) real number to be fixed. The factors $|a^{\Gamma}_1|^{-2} |a^{\Gamma}_2|^{-2}$ in our ansatz are due to the different exponents of $\partial^{w_i} \gamma$ in \eqref{eq:ib-sphere}. The factors $|z_i-z_j|^{-2}$ for $i,j=1,2,3$ in eq.~\eqref{eq:ansatz} are due to the usual Jacobian arising when fixing three points on the sphere, see \cite{Blumenhagen:2013fgp}. Finally, $A_{\Gamma}$ accounts for the overall normalization. 

Note that $C^{\Gamma}$ has worldsheet weight $-1$, see eq.~\eqref{dGamma} and $a_j^{\Gamma}$ has worldsheet weight $w_j$. Let us then compute the conformal weight on the worldsheet of the integrated correlator in eq.~\eqref{eq:ansatz}. It reads
\begin{equation}
    \Delta = \Delta(I_{\text f}) + \Delta(I_{\text{rest}}) -(w_1+w_2)-2 N - (n-3) +\sum_{j} w_j (-m_{w_j}-\tfrac{w_j-1}{2})+\sum_{j} w_j \,.
\end{equation}
On the other hand, the worldsheet conformal dimension of the right-hand-side of eq.~\eqref{eq:ansatz} is
\begin{equation}
    \Delta^{\prime} = \Delta(I_{\text f}) + \Delta(I_{\text{rest}}) -\alpha_C -(w_1+w_2) + \sum_j w_j \alpha_{w_j} +3 \,.
\end{equation}
Requiring that $\Delta=\Delta^{\prime}$ and using eq.~\eqref{eq:N-degree-covering}, we get
\begin{equation}
    \alpha_C = 2 \ , \quad \alpha_{w_j} = -m_{w_j}-\frac{w_j-1}{2} \,.
\end{equation}

\paragraph{Assembling all contributions} Combining all the factors and reintroducing the string coupling $g_s$ dependence, the integrated correlation function in eq.~(\ref{eq:hybrid-correlators-integrated}) becomes
\begin{equation}
    I = g_s^{n - 2} \, \sum_{\Gamma} |C^{\Gamma}| \prod_{a=1}^N |\xi^{\Gamma}_a|^{-1} \prod_{j} |a_j^{\Gamma}|^{-2h_j+\frac{w_j-1}{2}} \,, 
\end{equation}
where
\begin{equation}
    h_j = m_{w_j}+\frac{w_j-1}{2} \,,
\end{equation}
see eqs.~(\ref{eq:st-odd-w}) and (\ref{eq:m-w-odd}). Up to the overall normalization, this reproduces eq.~(\ref{eq:spacetime-correlator}).

\paragraph{When some $\boldsymbol{w_i}$ are even} Note that the same analysis also holds when some of the $w_i$'s are even. While the expression for $m_w^{\pm}$ differs from the one for $m_w$, see eqs.~\eqref{eq:m-w-odd} and \eqref{eq:m-w-even}, we still have
\begin{equation} \label{eq:hw-even}
    h_{j} = m^{\pm}_{w_j}+\frac{w_j-1}{2} \,,
\end{equation}
also for $w_i$ even, see eqs.~(\ref{eq:st-even-w}) and (\ref{eq:m-w-even}). Additionally, recall that when $w_i$ is even, the ground state on the worldsheet transforms as an $\mathfrak{su}(2)$ doublet. This is exactly what happens in the dual CFT. In particular, if we define
\begin{equation}
    f = f_1+f_2 \ ,
\end{equation}
then the vertex operators for $w$ can be written as
\begin{equation}
    \Phi_w^{\pm} = e^{\pm \frac{i f}{2}} \widetilde{\Phi}_w \,,
\end{equation}
where $\widetilde{\Phi}_w$ is the ground state~\eqref{eq:w-odd-twisted-delta} with $m_w$ replaced by $m_w \mapsto m_w^{\pm}$, cf.~eqs.~\eqref{eq:w-odd-twisted-delta} and \eqref{eq:w-even-twisted-delta}. In particular, notice that $f$ has a trivial OPE with $f_1-f_2, \beta$ and $\gamma$. Hence, when some of the $w$'s are even, one gets the extra contribution of the corresponding correlation function of the boson $f$. Since $e^{\pm \frac{i f}{2}}$ is the bosonization of the fermionic part of the R-sector ground states, exactly the same correlator arises in the dual CFT. This shows that the matching with the symmetric orbifold correlators can be extended to arbitrary parity of the $w_i$ and that the selection rules on the worldsheet exactly reproduce the ones of the boundary CFT.


\section{Conclusions, discussion and open questions}
\label{sec:conclusions}


In this paper we considered an alternative free-field realization of the $\mathfrak{psu}(1,1|2)_1$ current algebra \cite{Beem:2023dub} and explained how this simplifies the computation of spectrum and correlation functions for pure NS-NS tensionless strings on $\text{AdS}_3 \times \text{S}^3 \times \mathbb{T}^4$. We reviewed the relation of the alternative free-field realization with the one adopted in \cite{Eberhardt:2018ouy, Dei:2020zui} and explained how it is related to the Wakimoto realization of $\mathfrak{sl}(2,\mathbb R)_k$. We also clarified the geometric origin of the so-called `secret representations', which essentially are screening operators one needs to insert in correlation functions to prevent them from trivially vanishing. Expressing vertex operators in terms of delta functions of free fields, we showed how the localization of worldsheet correlation functions emerges from the path integral. Finally, we derived an exact match of tree level $n$-point functions of tensionless $\text{AdS}_3 \times \text{S}^3 \times \mathbb{T}^4$ strings with symmetric orbifold correlators. 

Let us discuss a number of points that we believe deserve further investigation and provide a list of potential applications of the results derived in this paper.

\paragraph{Higher-genus correlators and topological A model:} In Section~\ref{sec:correlation-functions} and Appendix~\ref{app:path-integral} we showed that tree level correlators of tensionless strings on $\text{AdS}_3 \times \text{S}^3 \times \mathbb T^4$ can be computed by relating them to correlators of the $\mathcal{N}=(2,0)$ topological A-model on $\mathbb{C}$. The connection between the two string models extends to higher genus and higher genus correlators of tensionless $\text{AdS}_3$ strings should hence be under good control. In fact, we are studying this in \cite{higher-genus-paper}. It would also be very interesting to understand whether there is any relation between tensionless AdS$_3$ $n$-point functions and the correlation functions of the topological models investigated in \cite{Gopakumar:2022djw}.

\paragraph{Correlation functions in the operator formalism:} We have computed tree-level correlators from the path integral making use of the delta function identities introduced in Section~\ref{sec:delta-function}. While these identities naturally enter path integral computations of superstring theory, it would be interesting to compute worldsheet correlators also from an operator formalism perspective. Building on \cite{Dei:2020zui}, where localization of correlators has been derived, one could adapt the strategy of \cite{Dei:2021yom, Dei:2022pkr} to the tensionless string and derive a Knizhnik-Zamolodchikov equation for hybrid formalism correlators. At tree level and at least experimentally for low values of the spectral flow parameters $w_i$, this should allow to reproduce from the worldsheet in the operator formalism the various factors entering eq.~\eqref{eq:spacetime-correlator}.

\paragraph{Normalization of correlators:} At various points in our computation of string correlators, we neglected overall $w_i$-dependent factors and did not compute the normalization of string $n$-point functions. However, at least at tree level, this can be fixed by requiring four-point functions to correctly factorize into products of three-point functions. Alternatively, similarly to what was done in \cite{Eberhardt:2021vsx, Dei:2022pkr}, the correct normalization of $n$-point functions can be deduced by computing the normalization of the gravitational path integral and of vertex operators, which in turn follow by imposing an exact match of three and four-point functions with the dual CFT. 

\paragraph{\boldmath Tensionless strings on $\text{AdS}_{3} \times \text{S}^{3} \times \text{S}^{3} \times \text{S}^{1} $:}
Another consistent worldsheet background for Type IIB string theory is $\text{AdS}_3\times\text{S}^3\times\text{S}^3\times\text{S}^1$. In the hybrid formalism, the string theory on the background $\text{AdS}_3\times\text{S}^3\times\text{S}^3$ is described by a sigma model on the supergroup $\text{D}(2,1;\alpha)$ \cite{Eberhardt:2019niq}. The bosonic subalgebra $\mathfrak{sl}(2,\mathbb{R})_k\oplus\mathfrak{su}(2)_{k^+}\oplus\mathfrak{su}(2)_{k^-}$ has levels satisfying
\begin{equation}
\frac{1}{k}=\frac{1}{k^+}+\frac{1}{k^-}\,.
\end{equation}
For $k^+=k^-=1$ and $k=1/2$, there is a free-field realization of the worldsheet theory based on a single pair of symplectic bosons and two pairs of free fermions. While it has been conjectured that this worldsheet theory is dual to the symmetric orbifold of the so-called $\mathcal{S}_{\kappa}$ theory with $\kappa=0$ \cite{Eberhardt:2017pty}, a careful analysis of the correlation functions has never been performed. It would be interesting to explore whether some of the technology developed in this paper could be useful in exploring the $\mathfrak{d}(2,1;\alpha)$ theory in more detail, and specifically whether it would be possible to compute the correlators of the worldsheet theory explicitly as we did in Section \ref{sec:correlation-functions}.

\paragraph{\boldmath Strings on $\text{AdS}_{5} \times \text{S}^{5}$ and an alternative free-field realization for $\mathfrak{psu}(2,2|4)$:}

In \cite{Gaberdiel:2021qbb,Gaberdiel:2021jrv} a worldsheet dual to free $\mathcal{N}=4$ super Yang-Mills theory was proposed, which relied in one chiral half of a free-field realization of $\mathfrak{psu}(2,2|4)_1$. This free-field realization is completely analogous to the symplectic boson realization of $\mathfrak{psu}(1,1|2)_1$ \cite{Eberhardt:2018ouy,Dei:2020zui} but with twice the number of bosonic and fermionic degrees of freedom. One may suspect that the alternative free-field realization of $\mathfrak{psu}(1,1|2)_1$ can be extended to a realization of $\mathfrak{psu}(2,2|4)_1$. Since the theory of \cite{Gaberdiel:2021qbb,Gaberdiel:2021jrv} naively resembles a twistor string in four dimensions, the natural guess for such a realization would be given by three $\beta\gamma$ systems and four $p\theta$ systems which parametrize the super-twistor space $\mathbb{CP}^{3|4}$ in local coordinates. In fact, such a realization has been found in \cite{Beem:2023dub}. It might hence be possible to perform analogous computations to those detailed in this work, and try to extract correlation functions of free $\mathcal{N}=4$ super Yang-Mills directly from the proposed worldsheet theory.

\paragraph{AdS$_{\boldsymbol 3}$ strings at generic tension:}

One of the main ingredients in our calculations of correlation functions in the $k=1$ worldsheet theory was expressing vertex operators directly in terms of Wakimoto variables. Specifically, writing spectrally-flowed vertex operators in terms of delta-function operators in the $\beta\gamma$ system allowed us to compute the path integral directly and compare the result to symmetric orbifold correlation functions. However, not only the $k=1$ theory, but the $\text{SL}(2,\mathbb{R})_k$ WZW model at any value of $k$ admits a Wakimoto representation. Thus, one would expect that our expressions for spectrally-flowed vertex operators could be immediately generalized to $k>1$. While something similar has already been attempted in the past \cite{Giribet:1999ft, Giribet:2000fy, Giribet:2001ft, Satoh:2001bi, Iguri:2007af, Iguri:2009cf, Giribet:2015oiy}, the techniques developed here might pave a path forward to understanding the perturbative CFT dual to bosonic string theory on $\text{AdS}_3$ \cite{Eberhardt:2021vsx} directly from the path integral perspective. It might also provide a first principle proof for the closed-form formulae proposed in \cite{Dei:2021xgh, Dei:2021yom} for correlators of AdS$_3$ spectrally flowed vertex operators.\footnote{A proof for the three-point function correlator formula was given in \cite{Bufalini:2022toj}, while for the four-point function a first principle derivation is still missing.}

\paragraph{Non-critical strings on AdS$_{\boldsymbol 3} \times \text S^{\boldsymbol 3}$ and twisted holography:} The central charges of the worldsheet $\mathfrak{psu}(1,1|2)_1$ algebra and of the $\rho$ and $\sigma$ ghosts add up to zero, $c(\mathfrak{psu}(1,1|2)_1) + c(\rho) + c(\sigma) = 0$. This suggests that a consistent six-dimensional string theory on AdS$_{3} \times \text S^{3}$ can be constructed. In the parlance of \cite{Kutasov:1991pv, Giveon:1999zm, Murthy:2004zx, Murthy:2006eg}, it would be described as a non-critical string theory at $k=1$. If such a string theory exists, it would be interesting to understand whether it is related to the $k<1$ non-critical string theory of \cite{Balthazar:2021xeh}. Moreover, one may suspect that such a $k=1$ non-critical string theory defines a closed subsector of the ten-dimensional $k=1$ string theory on AdS$_{3} \times \text S^{3} \times \mathbb T^4$ and that a similar dual subsector can  be identified in the boundary CFT$_2$. This putative holographic pair may be described as a lower-dimensional analogue of the twisted holography of \cite{Costello:2018zrm} and have interesting relation with the topological duality of \cite{Lerche:2023wkj}.

\paragraph{Acknowledgements} We thank Soumangsu Chakraborty, Cassiano Daniel, Lorenz Eberhardt, Matthias Gaberdiel, Francesco Galvagno, Jeffrey Harvey, Shota Komatsu, Emil Martinec, Edward Mazenc, Nathan McStay, Beat Nairz, Ron Reid-Edwards, Savdeep Sethi, Vit Sriprachyakul, and Jacob Vo\v{s}mera for useful discussions. We thank Lorenz Eberhardt, Matthias Gaberdiel, Nathan McStay, and Ron Reid-Edwards for helpful comments on a draft of this paper. We also thank the organizers of the CERN workshop `Precision Holography', where many of these ideas were refined. A.D.~acknowledges support from the Mafalda \& Reinhard Oehme Fellowship. The work of BK and KN was supported by the Swiss National Science Foundation through a personal grant and via the NCCR SwissMAP. The work of BK was in addition supported by STFC consolidated grants ST/T000694/1 and ST/X000664/1.

\appendix

\section{\boldmath \texorpdfstring{$\beta\gamma$}{beta-gamma} systems} \label{app:beta-gamma}

In this appendix, we provide a brief introduction to bosonic first-order systems ($\beta\gamma$ systems) and list the conventions used throughout the main text.

\subsection*{Action and symmetries}

A $\beta\gamma$ system is a free conformal field theory with action
\begin{equation}
S=\frac{1}{2\pi}\int\beta\overline{\partial}\gamma\,.
\end{equation}
The fields $\beta,\gamma$ are taken to be chiral with holomorphic scaling dimension $\Delta(\beta)=\Lambda$ and $\Delta(\gamma)=1-\Lambda$, so that the Lagrangian has scaling dimension $(1,1)$.\footnote{In the main text, we take $\Lambda=1$. However, in this appendix, we will largely be agnostic about the value of $\Lambda$.} These fields satisfy the OPEs
\begin{equation}
\beta(z)\gamma(w)\sim-\frac{1}{z-w}\,,
\end{equation}
and the stress tensor of the theory is given by
\begin{equation}
T=-\Lambda(\beta\partial\gamma)+(1-\Lambda)(\partial\beta\,\gamma)\,.
\end{equation}
The global $\mathbb{C}^*$ symmetry $\beta\to \alpha \beta$, $\gamma\to \alpha^{-1} \gamma$ is generated by the Noether current
\begin{equation}
J(z)=(\beta\gamma)\,.
\end{equation}
From this definition we can read off the OPEs
\begin{equation}
J(z)\beta(w)\sim \frac{\beta(w)}{z-w}\,,\quad J(z)\gamma(w)\sim -\frac{\gamma(w)}{z-w}\,,
\end{equation}
so that $\beta$ and $\gamma$ have $J_0$ charge $+1$ and $-1$, respectively. We also note the OPE
\begin{equation}
J(z)J(w)\sim-\frac{1}{(z-w)^2}\,.
\end{equation}

An important property of $\beta\gamma$ systems is that the current $J$ is not a conformal primary of the theory unless $\Lambda=1/2$. Indeed, we have
\begin{equation}
T(z)J(w)\sim\frac{Q_J}{(z-w)^3}+\frac{J(w)}{(z-w)^2}+\frac{\partial J(w)}{z-w}\,,
\end{equation}
where $Q_J=1-2\Lambda$ is the `background charge' of $J$. A consequence of this background charge is that in any correlation function, the sum of $J_0$-charges will not vanish, but rather must equal $Q_J(g-1)$ if the worldsheet has genus $g$. A concrete way of putting this is that if one considers the correlation function
\begin{equation}
\Braket{J(z)\prod_{i=1}^{n}\Phi_{i}(z_i)}
\end{equation}
as a holomorphic one-form in $z$, then we must have
\begin{equation}
\sum_{i=1}^{n}\underset{z=z_i}{\text{Res}}\Braket{J(z)\prod_{i=1}^{n}\Phi_{i}(z_i)}=Q_J(g-1)\,.
\end{equation}

\subsection*{Bosonization}

Here we will briefly review the bosonization of $\beta\gamma$ systems. For a more in-depth explanation, see, for example, Chapter 13 of \cite{Blumenhagen:2013fgp}.

Given the current $J=(\beta\gamma)$, we can introduce a scalar field $\phi$ such that $J=-\partial\phi$, which satisfies the OPE
\begin{equation}
\phi(z)\phi(w)\sim-\log(z-w)\,.
\end{equation}
The background charge of $\phi$ is $Q_{\phi}=1-2\Lambda$ and its stress tensor is
\begin{equation}
T_{\phi}=-\frac{1}{2}(\partial\phi)^2-\frac{Q_{\phi}}{2}\partial^2\phi\,,
\end{equation}
which has central charge $c(\phi)=1+3Q^2_{\phi}$.

Given the scalar $\phi$, we can reconstruct the $\beta\gamma$ system by introducing a fermionic pair $\xi,\eta$ of weights $\Delta(\eta)=1$ and $\Delta(\xi)=0$ with OPE
\begin{equation}
\eta(z)\xi(w)\sim\frac{1}{z-w}\,.
\end{equation}
In terms of $\phi,\xi,\eta$, we have
\begin{equation}\label{eq:bosonization-xi-eta}
\beta=e^{\phi}\partial\xi\,,\quad\gamma=\eta\,e^{-\phi}\,.
\end{equation}
Furthermore, we can bosonize $\eta,\xi$ as
\begin{equation}
\xi=e^{i\kappa}\,,\quad\eta=e^{-i\kappa}
\end{equation}
where $i \kappa$ is a scalar with background charge $Q_{\kappa}=-1$ and which satisfies the OPE
\begin{equation}
\kappa(z)\kappa(w)\sim-\log(z-w)\,.
\end{equation}
In particular, the stress-tensor of $\kappa$ is
\begin{equation}
    T_{\kappa} = -\frac{1}{2} (\partial \kappa)^2 + \frac{i}{2} \partial^2 \kappa \,.
\end{equation}
Thus, $\beta,\gamma$ can be written in terms of $\phi,\kappa$ as
\begin{equation}
\beta=e^{\phi+i\kappa}\partial(i\kappa)\,,\quad\gamma=e^{-\phi-i\kappa}\,.
\end{equation}

It is important to note that the map from the scalars $(\phi,\kappa)$ to the fields $(\beta,\gamma)$ is not invertible. Indeed, note that, by equation \eqref{eq:bosonization-xi-eta}, the fields $(\beta,\gamma)$ are unaffected by the transformation $\xi\to\xi+\varepsilon$, while the $(\phi,\kappa)$ system certainly is. In the superstring theory literature, one often speaks of a \textit{small} Hilbert space $\mathcal{H}_{s}$ (generated by $\beta,\gamma$) and a \textit{large} Hilbert space $\mathcal{H}_{\ell}$ (generated by $\phi,\kappa$). Noting that the transformation $\xi\to\xi+\varepsilon$ is generated by $\eta_0$, and furthermore noting that the small Hilbert space consists of states invariant under this symmetry, we can describe the small Hilbert space as
\begin{equation}
\mathcal{H}_s=\text{ker}(\eta_0:\mathcal{H}_{\ell}\to\mathcal{H}_{\ell})\,.
\end{equation}
That is to say, the small Hilbert space is obtained by only keeping those states which are annihilated by $\eta_0$.

Let us finally discuss two points in Section~\ref{sec:delta-function} that were referred to this appendix. First, we provide an example of the identifications in eq.~\eqref{eq:fourier-deltas} and eq.~\eqref{eq:delta-function-cases}. The OPE between $\gamma$ and $\delta(\beta)$ can be read off as
\begin{equation}
\begin{split}
\gamma(z)\delta(\beta)(0)&=\int\frac{\mathrm{d}\zeta}{2\pi}\,\gamma(z)\,e^{i\zeta\beta(0)}\\
&=\int\frac{\mathrm{d}\zeta}{2\pi}\,\sum_{n=0}^{\infty}\frac{(i\zeta)^n}{n!}\gamma(z)(\beta^n)(0)\\
&\sim\int\frac{\mathrm{d}\zeta}{2\pi}\,\sum_{n=0}^{\infty}\frac{(i\zeta)^n}{n!}\frac{n}{z}(\beta^{n-1})(0)\\
&=\frac{1}{z}\int\frac{\mathrm{d}\zeta}{2\pi}\,i\zeta e^{i\zeta\beta(0)}=\frac{1}{z}\delta'(\beta)(0)\,,
\end{split}
\end{equation}
where in the last line we used the integral representation of the derivative $\delta'$ of the delta function, see \cite{Witten:2012bh} for more details.

Second, we provide a derivation of eq.~\eqref{eq:delta-function-m}. This can be shown most easily via an inductive argument. Using $\gamma=e^{-\phi-i\kappa}$, we have
\begin{equation}
\gamma(z)e^{(m+w)\phi+im\kappa}(y)\sim(z-y)^{w}e^{(m+w-1)\phi+i(m-1)\kappa}(y)+\cdots\,,
\end{equation}
so that
\begin{equation}
\frac{\partial^w\gamma(z)}{w!}e^{(m+w)\phi+im\kappa}(y)\sim e^{(m+w-1)\phi+i(m-1)\kappa}(y)+\cdots\,.
\end{equation}
Thus, by the definition of radial normal ordering, we have
\begin{equation}
\Big(\frac{\partial^w\gamma}{w!}e^{(m+w)\phi+im\kappa}\Big)=e^{(m+w-1)\phi+i(m-1)\kappa}\,.
\end{equation}
Treating this as a recursion relation with initial condition $e^{w\phi}=\delta_{w}(\gamma)$, we derive the expression \eqref{eq:delta-function-m}. As a sanity check we can compute its conformal weight, and we find
\begin{equation}
\begin{split}
\Delta\left(\left(\frac{\partial^w\gamma}{w!}\right)^{-m}\delta_{w}(\gamma)\right)&=-\frac{w(w+Q_{\phi})}{2}-m(w+1-\Lambda)\\
&=-\frac{(w+m)(w+m+Q_{\phi})}{2}+\frac{m(m-1)}{2}\,,
\end{split}
\end{equation}
which is indeed the conformal weight of $e^{(m+w)\phi+im\kappa}$.


\section{Evaluating the path integral}
\label{app:path-integral}

In the main text, we need to evaluate path integrals of the form
\begin{equation}\label{eq:appendix-goal-integral}
I=\int\mathcal{D}(\beta,\gamma)\int_{\Sigma}\delta(\beta(\lambda_a))\prod_{i=1}^{n}\delta_{w_i}(\gamma(z_i)-x_i)\,\mathcal{O}(\gamma)\,,
\end{equation}
where $\mathcal{O}(\gamma)$ is some (not necessarily local) operator depending only on $\gamma$ and the points $z_i,\lambda_a$, and we have chosen $N=1-g+\sum_{i}(w_i-1)/2$. In this appendix, we will argue that for genus $g=0$, this path integral evaluates to
\begin{equation}\label{eq:path-integral-evaluation-ansatz}
I=z_{12}^{-1}z_{23}^{-1}z_{13}^{-1}\sum_{\substack{\Gamma:\Sigma\to\mathbb{CP}^1\\\text{deg}(\Gamma)=N}}(C^{\Gamma})^{2}\prod_{i=1}^{n}(a_i^{\Gamma})^{-\frac{w_i+1}{2}}\,\mathcal{O}(\Gamma)\,\delta^{(n-3)}(\Gamma)\delta^{(N)}(\lambda_a,\lambda^{\star}_a)\,,
\end{equation}
up to an overall $w_i$-dependent constant. Here, the sum is over all branched holomorphic covering maps $\Gamma:\mathbb{CP}^1\to\mathbb{CP}^1$ of degree $N$, branched over $x_i\in\mathbb{CP}^1$ with order $w_i$, i.e.\ those maps which satisfy
\begin{equation}
\Gamma(z)\sim x_i+a_i^{\Gamma}(z-z_i^{\star})^{w_i}+\cdots\,,\quad z\to z_i^{\star}\,.
\end{equation}
Furthermore, we denote by $\lambda_a^\star$ the poles of $\Gamma$. The prefactor $C^{\Gamma}$ is defined so that
\begin{equation}
\partial\Gamma(z)=C^{\Gamma}\prod_{i=1}^{n}(z-z_i^{\star})^{w_i-1}\prod_{a=1}^{N}(z-\lambda_a^{\star})^{-2}\,.
\end{equation}
Finally, the delta functions $\delta^{(n-3)}(\Gamma)$ are defined by first restricting to covering maps which satisfy $\Gamma(z_i)=x_i$ for $i=1,2,3$.\footnote{This can always be done by acting on $\Gamma$ with a M\"obius transformation.} The delta function is then simply \cite{Eberhardt:2019ywk,Dei:2020zui}
\begin{equation}
\delta^{(n-3)}(\Gamma):=\prod_{i=4}^{n}\delta(z_i-z_i^{\star})\,.
\end{equation}
For notational convenience in the rest of the appendix, we will define
\begin{equation}\label{eq:tilded-coefficients}
\begin{split}
\tilde\xi_a&=\prod_{i=1}^{n}(\lambda_a-z_i)^{w_i-1}\prod_{b\neq a}(\lambda_a-\lambda_b)^{-2}\\
\tilde a_i&=\prod_{j\neq i}(z_i-z_j)^{w_j-1}\prod_{a=1}^{N}(z_i-\lambda_a)^{-2}\,.
\end{split}
\end{equation}
When $z_i=z_i^{\star}$ and $\lambda_a=\lambda_a^{\star}$, these are related to the covering map residues $\xi_a^{\Gamma}$ and Taylor coefficients $a_i^{\Gamma}$ (see Section \ref{sec:correlation-functions}) by
\begin{equation}\label{eq:tilde-covering-relation}
\tilde\xi_a=-(C^{\Gamma})^{-1}\xi_a^{\Gamma}\,,\quad\tilde{a}_i=\left(\frac{C^{\Gamma}}{w_i}\right)^{-1}a_i^{\Gamma}\,.
\end{equation}

We will derive \eqref{eq:path-integral-evaluation-ansatz} by comparison with correlation functions of a topological string theory for which integrals like \eqref{eq:path-integral-evaluation-ansatz} are required, and which result in known differential forms on $\mathcal{M}_{g,n}$. We compare to the string theories of \cite{Witten:1988xj,Cordes:1994sd,Witten:2005px,Frenkel:2005ku}. Specifically, we consider a 2D TCFT with action
\begin{equation}
S_{\text{top}}=\frac{1}{2\pi}\int_{\Sigma}\left(\beta\overline{\partial}\gamma+\chi\overline{\partial}\psi\right)\,,
\end{equation}
where $(\beta,\gamma)$ are the usual bosonic free fields of conformal dimensions $\Delta(\beta) =1$ and $\Delta(\gamma)=0$, and $(\chi,\psi)$ are their fermionic counterparts, i.e.\ topologically twisted fermions with $\Delta(\chi)=1$ and $\Delta(\psi)=0$. This action describes the infinite-volume limit of the $\mathcal{N}=(2,0)$ topological A-model on $\mathbb{C}$ and admits a twisted $\mathcal{N}=(2,0)$ algebra with generators
\begin{equation}
\begin{gathered}
T_{\text{top}}=-\beta\partial\gamma-\chi\partial\psi\,,\\
G^+_{\text{top}}=\beta\psi\,,\quad G^-_{\text{top}}=\chi\partial\gamma\,,\\
J_{\text{top}}=\psi\chi\,.
\end{gathered}
\end{equation}
Given the twisted algebra, we can define a string theory via the BRST charge $Q_{\text{top}}=(G^+_{\text{top}})_0$. Indeed, the theory is topological with respect to $Q_{\text{top}}$ since the action is BRST-exact:
\begin{equation}
S_{\text{top}}=\frac{1}{2\pi}\int_{\Sigma}\left[Q_{\text{top}},\chi\overline{\partial}\gamma\right]\,.
\end{equation}

Correlation functions of this topological string will count holomorphic curves $\Sigma\to\mathbb{C}$. In order to count curves into $\mathbb{CP}^1$, we have to compactify by adding a vertex operator which represents the point at infinity, similar to the procedure in Section \ref{sec:point at infinity}. As shown in \cite{Frenkel:2005ku}, the appropriate deformation operator in the topological model takes the form
\begin{equation}\label{eq:dtop}
D_{\text{top}}=\chi\partial\chi\left(\oint\gamma\right)\delta(\beta)=\delta_2(\chi)\left(\oint\gamma\right)\delta(\beta)=(G^{-}_{\text{top}})_{-1}\delta(\chi)\delta(\beta)\,,
\end{equation}
where we have defined
\begin{equation}
\delta_w(\chi)=\prod_{i=0}^{w-1}\partial^i\chi
\end{equation}
to be the fermionic version of the delta function operators defined in Section \ref{sec:delta-function}. The contour integral in \eqref{eq:dtop} is taken over a small circle around the insertion point.

Once defined, $D_{\text{top}}$ can be used to deform the free topological theory, just as in Section~\ref{sec:point at infinity}. The states whose correlation functions we are interested in are the analogues of the spectrally-flowed states constructed in Section \ref{sec:physical-states}. They take the form
\begin{equation}
\mathcal{O}_{w}(z,x)=\delta_{w}(\psi(z))\delta_{w}(\gamma(z)-x)\,.
\end{equation}
Indeed, $\mathcal{O}_w(x)$ is physical, since it in annihilated by the BRST operator $G^+_{\text{top}}$. We thus consider the amplitude
\begin{equation}
\int_{\mathcal{M}_{g,n}}\Braket{\prod_{\alpha=1}^{3g-3}\braket{G^-_{\text{top}},\mu_\alpha}\prod_{i=1}^{n}((G^-_{\text{top}})_{-1}\mathcal{O}_{w_i})(z_i,x_i)}
\end{equation}
where $\mu_\alpha$ are the $3g-3$ Beltrami differentials on $\text{T}_{\Sigma}\mathcal{M}_{g,n}$. Here, we use the notation
\begin{equation}
\braket{G_{\text{top}}^-,\mu_\alpha}:=\int_{\Sigma}G^-_{\text{top}}\,\mu_\alpha\,.
\end{equation}
As a topological string correlator, the above correlation function will define a differential form on the moduli space $\mathcal{M}_{g,n}$.

In computing the above correlation function, we know that the $\beta,\gamma$ integral will localize to the space of holomorphic maps $\Sigma\to\mathbb{CP}^1$ which are branched covers satisfying $\gamma(z_i)=x_i$. Upon integrating over $\mathcal{M}_{g,n}$, this space is extended to the moduli space of \textit{all} genus-$g$ branched covers over $\mathbb{CP}^1$ branched with index $w_i$ at $x_i$. This moduli space, which we call $S$, is a discrete (and in fact finite) subset of $\mathcal{M}_{g,n}$. The above path integral should thus reduce to a sum over $S$, weighted by an appropriate Jacobian coming from the path integral delta functions. However, since everything is supersymmetric under $G_{\text{top}}^+$, one would expect the bosonic and fermionic determinants to cancel, as is usual in supersymmetric field theories. Thus, a natural ansatz for the topological amplitude is
\begin{equation}\label{eq:hurwitz-counting}
\int_{\mathcal{M}_{g,n}}\Braket{\prod_{\alpha=1}^{3g-3}\braket{G^-_{\text{top}},\mu_\alpha}\prod_{i=1}^{n}((G^-_{\text{top}})_{-1}\mathcal{O}_{w_i})(z_i,x_i)}=|S|\,.
\end{equation}
That is, the topological string amplitude calculates the number of branched covering maps $\Sigma\to\mathbb{CP}^1$ branched over $x_i$ with index $z_i$, also known as Hurwitz numbers. This is natural from a topological string perspective, since the topological A-model computes Gromov-Witten invariants \cite{Witten:1988xj}, and it is known that Gromov-Witten theory is equivalent to Hurwitz theory with a Riemann surface as the target \cite{Okounkov:2002cja}.

Given that before integrating over $\mathcal{M}_{g,n}$ the worldsheet path integral has delta-function support at the points in $\mathcal{M}_{g,n}$ for which a holomorphic covering map $\Gamma$ exists (see the arguments of Section \ref{sec:point at infinity}), we can use \eqref{eq:hurwitz-counting} to determine that the coefficients must all be equal, i.e.
\begin{equation}\label{eq:top-correlator-localization}
\Braket{\prod_{\alpha=1}^{3g-3}\Braket{G^-_{\text{top}},\mu_{\alpha}}\prod_{i=1}^{n}((G^-_{\text{top}})_{-1}\mathcal{O}_{w_i})(z_i,x_i)}=\sum_{\Gamma}\delta^{(n+3g-3)}(\Gamma)\,.
\end{equation}
Here, the sum is over all branched covering maps $\Gamma$ of genus $g$, and $\delta^{(n+3g-3)}(\Gamma)$ is a delta function on $\mathcal{M}_{g,n}$ with support in $S$.

How would we have obtained the result \eqref{eq:top-correlator-localization} from the path integral? First, we note that
\begin{equation}
((G^-_{\text{top}})_{-1}\mathcal{O}_w)(z,x)=\delta_{w-1}(\psi(z))\left(w\frac{\partial^w\gamma(z)}{w!}\right)\delta_{w}(\gamma(z)-x)\,.
\end{equation}
Now, to compute the correlator from the path integral formalism, as explained above, we need to insert $N$ copies of the deforming field $D_{\text{top}}$, where $N$ is restricted by the fermionic charge. Since $G^-_{\text{top}}=\chi\partial\gamma$ the anomalous conservation of $J_{\text{top}}$ reads
\begin{equation}
2N+3g-3-\sum_{i=1}^{n}(w_i-1)=g-1\implies N=1-g+\sum_{i=1}^{n}\frac{w_i-1}{2}\,.
\end{equation}
That is, $N$ is determined to be the degree of the covering maps $\Gamma$. Once we have inserted the $N$ copies of $D_{\text{top}}$, we must compute
\begin{equation}
\frac{1}{N!}\int_{\Sigma^N}\mathrm{d}^{2N}\lambda_a\Braket{D_{\text{top}}\prod_{\alpha=1}^{3g-3}\braket{G^-_{\text{top}},\mu_{\alpha}}(\lambda_a)\prod_{i=1}^{n}((G^-_{\text{top}})_{-1}\mathcal{O}_{w_i})(z_i,x_i)}_0\,,
\end{equation}
where the subscript $0$ means that we are calculating the correlator with respect to the `undeformed' action, i.e.~without an insertion of $D_{\text{top}}$. Writing
\begin{equation}
\braket{G^{-}_{\text{top}},\mu_{\alpha}}=\int_{\Sigma}\mathrm{d}^2u_{\alpha}\,\chi\partial\gamma\,\mu_{\alpha}\,,
\end{equation}
we can compute this path integral by first computing the correlator
\begin{equation}\label{eq:top-ferm-bos-factorize}
\begin{split}
&\Braket{\prod_{a=1}^{N}\delta_2(\chi(\lambda_a))\prod_{\alpha=1}^{3g-3}\chi(u_{\alpha})\prod_{i=1}^{n}\delta_{w-1}(\psi(z_i))}_{\chi,\psi}\\
&\hspace{0.25cm}\times\Braket{\prod_{a=1}^{N}\left(\oint_{\lambda_a}\gamma\right)\delta(\beta(\lambda_a))\prod_{\alpha=1}^{3g-3}\partial\gamma(u_{\alpha})\prod_{i=1}^{n}\left(w_i\frac{\partial^{w_i}\gamma(z_i)}{w_i!}\right)\delta_{w_i}(\gamma(z_i)-x_i)}_{\beta,\gamma}\,,
\end{split}
\end{equation}
then 1) integrating over the points $\lambda_a$ and 2) integrating over the points $u_{\alpha}$, weighted by the Beltrami differential $\mu_{\alpha}$.

In this paper, we are only concerned with correlators at genus zero. In this case, the worldsheet is rigid (i.e.~there are no Beltrami differentials to integrate over), and there is a residual M\"obius symmetry that needs to be fixed. As is standard in string theory, we do this by fixing the points $z_1,z_2,z_3$ and only integrating over $z_i$ for $i\geq 4$. Since the operators at $z_1,z_2,z_3$ are not integrated, they will not be dressed with the factor of $(G^{-}_{\text{top}})_{-1}$.\footnote{This is analogous to the case of bosonic string theory, where `unintegrated' operators at genus zero are not dressed with $b$-ghosts.} Thus, the correlator we consider is of the form
\begin{equation}
\Braket{\prod_{i=1}^{3}\mathcal{O}_{w_i}(z_i,x_i)\prod_{i=4}^{n}((G_{\text{top}}^{-})_{-1}\mathcal{O}_{w_i})(z_i,x_i)}\,,
\end{equation}
which should reduce to a delta function of the form $\sum_{\Gamma}\delta^{(n-3)}(\Gamma)$ which fixes the $n-3$ undetermined cross-ratios on the worldsheet so that a holomorphic covering map to the base sphere exists. Bringing down $N$ copies of the operator $D_{\text{top}}$ gives (before integrating over the positions of $D_{\text{top}}$)
\begin{equation}
\begin{split}
&\Braket{\prod_{a=1}^{N}\delta_2(\chi(\lambda_a))\prod_{i=1}^{3}\delta_{w_i}(\psi(z_i))\prod_{i=4}^{n}\delta_{w_i-1}(\psi(z_i))}_{\chi,\psi}\\
&\hspace{0.5cm}\Braket{\prod_{a=1}^{N}\left(\oint_{\lambda_a}\gamma\right)\delta(\beta(\lambda_a))\prod_{i=1}^{3}\delta_{w_i}(\gamma(z_i)-x_i)\prod_{i=4}^{n}\left(w_i\frac{\partial^{w_i}\gamma(z_i)}{w_i!}\right)\delta_{w_i}(\gamma(z_i)-x_i)}_{\beta,\gamma}\,.
\end{split}
\end{equation}
Again, $N$ is determined to be $N=1+\sum_{i=1}^{n}(w_i-1)/2$ by the charge conservation of the $\chi\psi$ system. The fermionic correlator can be computed by Wick contractions,\footnote{This calculation is similar to that done in Section \ref{sec:correlation-functions}.} and the result is\footnote{We will drop the factors of $w_i$ since we are not interested in the overall $w_i$-dependent normalization.}
\begin{equation}
\Braket{\prod_{a=1}^{N}\delta_2(\chi(\lambda_a))\prod_{i=1}^{3}\delta_{w_i}(\psi(z_i))\prod_{i=4}^{n}\delta_{w_i-1}(\psi(z_i))}_{\chi,\psi}=z_{12}z_{23}z_{13}\prod_{a=1}^{N}\tilde\xi_a^{-1}\prod_{i=1}^{3}\tilde a_i^{\frac{w_i+1}{2}}\prod_{i=4}^{n}\tilde a_i^{\frac{w_i-1}{2}}\,,
\end{equation}
where $\tilde a_i,\tilde \xi_a$ are defined in equation \eqref{eq:tilded-coefficients}. The $\beta\gamma$ contribution can be computed by noting that it will localize onto covering maps $\Gamma$, and so we can simply pull out the residues $\oint\gamma$ and Taylor coefficients $\partial^w\gamma/w!$, so that
\begin{equation}
\begin{split}
&\Braket{\prod_{a=1}^{N}\left(\oint_{\lambda_a}\gamma\right)\delta(\beta(\lambda_a))\prod_{i=1}^{3}\delta_{w_i}(\gamma(z_i)-x_i)\prod_{i=4}^{n}\left(w_i\frac{\partial^{w_i}\gamma(z_i)}{w_i!}\right)\delta_{w_i}(\gamma(z_i)-x_i)}\\
&\hspace{2cm}=\prod_{a=1}^{N}\xi^{\Gamma}_a\prod_{i=4}^{n}a^{\Gamma}_i\Braket{\prod_{a=1}^{N}\delta(\beta(\lambda_a))\prod_{i=1}^{n}\delta_{w_i}(\gamma(z_i)-x_i)}
\end{split}
\end{equation}
in the neighborhood of a particular covering map $\Gamma$. Thus, demanding that the full topological correlator (before integrating over $\lambda_a$, $z_i$) localizes onto the branch points $z_i^{\star}$ and poles $\lambda_a^{\star}$ of covering maps $\Gamma$, we find
\begin{equation}
\Braket{\prod_{a=1}^{N}\delta(\beta(\lambda_a))\prod_{i=1}^{n}\delta_{w_i}(\gamma(z_i)-x_i)}=z_{12}^{-1}z_{23}^{-1}z_{13}^{-1}\sum_{\Gamma}(C^{\Gamma})^2\prod_{i=1}^{n}a_i^{-\frac{w_i+1}{2}}\delta^{(n+3g-3)}(\Gamma)\delta^{(N)}(\lambda_a^{\star})\,.
\end{equation}
The factor of $(C^{\Gamma})^2$ comes from evaluating the fermionic path integral at $z_i=z_i^{\star}$ and $\lambda_a=\lambda_a^{\star}$, see equation \eqref{eq:tilde-covering-relation}. Here, we find that the path integral contains the Jacobian factor $z_{12}^{-1}z_{23}^{-1}z_{13}^{-1}$, which is an artifact of fixing $z_1,z_2,z_3$ while integrating over the rest of the points.\footnote{In bosonic string theory, this factor comes from the three-point function $\braket{c(z_1)c(z_2)c(z_3)}$ of the $c$-ghosts attached to unintegrated operators.}

Given an operator $\mathcal{O}(\gamma)$, we can also throw it into the correlation function. The result will be that $\gamma$ will simply be replaced with the covering map $\Gamma$ upon integration, so that
\begin{multline}
\Braket{\prod_{a=1}^{N}\delta(\beta(\lambda_a))\prod_{i=1}^{n}\delta_{w_i}(\gamma(z_i)-x_i)\mathcal{O}(\gamma)}\\
=z_{12}^{-1}z_{23}^{-1}z_{13}^{-1}\sum_{\Gamma}(C^{\Gamma})^{2}\prod_{i=1}^{n}(a_i^{\Gamma})^{-\frac{w_i+1}{2}}\,\mathcal{O}(\Gamma)\,\delta^{(n-3)}(\Gamma)\delta^{(N)}(\lambda_a,\lambda^{\star}_a)\,,
\end{multline}
which is the integral we set out to derive.


\section{Relation to the symplectic boson theory}
\label{app:ours-to-symp}

Following \cite{Beem:2023dub}, let us discuss how the free fields $\beta, \gamma$ and $p_a, \theta^a$ introduced in Section~\ref{sec:worldsheet theory} are related to the symplectic boson worldsheet theory of \cite{Eberhardt:2018ouy, Dei:2020zui}. It was shown in \cite{Gaiotto:2017euk, Eberhardt:2018ouy} that two pairs of symplectic bosons 
\begin{equation}
[\xi^\alpha_r , \eta^\beta_s ] = \epsilon^{\alpha \beta} \delta_{r, -s} \,, \qquad \alpha, \beta \in \{ +, - \} \,,
\label{sympl-bosons-algebra}
\end{equation}
together with two pairs of free fermions 
\begin{align}
    \{\psi^\alpha_r,\chi^\beta_s\}=\epsilon^{\alpha\beta} \delta_{r, -s}\,, \qquad \alpha, \beta \in \{ +, - \} \,,
\end{align}
generate the $\mathfrak{u}(1,1|2)_1$ algebra \eqref{u(1,1|2)}\footnote{We follow the conventions of \cite{Naderi:2022bus}. Our conventions are related to those of \cite{Eberhardt:2019qcl} by $S^{\alpha \beta -}_m \to -S^{\alpha \beta -}_m$.}
\begin{subequations}
\begin{align}
J^3_m&=-\tfrac{1}{2} (\eta^+\xi^-)_m-\tfrac{1}{2} (\eta^-\xi^+)_m\,, \qquad  K^3_m =-\tfrac{1}{2} (\chi^+\psi^-)_m-\tfrac{1}{2} (\chi^-\psi^+)_m\,, \\
J^\pm_m&=(\eta^\pm\xi^\pm)_m\,, \hspace{99pt}  K^\pm_m =\pm(\chi^\pm\psi^\pm)_m\,, \\
S_m^{\alpha\beta+}&=(\chi^\beta \xi^\alpha)_m\,, \hspace{93pt}  S_m^{\alpha\beta-} =(\eta^\alpha\psi^\beta)_m \,,  \\
Z_m & = -\tfrac{1}{2} (\eta^+\xi^-)_m+\tfrac{1}{2} (\eta^-\xi^+)_m -\tfrac{1}{2} (\chi^+\psi^-)_m+\tfrac{1}{2} (\chi^-\psi^+)_m \,, \\
Y_m & = -\tfrac{1}{2} (\eta^+\xi^-)_m+\tfrac{1}{2} (\eta^-\xi^+)_m + \tfrac{1}{2} (\chi^+\psi^-)_m - \tfrac{1}{2} (\chi^-\psi^+)_m \,. 
\end{align}
\label{old-free-field-realization}%
\end{subequations}
The chiral algebra $\mathfrak{psu}(1,1|2)_1$ can be obtained from \eqref{old-free-field-realization} by gauging the U$(1)$'s generated by the null currents $Y$ and $Z$. When expressing vertex operators in terms of symplectic bosons, the physical states must also obey
\begin{equation}
    Z_n \phi = 0 \,, \qquad n \geq 0 \,, 
    \label{Zn=0}
\end{equation}
together with eq.~\eqref{eq:physical}. Symplectic bosons and free fermions can be bosonized as \cite{Naderi:2022bus}
\begin{subequations}
\begin{align}
\xi^-& =-e^{-\phi_1-i\kappa_1} \,, & \eta^+ &= e^{\phi_1+i\kappa_1}\partial(i\kappa_1) \,, \\
\xi^+& =e^{\phi_2+i\kappa_2} \,, & \eta^-&= -e^{-\phi_2-i\kappa_2}\partial(i\kappa_2)  \,, \\
\chi^+& =e^{iq_1} \,, & \psi^-&=-e^{-iq_1} \,, \\
\psi^+& =e^{iq_2} \,, & \chi^-& =e^{-iq_2} \,, 
\end{align}
\label{xi-eta-bosonization}%
\end{subequations}
where $\phi_i$, $\kappa_i$ and $q_i$ for $i= 1, 2$ obey the OPEs
\begin{subequations}
\begin{align}
\phi_i(z)\phi_j(w) &\sim -\delta_{ij}\log(z-w)\,, \\
\kappa_i(z)\kappa_j(w) &\sim -\delta_{ij}\log(z-w)\,, \\
q_i(z)q_j(w) &\sim -\delta_{ij}\log(z-w) \,. 
\end{align}%
\label{phii-kappai-qi-OPEs}
\end{subequations}
As can be read off from the stress tensor
\begin{equation}
T_{\text{sym}} =-\mfrac{1}{2} \sum_{i=1}^2 \Bigl( (\partial\phi_i)^2 + (\partial\kappa_i)^2 + (\partial q_i)^2 \Bigr)  +\mfrac{i}{2}\partial^2 \kappa_1 -\mfrac{i}{2} \partial^2 \kappa_2 \,, 
\end{equation}
$\phi_1$, $\phi_2$, $q_1$ and $q_2$ have vanishing background charge while $\kappa_1$ and $\kappa_2$ have background charges $+1$ and $-1$ respectively. The two null currents $Z$ and $Y$ can be expressed in terms of the bosons \eqref{phii-kappai-qi-OPEs} as
\begin{align}
   Z &  =-\mfrac{1}{2}\partial \phi_1 + \mfrac{1}{2} \partial \phi_2 + \mfrac{i}{2}\partial q_1-\mfrac{i}{2}\partial q_2 \,,  \\
    Y &  =-\mfrac{1}{2}\partial \phi_1 + \mfrac{1}{2}\partial \phi_2 - \mfrac{i}{2}\partial q_1 + \mfrac{i}{2}\partial q_2 \,. 
\end{align}
The fields $\phi$, $\kappa$, $f_1$ and $f_2$, entering the bosonization of the free-fields of Section~\ref{sec:worldsheet theory}, see eq.~\eqref{new-free-fields-bosonization}, are related to the bosons \eqref{phii-kappai-qi-OPEs} by
\begin{subequations} \label{eq:bosonization-equivalence}
\begin{align}
i \partial f_1 & = -\tfrac{1}{2} \partial \phi_1  - \tfrac{1}{2} \partial \phi_2  - i \partial \kappa_2 + \tfrac{i}{2} \partial  q_1 + \tfrac{i}{2} \partial q_2 \,, \\ 
i \partial f_2 & = \tfrac{1}{2} \partial \phi_1 + \tfrac{1}{2} \partial \phi_2 + i \partial \kappa_2 + \tfrac{i}{2} \partial q_1 + \tfrac{i}{2} \partial q_2 \,, \\
\partial \phi & =   \partial \phi_1 + \partial \phi_2 + i\partial \kappa_2 \,, \\ 
i \partial \kappa & = i \partial \kappa_1\,.
\end{align}
\end{subequations}
Notice that the fields $i \partial f_1 $, $i \partial f_2 $, $\partial \phi$, $i \partial \kappa$ --- and hence the free fields $\beta$, $\gamma$, $p_a$, $\theta^a$ of Section~\ref{sec:worldsheet theory} --- commute with \emph{both} $Z$ and $Y$. This makes it clear that the two null currents $Y$ and $Z$ are now completely decoupled and when expressing vertex operators in terms of the free fields $\beta$, $\gamma$, $p_a$, $\theta^a$, eq.~\eqref{Zn=0} is automatically satisfied. Moreover, since the operator $Q$ in eq.~\eqref{eq:q} carries no $Y$ charge, there is no longer any need to insert in correlation functions the $W$ fields of \cite{Dei:2020zui}.

\subsection*{As a gauge choice}

We can think of the free field realization \eqref{eq:free-field-action} as a `gauge-fixed' form of the free field realization of \cite{Eberhardt:2018ouy,Dei:2020zui}. Specifically, we can group the symplectic bosons and free fermions into vectors
\begin{equation}
\mathcal{Y}=\begin{pmatrix}
\eta^- & \eta^+ & \chi^- & \chi^+
\end{pmatrix}\,,\quad
\mathcal{Z}=\begin{pmatrix}
\xi^+ \\ -\xi^- \\ \psi^+ \\ -\psi^-
\end{pmatrix}\,.
\end{equation}
The action of the theory is
\begin{equation}
S=\frac{1}{2\pi}\int\mathcal{Y}\overline{\partial}\mathcal{Z}\,.
\end{equation}
Then the $Z$ symmetry which needs to be gauged acts on $\mathcal{Y}$ and $\mathcal{Z}$ as
\begin{equation}
\mathcal{Z}\to\alpha\mathcal{Z}\,,\quad\mathcal{Y}\to\alpha^{-1}\mathcal{Y}\,,
\end{equation}
where $\alpha$ is a local function on the worldsheet. Using $\alpha=1/\xi^+$, we find
\begin{equation}
\mathcal{Z}\to
\begin{pmatrix}
1 \\ -\xi^-/\xi^+ \\ \psi^+/\xi^+ \\ -\psi^-/\xi^+
\end{pmatrix}\,,\quad\mathcal{Y}\to
\begin{pmatrix}
\eta^-\xi^+ & \eta^+\xi^+ & \chi^-\xi^+ & \chi^+\xi^+
\end{pmatrix}\,.
\end{equation}
Defining $\gamma=-\xi^-/\xi^+$, $\beta=\eta^+\xi^+$, $\theta^1=\psi^+/\xi^+$, $\theta^2=-\psi^-/\xi^+$, $p_1=\chi^-\xi^+$, and $p_2=\chi^+\xi^+$, the action becomes
\begin{equation}
S=\frac{1}{2\pi}\int\left(\beta\overline{\partial}\gamma+p_a\overline{\partial}\theta^a\right)\,,
\end{equation}
which is the chiral part of the action \eqref{eq:free-field-action}. Indeed, one can show that the above definitions of $\beta,\gamma,p_a,\theta^a$ are consistent with the bosonization eqs.~\eqref{new-free-fields-bosonization}, \eqref{xi-eta-bosonization} and \eqref{eq:bosonization-equivalence}.

The choice of gauge so that $\mathcal{Z}_0=1$ is somewhat worrying, since it is not well-defined everywhere. Specifically, when $\xi^+=0$, the above procedure breaks down. This is signaled in the fact that there are locations in the $\beta,\gamma,p_a,\theta^a$ theory for which $\gamma$ and $\theta^a$ have poles, despite there being no operator inserted there. This technicality is accounted for in the $\beta,\gamma,p_a,\theta^a$ theory via the deformation described in Section \ref{sec:point at infinity}.


\section{The hybrid formalism} 
\label{app:hybrid}
In this appendix, we briefly review the hybrid formalism of superstring on $\text{AdS}_3\times \text{S}^3 \times \mathbb{T}^4$ with pure NS-NS flux. This formalism is developed in \cite{Berkovits:1999im}. For reviews see e.g.\ \cite{Gerigk:2012lqa,Gaberdiel:2022als,Naderi:2022bus}. As we discussed in Section~\ref{sec:worldsheet-theory}, the left-moving part of the theory consists of the following ingredients:
\begin{itemize}
    \item the WZW model on $\mathfrak{psu}(1,1|2)_k$,
    \item $\rho$ and $\sigma$ bosons,
    \item a topologically twisted $\mathbb{T}^4$ theory.
\end{itemize}
The first item in the list is described in details in Section~\ref{sec:new-free-field-realization}. The second consists of two free bosons $\rho$ and $\sigma$ with non-zero background charges, see eqs.~(\ref{eq:rho-sigma}). The fields with the subscript $C$ below denote the generators of the last item, a topologically twisted $\mathcal{N}=4$ algebra of $\mathbb{T}^4$, see Appendix~\ref{app:topologically twisted algebra} for our conventions. These three ingredients give rise to a topologically twisted $\mathcal{N}=4$ algebra on the worldsheet with $c=6$. In fact, the generators have the following form
\begin{subequations} \label{eq:n=4-worldsheet}
\begin{equation}
    T=T_{\text{free}}-\frac{1}{2} [(\partial\rho)^2 + (\partial \sigma)^2] + \frac{3}{2} \partial^2(\rho+i\sigma) + T_C \,,
\end{equation}
\begin{equation}
    G^+ = e^{-\rho} Q + e^{i\sigma} T - \partial(e^{i\sigma} \partial(\rho+iH)) + G^+_C \,,
\end{equation}
\begin{equation}
    G^- = e^{-i\sigma} \,,
\end{equation}
\begin{equation}
    J = \tfrac{1}{2}\partial(\rho+i\sigma+iH) \,,
\end{equation}
\begin{equation}
    J^{\pm\pm}=e^{\pm(\rho+i\sigma+iH)} \,,
\end{equation}
\begin{equation}
    \widetilde{G}^+ = e^{\rho+iH} \,,
\end{equation}
\begin{equation}
    \widetilde{G}^- = e^{-2\rho-i\sigma-i H} Q - e^{-\rho-i H} T - e^{-\rho-i\sigma} \widetilde{G}^-_C + e^{-\rho-iH} [\partial(i\sigma) \partial(\rho+iH)+\partial^2(\rho+iH)] \,.
\end{equation}
\end{subequations}
Note that this theory is topologically twisted. For example, $T$ is a primary of weight $2$. However, the central charge still shows itself in the algebra, for example, we have
\begin{equation}
    J(z) J(w) = \frac{c/12}{(z-w)^2} \,.
\end{equation}
Also note that this is the algebra before the similarity transformation of \cite{Berkovits:1999im}, see \cite{Gaberdiel:2021njm,Naderi:2022bus} for more comments on this.

Finally in order to define pure NS-NS flux string theory on $\text{AdS}_3 \times \text{S}^3 \times \mathbb{T}^4$, we need a realization of $\mathfrak{psu}(1,1|2)_k$. As we mentioned in Section~\ref{sec:new-free-field-realization}, we have realized $\mathfrak{psu}(1,1|2)_1$ in terms of $(\beta,\gamma)$ and $(p_a,\theta^a)$ for $a\in \{1,2\}$. Therefore, we can write down the worldsheet $\mathcal{N}=4$ algebra whose cohomology describes physical states of string theory on $\text{AdS}_3 \times \text{S}^3 \times \mathbb{T}^4$ with $k=1$, see eq.~\eqref{eq:physical}. Here $T_{\text{free}}$ is the stress-tensor of $\mathfrak{psu}(1,1|2)_1$ defined in eq.~(\ref{eq:system-stress-tensor}) with central charge $c=-2$. Note that we are using the result of eq.~(\ref{eq:t-free-t-psu}). $Q$ reads
\begin{equation} \label{eq:q}
    Q = p_1 p_2 \partial \gamma \,.
\end{equation}
What is called the $P$ operator in \cite{Berkovits:1999im,Gaberdiel:2022als} is proportional to
\begin{equation}
    P(z) = V(P_{-4} \ket{0},z) = \alpha V(S^{+++}_{-1} S^{+-+}_{-1} S^{-++}_{-1} S^{--+}_{-1} \ket{0},z) \,,
\end{equation}
where $\ket{0}$ is the vacuum and $\alpha$ is some real number. By a direct calculation, we can confirm that the state in the most right hand side vanishes, so we have $P=0$, as it was the case in \cite{Dei:2020zui}.

\section{Algebras and representations}
\label{app:algebras-and-reps}

In this appendix we list (anti-)commutation relations of various algebras appearing in the main text. 

\subsection*{$\boldsymbol{\mathcal N=4}$ spacetime algebra}

The small $\mathcal N = 4$ algebra with $c=6$ reads (listing only non-zero (anti-)commutators)
\begin{subequations} \label{eq:n=4-commutation-relations}
\begingroup
\allowdisplaybreaks
\begin{align}
[\mathcal L_m , \mathcal L_n] &= (m-n) \, \mathcal L_{m+n} + \tfrac{1}{2} \, \mathcal I \, m (m^2-1) \, \delta_{m+n,0} \,, \\
[ \mathcal L_m , \mathcal G^{\pm}_r] &= (\tfrac{1}{2}m-r) \, \mathcal G^{\pm}_{m+r} \,, \\
[ \mathcal L_m , \widetilde{\mathcal G}^{\pm}_r] &= (\tfrac{1}{2}m-r) \, \widetilde{\mathcal G}^{\pm}_{m+r}  \,, \\
[\mathcal L_m, \mathcal J^{\pm \pm}_n] &= - n \mathcal J^{\pm \pm}_{m+n} \,, \\
[\mathcal L_m, \mathcal J_n] &= - n \mathcal J_{m+n} \,, \\
[\mathcal J_m, \mathcal J_n] &= \tfrac{1}{2}  \, \mathcal I  \, m \, \delta_{m+n,0} \,, \\
[\mathcal J_m, \mathcal J^{\pm \pm}_n] &= \pm \mathcal J^{\pm \pm}_{m+n} \,, \\
[\mathcal J_m, \mathcal G^{\pm}_r] & = \pm \tfrac{1}{2} \, \mathcal G^{\pm}_{m+r} \,, \\
[\mathcal J_m, \widetilde{\mathcal G}^{\pm}_r] & = \pm \tfrac{1}{2} \widetilde{\mathcal G}^{\pm}_{m+r} \,, \\
[\mathcal J^{++}_m, \mathcal J^{--}_n] &= \mathcal I \, m \, \delta_{m+n,0} + 2 \, \mathcal J^3_{m+n} \,,  \\
[\mathcal J^{\pm \pm}_m , \mathcal G^{\mp}_r] & = \pm \widetilde{\mathcal G}^{\pm}_{m+r} \,,  \\
[\mathcal J^{\pm \pm}_m , \widetilde{\mathcal G}^{\mp}_r] & = \mp \mathcal G^{\pm}_{m+r} \,, \\
\{ \mathcal G^+_r, \mathcal G^-_s \} &= (r^2 -\tfrac{1}{4}) \, \delta_{r+s,0} \, \mathcal I + (r-s) \, \mathcal J_{r+s} + \mathcal L_{r+s}  \,,  \label{G+-G--spacetime-N=4}\\
\{ \widetilde{\mathcal G}^+_r, \widetilde{\mathcal G}^-_s \} &= (r^2 -\tfrac{1}{4}) \, \delta_{r+s,0} \, \mathcal I + (r-s) \, \mathcal J_{r+s} + \mathcal L_{r+s}  \,,\\
\{G^{\pm}_r,\widetilde{G}^{\pm}_s\} &= \mp (r-s) J^{\pm\pm}_{r+s} \,.
\end{align}
\endgroup
\end{subequations}

\subsection*{\boldmath Topologically twisted $\mathcal N = 4$ algebra on $\mathbb T^{\boldsymbol 4}$}
\label{app:topologically twisted algebra}

The theory on $\mathbb{T}^4$ is generated by $4$ free bosons and $4$ free fermions satisfying
\begin{subequations} \label{eq:bos-ferm-t4}
\begin{align}
[\partial \mathcal X^i_n , \partial \bar{\mathcal X}^j_m] & = n \, \delta_{ij} \, \mathcal I \, \delta_{n+m,0} \,,  \\
\{ \Psi^{\alpha, j}_r, \Psi^{\beta, l}_s \} & = \epsilon^{\alpha \beta} \, \epsilon^{l j} \, \mathcal I \, \delta_{r+s,0} \,,
\end{align}  
\end{subequations}
where $\mathcal{I}$ is the identity, $j\in\{1,2\}$ and $\alpha,\beta\in\{\pm\}$. Also, note that we have $\epsilon^{+-}=-\epsilon^{-+}=1$ and $\epsilon^{12}=-\epsilon^{21}=1$ while the other combinations vanish. These fields form an $\mathcal{N}=4$ algebra with $c=6$ via the following generators
\begin{subequations}\label{eq:t4-n=4}
\begingroup
\allowdisplaybreaks
   \begin{align}
    T &= \partial \mathcal X^j \partial \bar{\mathcal X}^j + \frac{1}{2} \epsilon^{\alpha\beta} \epsilon^{jl} \Psi^{\alpha,j} \partial \Psi^{\beta,l}  \,, \\
    J &= -\tfrac{1}{4} \delta^{\alpha,-\beta} \epsilon^{jl} \Psi^{\alpha,j} \Psi^{\beta,l} \,, \\
    J^{++} & = \Psi^{+,1} \Psi^{+,2} \,, \\
    J^{--} &= \Psi^{-,2} \Psi^{-,1} \,, \\
    G^+ &= \partial \bar{\mathcal X}^j \Psi^{+,j} \,, \\
    G^- &= - \epsilon^{jk} \partial \mathcal X^j \Psi^{-,k} \,, \\
    \widetilde{G}^+ &= -\epsilon^{jk} \partial \mathcal X^j \Psi^{+,k} \,, \\
    \widetilde{G}^- &= - \partial \bar{\mathcal X}^j \Psi^{-,j} \,.
\end{align} 
\endgroup
\end{subequations}
The topological twist demands setting
\begin{equation} \label{eq:tc}
    T_C = T + \partial J \,.
\end{equation}
Then the central term in $T_C$ is absent. In the worldsheet description (see Appendix~\ref{app:hybrid}), the topologically twisted $\mathbb{T}^4$ appears and we denote its generators with a subscript $C$. In fact, the generators are exactly the same as in eqs.~(\ref{eq:t4-n=4}) with the only difference being that $T$ is replaced by $T_C$ in eq.~(\ref{eq:tc}). In this theory, the conformal dimensions are shifted according to the charge under $J_C$: $J^{++}_C$ has weight $0$, $G^{+}_C$ and $\widetilde{G}^+_C$ have weight $1$, and $J^{--}_C$, $G^-_C$ and $\widetilde{G}^{-}_C$ have weight $2$. In fact, they satisfy the (anti-)commutation relations listed in eqs.~(\ref{eq:n=4-commutation-relations}) with the following modifications\footnote{The commutation relations with $\mathcal{L}_n$ that only get shifted by the conformal dimensions (as we just discussed the shift above) are omitted for brevity.}
\begin{subequations}
\begin{align}
[(\mathcal L_C)_m , (\mathcal L_C)_n] &= (m-n) \, (\mathcal L_C)_{m+n} \,, \\
[(\mathcal L_C)_m, (\mathcal J_C)_n] &= - n (\mathcal J_C)_n -\tfrac{1}{2} m (m+1) \delta_{n+m,0} \mathcal{I} \,, \\
\{ (\mathcal G^+_C)_m, (\mathcal G^-_C)_n \} &= m(m-1) \, \delta_{r+s,0} \, \mathcal I + 2 m \, (\mathcal J_C)_{m+n} + (\mathcal L_C)_{m+n}  \,,  \\
\{ (\widetilde{\mathcal{G}}^+_C)_m, (\widetilde{\mathcal{G}}^-_C)_n \} &= m(m-1) \, \delta_{r+s,0} \, \mathcal I + 2 m \, (\mathcal J_C)_{m+n} + (\mathcal L_C)_{m+n}  \,.
\end{align}
\end{subequations}
In the main text, we have also used a bosonization of the fermions of the topologically twisted $\mathbb{T}^4$ for which we have
\begin{equation}
    \Psi^{+,1} = e^{iH^1} \ , \quad \Psi^{+,2} = e^{i H^2} \ , \quad \Psi^{-,1} = e^{-iH^2} \ , \quad \Psi^{-,2} = - e^{-iH^1} \,.
\end{equation}
In our conventions, 
\begin{equation}
    H^j(z) H^k(w) \sim -\ln{(z-w)} \ , \quad j,k\in\{1,2\} \,.
\end{equation}
In this language, for the Cartan generator of the $\mathcal{R}$-symmetry currents we get
\begin{equation}
    J_C = \tfrac{1}{2} \partial(iH) \ , \quad H = H^1+H^2 \,. 
\end{equation}
Note that we have
\begin{equation}
    H(z) H(w) \sim -2 \ln{(z-w)} \ .
\end{equation}

\subsection*{\boldmath $\mathfrak{psu}(1,1|2)_1$ and $\mathfrak{u}(1,1|2)_1$ algebras on the worldsheet}
\label{app:psu-u-algebras}

In our conventions the $\mathfrak{psu}(1,1|2)_1$ current algebra reads
\begin{subequations}
\begingroup
\allowdisplaybreaks
\begin{align}
    [J^3_m, J^3_n] &= -\tfrac{1}{2} m \delta_{m+n,0} \,, \\
    [J^3_m, J^\pm_n] &= \pm J^\pm_{m+n}  \,, \\
    [J^+_m, J^-_n] &= m \delta_{m+n, 0} - 2 J^3_{m+n} \,, \\
    [K^3_m, K^3_n] &= \tfrac{1}{2} m \delta_{m+n,0} \,, \\
    [K^3_m, K^\pm_n] &= \pm K^\pm_{m+n}  \,, \\
    [K^+_m, K^-_n] &= m \delta_{m+n, 0} + 2 K^3_{m+n} \,, \\
    [J^a_m, S_n^{\alpha \beta  \gamma}] &= \tfrac{1}{2} c_a (\sigma^a)^\alpha{}_\mu S^{\mu \beta \gamma}_{m+n} \,, \\
     [K^a_m, S_n^{\alpha \beta  \gamma}] &= \tfrac{1}{2} (\sigma^a)^\beta{}_\mu S^{\alpha \nu \gamma}_{m+n} \,, \\
     \{S_m^{\alpha \beta \gamma}, S_n^{\mu \nu \rho}\} &=- m \epsilon^{\alpha \mu} \epsilon^{\beta \nu} \epsilon^{\gamma \rho} \delta_{m+n,0} + \epsilon^{\beta \nu} \epsilon^{\gamma \rho} c_a (\sigma_a)^{\alpha \mu}J^a_{m+n} - \epsilon^{\alpha \mu} \epsilon^{\gamma \rho} (\sigma_a)^{\beta \nu} K^a_{m+n} \,. \label{SS-psu(1,1|2)}
\end{align}
\endgroup
\label{psu(1,1|2)}%
\end{subequations}
We reserved Greek letters $\alpha, \beta, \dots$ etc.~for spinor indices, taking values in $\{ +,-\}$. The adjoint index $a$ takes values in $\{+, -, 3 \}$. The constant $c_a$ reads
\begin{equation}
    c_{-} = -1 \,, \qquad  c_{3} = 1 \,, \qquad  c_{+} = 1 \,. 
\end{equation}
and $\epsilon^{+-} = -\epsilon^{-+}=1$. The non-vanishing entries of the $\sigma$-matrices are 
\begin{align}
    (\sigma^-)^+{}_- & = 2 \,, &  (\sigma^3)^-{}_- & = -1 \,,   &  (\sigma^3)^+{}_+ & = 1 \,,  &  (\sigma^+)^-{}_+ & = 2 \,,  \\
    (\sigma_-)^{--} & = 1 \,, &  (\sigma_3)^{-+} & = 1 \,,   &  (\sigma_3)^{+-} & = 1 \,,  &  (\sigma_+)^{++} & = -1 \,,  \\
    (\sigma^-)_{--} & = 2 \,, &  (\sigma^3)_{+-} & = 1 \,,   &  (\sigma^3)_{-+} & = 1 \,,  &  (\sigma^+)_{++} & = -2 \,. 
\end{align}
The $\mathfrak{u}(1,1|2)_1$ algebra has commutation relations similar to \eqref{psu(1,1|2)},
\begin{subequations}
\begingroup
\allowdisplaybreaks
\begin{align}
    [J^3_m, J^3_n] &= -\tfrac{1}{2} m \delta_{m+n,0} \,, \\
    [J^3_m, J^\pm_n] &= \pm J^\pm_{m+n}  \,, \\
    [J^+_m, J^-_n] &= m \delta_{m+n, 0} - 2 J^3_{m+n} \,, \\
    [K^3_m, K^3_n] &= \tfrac{1}{2} m \delta_{m+n,0} \,, \\
    [K^3_m, K^\pm_n] &= \pm K^\pm_{m+n}  \,, \\
    [K^+_m, K^-_n] &= m \delta_{m+n, 0} + 2 K^3_{m+n} \,, \\
    [J^a_m, S_n^{\alpha \beta  \gamma}] &= \tfrac{1}{2} c_a (\sigma^a)^\alpha{}_\mu S^{\mu \beta \gamma}_{m+n} \,, \\
     [K^a_m, S_n^{\alpha \beta  \gamma}] &= \tfrac{1}{2} (\sigma^a)^\beta{}_\mu S^{\alpha \nu \gamma}_{m+n} \,, \\
     \{S_m^{\alpha \beta \gamma}, S_n^{\mu \nu \rho}\} &=- m \epsilon^{\alpha \mu} \epsilon^{\beta \nu} \epsilon^{\gamma \rho} \delta_{m+n,0} + \epsilon^{\beta \nu} \epsilon^{\gamma \rho} c_a (\sigma_a)^{\alpha \mu}J^a_{m+n} \nonumber \\
     & \hspace{110pt} - \epsilon^{\alpha \mu} \epsilon^{\gamma \rho} (\sigma_a)^{\beta \nu} K^a_{m+n} - \epsilon^{\alpha \mu} \epsilon^{\beta \nu} \delta^{\gamma, - \rho} Z_{m+n} \,, \label{SS-u(1,1|2)} \\
     [Z_m, Y_n] & = -m \delta_{m+n,0} \,.
\end{align}
\endgroup
\label{u(1,1|2)}%
\end{subequations}
Notice the two additional generators $Y_n$ and $Z_m$ and compare eq.~\eqref{SS-u(1,1|2)} with \eqref{SS-psu(1,1|2)}.

\paragraph{The short representation of $ \mathfrak{psu}\boldsymbol{(1,1|2)_{1}}$} Labelling the states in the short representation of $\mathfrak{psu}(1,1|2)_1$ as 
\begin{align}
    & \ket{m, \uparrow, 0} \,,  \ \ket{m, \downarrow, 0} \in (\mathcal C_\lambda^{\frac{1}{2}}, \boldsymbol{2}) \,, \\
    & \ket{m, 0 ,\uparrow } \in (\mathcal C_{\lambda+\frac{1}{2}}^0, \boldsymbol{1}) \,, \quad \ket{m, 0 ,\downarrow } \in (\mathcal C_{\lambda+\frac{1}{2}}^1, \boldsymbol{1}) \,,
\end{align}
in our conventions the bosonic generators act as
\begin{subequations}
\begingroup
\allowdisplaybreaks
\begin{align}
    J^3_0 \ket{m, \updownarrow, 0} &= m \ket{m, \updownarrow, 0} \,, &   J^3_0 \ket{m, \updownarrow, 0} &= m \ket{m, 0, \updownarrow} \,, \\
J^+_0 \ket{m, \updownarrow, 0} &= (m+\tfrac{1}{2}) \ket{m+1, \updownarrow, 0} \,, &   J^+_0 \ket{m, 0, \uparrow} & = m \ket{m+1, 0, \uparrow} \,, \\    
J^+_0 \ket{m, 0, \downarrow} &= (m+1) \ket{m+1, 0,  \downarrow} \,, &   J^-_0 \ket{m, \updownarrow, 0}&= (m-\tfrac{1}{2}) \ket{m-1, \updownarrow, 0} \,, \\    
J^-_0 \ket{m, 0, \uparrow} &= m \ket{m-1, 0,  \uparrow} \,, &   J^-_0 \ket{m, 0, \downarrow}&= (m-1) \ket{m-1, 0, \downarrow} \,, \\  
K^3_0 \ket{m, \uparrow, 0} &= \tfrac{1}{2} \ket{m, \uparrow, 0} \,, &   K^3_0 \ket{m, \downarrow, 0} &= -\tfrac{1}{2} \ket{m, \downarrow, 0} \,, \\
K^+_0 \ket{m, \downarrow, 0} & = \ket{m, \uparrow, 0} \,, &   K^-_0 \ket{m, \uparrow, 0} &=  \ket{m, \downarrow, 0} \,. 
\end{align}
\endgroup
For the supercharges we have
\begingroup
\allowdisplaybreaks
\begin{align}
    S_0^{---}\ket{m, \uparrow, 0} & = -(m-\tfrac{1}{2}) \ket{m -\tfrac{1}{2}, 0, \downarrow} \,, & S_0^{---}\ket{m, 0,  \uparrow} & = -m \ket{m -\tfrac{1}{2}, \downarrow, 0} \,,  \\
    S_0^{+--}\ket{m, \uparrow, 0} &= (m+\tfrac{1}{2}) \ket{m +\tfrac{1}{2}, 0, \downarrow} \,, & S_0^{+--}\ket{m, 0, \uparrow} &= m \ket{m +\tfrac{1}{2}, \downarrow, 0} \,, \\
    S_0^{--+}\ket{m, \uparrow, 0} & = - \ket{m -\tfrac{1}{2}, 0, \uparrow} \,, & S_0^{--+}\ket{m, 0, \downarrow} & = \ket{m -\tfrac{1}{2}, \downarrow, 0} \,,  \\
    S_0^{+-+}\ket{m, \uparrow, 0} &= \ket{m +\tfrac{1}{2}, 0, \uparrow} \,, & S_0^{+-+}\ket{m, 0, \downarrow} &= -\ket{m +\tfrac{1}{2},  \downarrow, 0} \,  \\
    S_0^{-+-}\ket{m, \downarrow, 0} & = (m -\tfrac{1}{2})\ket{m -\tfrac{1}{2}, 0, \downarrow} \,, & S_0^{-+-}\ket{m, 0, \uparrow} & = -m \ket{m -\tfrac{1}{2}, \uparrow, 0} \,, \\
    S_0^{++-}\ket{m, \downarrow, 0} &= -(m +\tfrac{1}{2})\ket{m +\tfrac{1}{2}, 0, \downarrow} \,,  & S_0^{++-}\ket{m, 0, \uparrow} &= m \ket{m +\tfrac{1}{2}, \uparrow, 0} \,, \\
    S_0^{-++}\ket{m, \downarrow, 0} & = \ket{m -\tfrac{1}{2}, 0, \uparrow} \,, & S_0^{-++}\ket{m, 0, \downarrow} & = \ket{m -\tfrac{1}{2}, \uparrow, 0} \,,  \\
    S_0^{+++}\ket{m, \downarrow, 0} &= -\ket{m +\tfrac{1}{2}, 0, \uparrow} \,, & S_0^{+++}\ket{m, 0, \downarrow} &= -\ket{m +\tfrac{1}{2}, \uparrow, 0} \,,
\end{align}
\endgroup
\label{psu-rep}
\end{subequations}
while all other actions are zero. 

\bibliography{references.bib}
\bibliographystyle{utphys.bst}

\end{document}